\documentclass[a4paper,10pt]{article}
\pdfoutput=1 

\usepackage{jheppub} 
\usepackage[dvipsnames]{xcolor}
\usepackage[T1]{fontenc} 
\usepackage{braket}
\usepackage{nicematrix}
\usepackage{cancel}
\usepackage{graphicx}	
\usepackage[english]{babel}
\usepackage{amsmath,amsfonts,amsthm,amssymb, psfrag} 
\usepackage{tabularx,tikz}
\usetikzlibrary{knots}
\usepackage[fleqn,tbtags]{mathtools}
\usepackage{blkarray}
\usepackage{tikz,tikz-cd}
\usepackage{hyperref}
\usepackage{pgfplots}
\usepackage{colortbl}

\allowdisplaybreaks[1]

\newcommand{\bal}{\begin{equation}\begin{aligned}}
\newcommand{\eal}{\end{aligned} \end{equation}}

\newcommand{\EE}{\varepsilon_4}
\newcommand{\PP}{e^{i\mathcal{P}_4}}
\newcommand{\invPP}{e^{-i\mathcal{P}_4}}

\newcommand{\EEt}{\varepsilon_3}
\newcommand{\PPt}{e^{i\mathcal{P}_3}}
\newcommand{\invPPt}{e^{-i\mathcal{P}_3}}

\newcommand{\cP}{{\mathcal{P}_4}}
\newcommand{\cPt}{{\mathcal{P}_3}}

\def\be{\begin{equation}}
\def\ee{\end{equation}}
\def\bea{\begin{eqnarray}}
\def\eea{\end{eqnarray}}

\newcommand{\Cset}{{\,\,{{{^{_{\pmb{\mid}}}}\kern-.47em{\mathrm C}}}}}

\title{Long-range to the Rescue of Yang-Baxter II}

\author[a]{Deniz N. Bozkurt}
\author[a]{Juan Miguel Nieto Garc\'ia}
\author[b]{Ziwen Kong}
\author[b]{Elli Pomoni}

\affiliation[a]{II. Institut f\"ur Theoretische Physik, Universit\"at Hamburg, Luruper Chaussee 149,
22607 Hamburg, Germany}
\affiliation[b]{Deutsches Elektronen-Synchrotron DESY, Notkestr. 85, 22607 Hamburg, Germany}

\preprint{DESY 25-083, ZMP-HH/25-10}

\abstract{ 

\vspace*{0.5cm}

We study the spin chain model capturing the one-loop spectral problem of the simplest $\mathcal{N}=2$ superconformal quiver gauge theory in four dimensions, obtained from a marginal deformation of the $\mathbb{Z}_2$ orbifold of $\mathcal{N}=4$ SYM. In Part I of this work \cite{Bozkurt:2024tpz}, we solved for the three-magnon eigenvector and found that it exhibits long-range behavior, despite the Hamiltonian being of nearest-neighbor type.
In this paper, we extend the analysis to the four-magnon sector and construct explicit eigenvectors.
These solutions are compatible with both untwisted and twisted periodic boundary conditions, and they allow for the computation of anomalous dimensions of single-trace operators of the gauge theory. We validate our results by direct comparison with brute-force diagonalization of the spin chain Hamiltonian.
Additionally, we uncover a novel structural relation between eigenstates with different numbers of excitations. In particular, we show that the four-magnon eigenstates can be written in terms of the three-magnon solution, revealing a recursive pattern and hinting at a deeper underlying structure. Lastly, the four-magnon solution obeys an infinite tower of Yang-Baxter equations, as was the case for the three-magnon solution.

}
\begin{document} 
\maketitle
\flushbottom
\section{Introduction}
Integrability has played an essential role in advancing our understanding of the AdS/CFT correspondence, especially in the canonical example of the duality between four-dimensional $\mathcal{N}=4$ Super Yang-Mills (SYM) theory and Type IIB string theory on $AdS_5 \times S^5$ \cite{Beisert:2010jr}. In the planar limit, early studies showed that the one-loop dilatation operator in the scalar sector coincides with the Hamiltonian of an integrable SO(6) spin chain, providing the first clear manifestation of integrability in this framework \cite{Minahan:2002ve}.

A central open question in this line of research is whether four-dimensional gauge theories with less-than-maximal supersymmetry can still be described and solved via integrable spin chains in the planar limit \cite{Korchemsky:2010kj,Zoubos:2010kh}. The next natural step beyond $\mathcal{N}=4$ SYM involves its orbifolds, which, at the orbifold point of the conformal manifold, lead to integrable superconformal field theories (SCFTs) \cite{Beisert:2005he}. 
 Further marginal deformations away from this point give rise to quiver SCFTs, which span a significant portion of the landscape of Lagrangian SCFTs with known gravity duals \cite{Kachru:1998ys,Lawrence:1998ja}.
In this work, we focus on the simplest such $\mathcal{N}=2$ SCFT: the marginally deformed $\mathbb{Z}_2$ orbifold of $\mathcal{N}=4$ SYM, which has been the subject of extensive study~\cite{Gadde:2009dj, Gadde:2010zi,Liendo:2011xb,Gadde:2010ku,Pomoni:2011jj,Gadde:2012rv,Pomoni:2013poa,Mitev:2014yba,Mitev:2015oty,Niarchos:2019onf,Pomoni:2019oib,Niarchos:2020nxk,Pomoni:2021pbj,Galvagno:2020cgq,Galvagno:2021bbj,Gaberdiel:2022iot}. This theory interpolates between the $\mathbb{Z}_2$ orbifold of $\mathcal{N}=4$ SYM—an integrable theory with a well-understood AdS dual—and $\mathcal{N}=2$ superconformal QCD (SCQCD), conjectured to be dual to a non-critical string theory~\cite{Gadde:2009dj} (see also~\cite{Dei:2024frl}).

The spin chain model capturing the one-loop spectral problem of this $\mathcal{N}=2$ superconformal field theory was constructed and analyzed in~\cite{Gadde:2010zi,Liendo:2011xb}. The apparent absence of a Yang--Baxter equation (YBE) for the scattering coefficients in the scalar subsectors initially suggested that the model was non-integrable, rendering the computation of anomalous dimensions for single-trace operators seemingly intractable within an integrability-based framework. However, while the model is not integrable in the conventional sense, recent developments have uncovered hidden structures~\cite{Pomoni:2021pbj,Bertle:2024djm} indicating that it may belong to a novel class of integrable systems governed by a \emph{dynamical} Yang--Baxter equation (DYBE)~\cite{Felder:1994be,Felder:1996xym}, specifically of the type studied in~\cite{Felder:2020tct,Ren:2023mtn}. Importantly, the model is neither rational nor trigonometric;  it was shown to exhibit elliptic dependence, a $\mathbb{Z}_2$-dynamical structure, and an underlying algebroid symmetry---placing it outside the standard paradigms of integrability.

Insights into this structure came from~\cite{Pomoni:2021pbj}, where a quasi-Hopf symmetry algebra was uncovered in the quantum plane limit, defined by an $R$-matrix directly extracted from the Lagrangian of the quiver gauge theory. Remarkably, the R-symmetry generators broken by the orbifold projection are not lost, but rather persist as quantum generators of a novel deformed symmetry algebra. Moreover, the resulting spin chains are \emph{dynamical}, in the sense that their Hamiltonians depend on parameters that vary along the chain. The one-loop holomorphic $\text{SU}(3)$ scalar sector was shown to map to a dynamical 15-vertex model---an RSOS (Restricted Solid-on-Solid) model---whose adjacency graph is determined by the gauge theory's quiver diagram. In particular, one $\text{SU}(2)$ scalar subsector gives rise to an alternating (dynamical XXX) nearest-neighbor Hamiltonian, while another choice of $\text{SU}(2)$ subsector corresponds to a dynamical dilute Temperley--Lieb model.

Furthermore, in~\cite{Bertle:2024djm},
a mathematically elegant description of
the Hilbert space of spin chain states was uncovered for models arising from orbifolds of $\mathcal{N}=4$ SYM.
This vector space
is not a simple tensor product,
as for conventional spin chain models,
due to the different gauge group representations of the fundamental fields. Instead, the spin chain states were interpreted as elements of a path groupoid associated with the quiver diagram. This perspective elucidates the challenges in applying the standard coordinate Bethe ansatz (CBA). Moreover, together with the realization that the SU(4) R-symmetry of $\mathcal{N}=4$ SYM is not broken but repackaged into an SU(4) algebroid~\cite{Bertle:2024djm}, these results provide further motivation to explore the role of elliptic quantum groupoids~\cite{konnobook,Felder:2020tct,Ren:2023mtn} in understanding the integrable structure of the deformed theory.

Dynamical elliptic models differ fundamentally from rational and trigonometric integrable systems because the standard YBE is replaced by a DYBE, which involves non-trivial shifts with respect to a dynamical parameter distinct from the rapidities. While the connection between elliptic quantum groups and elliptic integrable models—particularly through their realization via RSOS models—has been extensively explored in various contexts~\cite{konnobook,Felder:1994be,Felder:1996xym,Felder:2020tct,Ren:2023mtn}, a solution of the eigenvalue problem in the spirit of the coordinate Bethe ansatz (CBA) remains unknown for most such systems.\footnote{In this direction the only relevant notable exception we are aware of is the Inozemtsev chain~\cite{Inozemtsev_1995} (see also~\cite{Serban:2013jua}), where a long-range CBA has recently been constructed~\cite{Klabbers:2020osb,Klabbers:2024abs}.} This  gap continues to obscure the structure of the spin chains related to quiver gauge theories under consideration and highlights the need for  new computational approaches.

To begin closing this gap, in Part I of this work~\cite{Bozkurt:2024tpz}, we constructed a long-range generalization of the CBA for the three-magnon problem. After demonstrating that no solution exists within the standard CBA framework, even after using a finite number of contact terms, we introduced a novel ansatz involving an infinite series of position-dependent corrections. These corrections, encoded in generating functions, were determined by solving the eigenvalue equations subject to the model’s symmetries. This approach successfully allowed for the computation of the three-magnon eigenstates in the infinite chain limit. A natural next step is to investigate whether this framework generalizes to higher magnon number and accommodates periodic boundary conditions. In the present work, we take precisely these steps by extending the formalism to the four-magnon sector. Importantly, while three-magnon spin chain states cannot be closed due to the mismatch of color indices, the four-magnon sector permits consistent color contractions, thereby allowing for closed, translationally invariant spin chain configurations.

Accordingly, this paper focuses on the four-magnon spin chain states in the scalar subsector previously studied in~\cite{Bozkurt:2024tpz}. We solve the corresponding eigenvalue problem under certain symmetry and analyticity assumptions dictated by the quiver, and construct a permutationally symmetric solution for the open spin chain. A central question we explore is whether a form of integrability persists in this long-range setting---specifically, whether a finite set of data (namely, the known scattering coefficients and position-dependent corrections from the three-magnon sector) suffices to reconstruct arbitrary eigenstates of the Hamiltonian.

To investigate this, we introduce a novel averaging procedure that can be interpreted as \emph{eliminating} one magnon via \emph{smearing} in an $(M{+}1)$-magnon eigenstate producing an $M$-magnon solution. We apply this limit to the two-, three-, and four-magnon problems, establishing explicit relations among their solutions. In particular, we use it to construct a four-magnon solution in terms of the three-magnon result of~\cite{Bozkurt:2024tpz}. Although extending this method to general $M$-magnon states remains an open and technically demanding challenge, our results for the four-magnon case suggest that such a generalization may be feasible.

To further investigate the presence of integrable structure, we employ the Yang operator formalism~\cite{YangPhysRev.168.1920}, which captures the relations between different permutations of plane wave coefficients across distinct kinematic regimes (i.e., orderings of momenta and positions). In our case, these coefficients are  position-dependent, reflecting the long-range or non-local structure of the wavefunction. As in the three-magnon case, we find that the four-magnon Yang operator satisfies an infinite tower of Yang--Baxter equations.

Finally, a key feature of the four-magnon states is their natural compatibility with periodic boundary conditions—both untwisted and twisted—which establishes a direct connection to single-trace operators in the marginally deformed $\mathbb{Z}_2$ orbifold theory. After analyzing the open-chain solution, we impose boundary conditions, enabling the computation of anomalous dimensions for single-trace operators with four bifundamental scalars. These results are validated through comparison with brute-force diagonalization of the spin chain Hamiltonian.
Our findings demonstrate that our long-range generalization of Bethe ansatz provides a viable framework for computing anomalous dimensions in this setting. 

\bigskip 

This paper is organized as follows:
\begin{itemize}
\item In Section \ref{sec:gaugetheory}, for completeness, we briefly recall previous results on the marginally deformed $\mathbb{Z}_2$ orbifold of $\mathcal{N}=4$ SYM and describe the spin chain arising from the computation of one-loop anomalous dimensions in the planar limit for the $XZ$ sector.
\item In Section \ref{sec:fourmagnon}, we apply the long-range Bethe ansatz to the four-magnon problem and derive the eigenvalue equations in position space. We then encode these equations into generating functions that capture the position-dependent corrections.
\item In Section \ref{sec:special}, following the symmetries of the problem and inspired by the special three-magnon solutions presented in \cite{Bozkurt:2024tpz}, we obtain a permutationally symmetric solution of the four-magnon eigenstate with factorized scattering coefficients.
\item In Section \ref{sec:smearing}, we introduce the smearing limit to establish a connection between open-infinite eigenstates with different numbers of excitations.
\item In Section \ref{sec:specialspecial}, we build on the intuition developed in the previous section and constrain the solution obtained in Section \ref{sec:special} further to present a final form of the four-magnon solution based on the three-magnon solution.
\item In Section \ref{sec:yang}, we apply the Yang operator formalism to the four-magnon solution we obtained in the previous section and draw a connection between the tower of modified YBE obtained for the three-magnon solution to the tower of modified YBE in four magnon solution.
\item In Section \ref{sec:periodic}, we impose periodic boundary conditions and focus on closed finite states. In particular, we apply the periodicity relations to two-magnon and four-magnon states, demonstrating that the long-range Bethe ansatz is compatible with periodic boundary conditions. We compute the eigenstates for length $4$, $5$, and $6$ spin chains, for both twisted and untwisted states.
\end{itemize}
Finally, in Section \ref{sec:conclusion} we conclude and outline future directions. Additionally, in Appendix \ref{sec:three}, we present a summary of the three-magnon solution given in \cite{Bozkurt:2024tpz} for completeness and in Appendix \ref{sec:cijkl} we collect the details of the four-magnon solution given in Section \ref{sec:specialspecial}.

\section{The $\mathbb Z_2$ Orbifold and its Spin Chains}
\label{sec:gaugetheory}
Orbifolding $\mathcal{N}=4$ SYM breaks the R-symmetry from $SU(4)_R$ to $SU(2)_L\times SU(2)_R\times U(1)_r$ while reshaping the field content. The fields of $\mathcal{N}=4$ SYM, originally residing in the adjoint representation of the $SU(2N)$ gauge group,  are projected onto different representations of the $SU(N)_1\times SU(N)_2$ gauge group. Specifically, the three complex adjoint scalars $X$, $Y$, and $Z$ of $\mathcal{N}=4$ SYM split under the orbifold projection as follows,
\begin{align}
 X\rightarrow Q=\begin{pmatrix}
        &Q_{12}\\Q_{21}&
    \end{pmatrix}\;,\quad\quad\quad Y\rightarrow \tilde Q=\begin{pmatrix}
        &\tilde Q_{12}\\\tilde Q_{21}&
    \end{pmatrix}\;,\quad\quad\quad Z\rightarrow  \phi=  \begin{pmatrix}
        \phi_1&\\&\phi_2
    \end{pmatrix}\;.\label{eq:fields}
\end{align}
which we obtained by using the $2N\times2N$ projection matrix
\begin{align}
    \gamma=\begin{pmatrix}
        \mathbf{1}&0\\0&-\mathbf{1}
    \end{pmatrix}\;,\label{eq:gamma}
\end{align}
and identifying each field with its conjugate
\begin{align}
    Z\sim \gamma Z\gamma^{-1}\;,\quad X\sim -\gamma X\gamma^{-1}\;,\quad Y\sim -\gamma Y\gamma^{-1}\;.
\end{align}
Therefore, the $\mathbb Z_2$ orbifold theory contains two adjoint scalar fields, $\phi_1$ and $\phi_2$, in $\textbf{Adj}_1 \times \mathbf{1}_2$ and $\mathbf{1}_1 \times \textbf{Adj}_2$ of the gauge group, respectively. The bifundamental fields $Q_{12}$ and $\tilde{Q}_{12}$ belong to $\square_1 \times \overline{\square}_2$, while $Q_{21}$ and $\tilde{Q}_{21}$ are in the $\overline{\square}_1 \times \square_2$ representation. In this notation, we use the subscripts of the scalar fields to keep track of the color indices of the two gauge groups.

Marginally deforming the $\mathbb Z_2$ orbifold theory away from the orbifold point generates a one-parameter family of SCFTs, parameterized by $\kappa=g_2/g_1$ with the superpotential given by
\begin{align}
    \mathcal{W}_{\mathbb Z_2}=ig_1 \text{tr}_2\left(\tilde Q_{21}\phi_1Q_{12}-Q_{21}\phi_1\tilde Q_{12}\right)-ig_2 \text{tr}_1\left( Q_{12}\phi_2\tilde Q_{21}-\tilde Q_{12}\phi_2Q_{21}\right)\;,
    \label{eq:Z2-superpotential}
\end{align}
without loss of generality, we take $\kappa\in(0,1]$. The resulting theory interpolates between the $\mathbb Z_2$ orbifold ($\kappa=1$) and the $\mathcal{N}=2$ SCQCD ($\kappa=0$). It features three distinct $SU(2)$ subsectors, which arise from the breaking of $SU(4)_R$ symmetry due to orbifolding. These subsectors, named after their $\mathcal{N}=4$ origins as given in \eqref{eq:fields}, are the $XY$-subsector, $XZ$-subsector, and $YZ$-subsector. In the $XY$-subsector, the $SU(2)$ symmetry remains unbroken and coincides with the $SU(2)_R$ symmetry. In contrast, the $XZ$- and $YZ$-subsectors have a broken $SU(2)$ symmetry. Since the breaking is identical in both cases, it suffices to study just one of them. Further details can be found in \cite{Gadde:2010zi,Pomoni:2019oib,Pomoni:2021pbj}.

\begin{figure}[t]
\begin{center}
\includegraphics[width=0.4\linewidth]{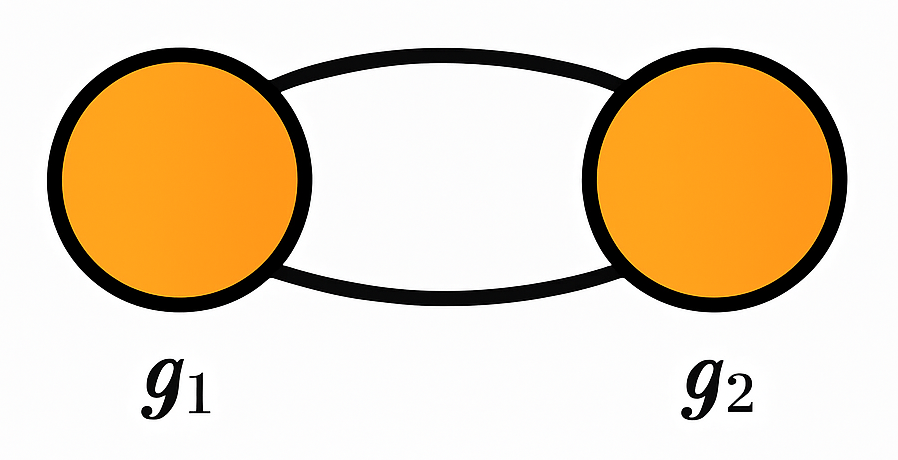}
\end{center}
\caption{\textit{The $\mathbb Z_2$ orbifold theory with $SU(N)_1\times SU(N)_2$}. The $\mathbb Z_2$ symmetry acts by reflecting the quiver across the vertical axis.}
\label{fig:quiver}
\end{figure}

In addition to its R-symmetry group, this theory also possesses a global discrete $\mathbb{Z}_2$ symmetry for all $\kappa$, that exchanges the two $SU(N)$ gauge groups depicted in Figure \ref{fig:quiver}. Along with the color contraction rules, this $\mathbb{Z}_2$ symmetry plays a crucial role in determining the spectrum of single-trace operators in the planar limit.
\subsection{Spin Chain Model}
The spectrum of the single-trace operators corresponds to the eigenvalue problem of a dynamical spin chain model described in \cite{Pomoni:2021pbj}, and further understood to be as a groupoid in \cite{Bertle:2024djm}. In this section, we briefly recall the spin chain model arising from the XZ subsector of this theory. The Hilbert space of these spin chains is constrained by the color contraction rules of scalar fields, which belong to different representations of the $SU(N)_1 \times SU(N)_2$ gauge group. The Hamiltonian of the $XZ$ sector is given by the computation of the one-loop renormalization mixing matrix of single-trace operators in planar limit \cite{Gadde:2010zi}. It is a nearest-neighbor type Hamiltonian,
\begin{align}
        \mathcal{H}=\sum_l \mathcal{H}_{l,l+1}\;,\label{eq:total-hamiltonian}
\end{align}
such that the Hamiltonian density reads
    \begin{equation}
				\label{Hamiltonian}
				\mathcal{H}_{\ell,\ell+1}=\begin{pNiceArray}{cccc|cccc}[first-row,first-col]
					&\phi_1 \phi_1& \phi_1Q_{12} & Q_{12}\phi_2 &Q_{12}Q_{21} & \phi_2\phi_2 & \phi_2Q_{21} &Q_{21}\phi_1 &Q_{21}Q_{12} \\
					\phi_1 \phi_1& 0 & 0 & 0 & 0 & 0 & 0 & 0 & 0 \\
					\phi_1Q_{12} & 0  & 1/\kappa & -1 & 0 & 0 & 0 & 0 & 0 \\
					Q_{12}\phi_2 & 0 & -1 &\kappa &0 & 0 & 0 & 0 & 0 \\
					Q_{12}Q_{21} & 0 & 0 & 0  &0 & 0 & 0 & 0 & 0 \\ \hline
					\phi_2\phi_2 & 0 & 0 & 0 & 0 & 0 &0& 0 & 0 \\
					\phi_2Q_{21} & 0 & 0 & 0 & 0 & 0 & \kappa & -1 & 0 \\
					Q_{21}\phi_1 & 0 & 0 & 0 & 0 & 0 & -1 & 1/\kappa & 0 \\
					Q_{21}Q_{12}  & 0 & 0 & 0 & 0 & 0 & 0 & 0 & 0
                    \end{pNiceArray}\;.
	\end{equation}
 \\ Note that the two-site basis vectors are organized according to the first color index of the two-site states and the Hamiltonian admits a block diagonal form. These two blocks can be exchanged by the transformation of $\kappa\to\kappa^{-1}$. This $\mathbb Z_2$ symmetry is described by the transformation,
\begin{align}
    \mathbb Z_2: \begin{pmatrix}
        \phi_1\\ Q_{12}
    \end{pmatrix}\leftrightarrow\begin{pmatrix}
        \phi_2\\ Q_{21}
    \end{pmatrix}\quad\text{together with}\quad \kappa\leftrightarrow \kappa^{-1}\;,\label{eq:Z2map}
\end{align}
which leads to the commutation relation
\begin{align}
    [\mathbb Z_2,\mathcal{H}]=0\;.\label{eq:z2commute}
\end{align}
The spin chain model inherits this $\mathbb Z_2$ symmetry from the $\mathbb Z_2$ symmetry of the gauge theory given in Figure \ref{fig:quiver}. We use this symmetry as the guiding principle to construct the eigenvectors of both closed and open chain states. Moreover, the Hamiltonian \eqref{eq:total-hamiltonian} commutes with the shift symmetry,
\begin{align}
    [U,\mathcal{H}]=0 \; .\label{eq:shiftcommute}
\end{align}
such that the shift operator maps each one-site state to one step further in the position state,
\begin{align}
    U\ket{\cdots\phi_1\phi_1Q_{12}(l_1)\phi_2\cdots\phi_2Q_{21}(l_2)\phi_1\cdots}=\ket{\cdots\phi_1\phi_1Q_{12}(l_1+1)\phi_2\cdots\phi_2Q_{21}(l_2+1)\phi_1\cdots}
\end{align}
This leads us to a CBA solution for one- and two-magnon states as presented in \cite{Bozkurt:2024tpz}.

Although the parity invariance is broken due to the restrictions coming from the color contractions, there exists another discrete symmetry that plays a similar role
\begin{align}
    [\left(\mathbb Z_2\right)^M\mathbb{P},\mathcal{H}]=0 \;,\label{eq:z2pcommute}
\end{align}
where $M$ is the total number of magnons and $\mathbb{P}$ is a parity operator that takes the color structure into account, and maps a given spin chain state to its reflection with respect to origin while preserving the color indices of $\phi$'s and exchanging $Q_{12}\leftrightarrow Q_{21}$ to ensure the correct order of color indices after reflection.

 The $\kappa\to1$ limit takes us to the orbifold point of the theory, where the Hamiltonian reduces to two copies of the Heisenberg XXX spin chain and the CBA is enough to diagonalize the model. However, for generic values of $\kappa$, the CBA fails to provide a solution and we need to use the long-range Bethe ansatz developed in \cite{Bozkurt:2024tpz}.

We consider the degenerate vacua constructed by the adjoint scalars with quantum numbers $\Delta=r$ and $R=0$ and define,\footnote{Note that the spin chain model admits other vacua, constructed using the bifundamental $Q$s, which was studied in \cite{Pomoni:2021pbj}.}
\begin{align}
\ket{\text{vac}}_{i}=\ket{\cdots\phi_i\phi_i\phi_i\cdots}\;,\quad\quad\text{for}\quad i=1,2\;.\label{eq:vacuum}
 \end{align}
The excitations around these vacua are given in terms of bifundamental $Q$s, which interpolate between the two distinct states. Specifically, the states $\ket{\phi_i}$ before and after the insertion of a magnon $\ket{Q_{ij}}$ must carry different color indices due to color contraction rules dictated by the gauge group representations. We step-by-step diagonalize the Hamiltonian in momentum space. Each eigenstate should have a fixed number of magnons and eigenvalues should be additive in terms of the one-magnon dispersion relation,
\begin{align}
    \mathcal{H}\ket{\Psi(p_1,\cdots,p_M)}=E_M\ket{\Psi(p_1,\cdots,p_M)} \;,
\end{align}
such that
\begin{align}
    E_M=\sum_{i=1}^M E(p_i)\;.
\end{align}
The one- and two-magnon CBA was solved in \cite{Gadde:2010zi}, yielding the dispersion relation for the model,
\begin{align}
    E(p)=\left(\kappa^{1/2}-\kappa^{-1/2}\right)^2+4\sin^2\left(\frac{p}{2}\right)\;.\label{eq:disprel}
\end{align}
The one-magnon solution has the plane wave form typical of the usual CBA solutions,
\begin{align}
    \ket{\Psi(p)}_{ij}=\sum_{l=-\infty}^\infty e^{ip l}\ket{Q_{ij}(l) }\;,\label{eq:one-mag12}
\end{align}
for $(ij)\in\{12,21\}$ and the position state is denoted by the single $Q_{ij}$ magnon, surrounded by $\phi_i$s on the left and $\phi_j$ states on the right. The two-magnon scattering coefficients are given as
\begin{align}
    S_\kappa(p_1,p_2)&=-\frac{a(p_2,p_1,\kappa)}{a(p_1,p_2,\kappa)}\;,\label{eq:defscat}\\
    S_{1/\kappa}(p_1,p_2)&=-\frac{a(p_2,p_1,\kappa^{-1})}{a(p_1,p_2,\kappa^{-1})}\;,\label{eq:defscat2}
\end{align}
such that
\begin{align}
    a(p_1,p_2,\kappa)=1+e^{ip_1+ip_2}-2\kappa e^{ip_1}\;,\label{eq:defa}
\end{align}
which is derived from the two-magnon eigenstate,
\begin{align}
    \ket{\Psi(p_1,p_2)}_{ii}=\sum_{l_1<l_2}\Psi_{ii}(p_1,p_2;l_1,l_2)\ket{Q_{ij}(l_1),Q_{ji}(l_2)}\label{eq:two-magnonCBA}\;,
\end{align}
for $i\in\{1,2\}$ and the wave function under a particular choice of normalization is expressed as
\begin{align}
    &\Psi_{11}(p_1,p_2;l_1,l_2)=(e^{il_1p_1+il_2p_2}+S_{\kappa}(p_1,p_2)e^{il_1p_2+il_2p_1})\;.\label{eq:two-wave1}\\
    &\Psi_{22}(p_1,p_2;l_1,l_2)=(e^{il_1p_1+il_2p_2}+S_{1/\kappa}(p_1,p_2)e^{il_1p_2+il_2p_1})\;.\label{eq:two-wave2}
\end{align}
We observe that $\kappa$ plays a crucial role inside the scattering coefficients depending on the order of $Q_{12}$ and $Q_{21}$ states. Furthermore, the CBA fails for the three-magnon problem, requiring a long-range ansatz to obtain a solution for three-magnon excitations and beyond \cite{Bozkurt:2024tpz}. Here, we apply the long-range Bethe ansatz from \cite{Bozkurt:2024tpz} to solve the four-magnon problem. As in the three-magnon case, we first identify a special solution before imposing periodic/anti-periodic boundary conditions to construct closed, finite spin chain states. This establishes the connection with the interpolating gauge theory. We review one of these special solutions in Appendix \ref{sec:three}.

\section{Four-Magnon Problem with Long-Range Bethe Ansatz}
 \label{sec:fourmagnon}
The four-magnon eigenvalue problem is given by
\begin{align}
    \mathcal{H}\ket{\Psi(p_1,p_2,p_3,p_4)}_{ii}=E_4(p_1,p_2,p_3,p_4)\ket{\Psi(p_1,p_2,p_3,p_4)}_{ii} \;, \label{eq:eig4}
\end{align}
for two configurations labeled by $i\in\{1,2\}$, which denote the color indices of the $\phi$'s in the left and right of the spin chain states. We restrict our search to solutions with additive energy property,
\begin{align}
    E_4(p_1,p_2,p_3,p_4)=\sum_{i=1}^4 E(p_i)\;,\label{en4}
\end{align}
with the dispersion relation given in \eqref{eq:disprel}. For the open infinitely long spin chains, the long-range Bethe ansatz is given as
\begin{align}
    \ket{\Psi(p_1,p_2,p_3,p_4)}_{11}&=\sum_{l_1<l_2<l_3<l_4}\Psi(l_1,l_2,l_3,l_4)\ket{Q_{12}(l_1),Q_{21}(l_2),Q_{12}(l_3),Q_{21}(l_4)} \;, \\
    \Psi(l_1,l_2,l_3,l_4)&=\sum_{\sigma\in\mathcal{S}_4}\left(A_\sigma+D_\sigma^{l_2-l_1-1,l_3-l_2-1,l_4-l_3-1}\right)e^{i \vec{p}_\sigma\cdot\vec{l}} \;, \label{four-magnon-wave}
\end{align}
such that 
\begin{align}
    e^{i \vec{p}_\sigma\cdot\vec{l}}=e^{i(p_{\sigma(1)}l_1+p_{\sigma(2)}l_2+p_{\sigma(3)}l_3+p_{\sigma(4)}l_4)} \;,
\end{align}
together with the $\mathbb Z_2$ conjugate state, $\ket{\Psi(p_1,p_2,p_3,p_4)}_{22}$, whose wave function the same up to the $\kappa\to \kappa^{-1}$ transformation. In this setup, each wave function as defined in \eqref{four-magnon-wave} consists of a sum over all permutations of four positions $\vec{l}$ and four momenta $\vec{p}$, denoted by using the symmetric group with four elements, $\mathcal{S}_4$.\footnote{In this article, when we need to refer to a specific permutation  $\sigma\in\mathcal{S}_4$ that maps $\sigma(1)=i$, $\sigma(2)=j$, $\sigma(3)=k$ and $\sigma(4)=l$, we will denote it as the string $(ijkl)$. This notation differs from the cyclic notation, so it is important to avoid any confusion between the two.} This will be discussed in more detail in Section \ref{sec:parityz2}. As in the three-magnon problem, we separate the scattering coefficients from the position-dependent corrections and adopt a general form that represents the Fourier transform of a function on a lattice. This approach allows us to project the four-magnon eigenvalue problem onto position space and obtain recursion relations for the position-dependent corrections while keeping the contributions from the scattering coefficients distinct. 
\subsection{Infinite Length Eigenvalue Equations}
\label{sec:eigvalueeqs}
We identify five types of equations that require solutions in terms of the scattering coefficients and position-dependent corrections. The shift symmetry of the wave function coming from \eqref{eq:shiftcommute} makes it convenient to switch from the position coordinates to the distance between the magnons to simplify the notation,
\begin{align}
    n=l_2-l_1-1\;,\quad m=l_3-l_2-1\;,\quad r=l_4-l_3-1\;.
\end{align}
These five types of equations correspond to all possible configurations of four magnons relative to each other. When all magnons are well separated, we obtain the \emph{non-interacting equations}. If two magnons are adjacent and the remaining two are far apart, we get \emph{two-magnon interaction equations}, this category includes three distinct cases depending on which pair is interacting. When there are two adjacent pairs that are separated from each other, we refer to them as \emph{two two-magnon interaction equations}. If three magnons are clustered together and the fourth is far apart, the result is \emph{three-magnon interaction equations}. Finally, when all four magnons are adjacent, we have \emph{four-magnon interaction equations}. These are listed as:
\begin{align}
    \text{non-interacting}\quad\quad &n,m,r>0\quad\quad &&\delta_{ijkl}(n,m,r)\nonumber\\
    \text{two-magnon interaction}\quad\quad &n,m>0, r=0\quad\quad &&\gamma_{r,ij}(n,m)\nonumber\\
    &m,r>0, n=0\quad\quad &&\gamma_{l,ij}(m,r)\nonumber\\
    &n,r>0, m=0\quad\quad &&\gamma_{m,ij}(n,m)\nonumber\\
    \text{two two-magnon interaction}\quad\quad &m>0, n,r=0\quad\quad &&\gamma_{rl,ij}(m)\nonumber\\
    \text{three-magnon interaction}\quad\quad &n>0, m,r=0\quad\quad &&\beta_{r,i}(n)\nonumber\\
    &r>0, m,n=0\quad\quad &&\beta_{l,i}(r)\nonumber\\
    \text{four-magnon interaction}\quad\quad &n,m,r=0\quad\quad &&\alpha\nonumber
\end{align}
We now proceed to introduce and analyze each of these cases in detail. While the derivations may appear repetitive, they are crucial for thoroughly addressing the intricacies of the four-magnon eigenvalue problem. Moreover, by exploring the relationships between these equations, shaped by the symmetries of the model, we will uncover special families of solutions.\\\\
\textbf{Non-interacting Equations}\\
We start by looking at the non-interacting equations, resulting from the well-separated magnons,
\begin{align}
    \delta(n,m,r)=&e^{-il_1\cP}\braket{l_1,l_2,l_3,l_4|\mathcal{H}-E_4|\Psi(p_1,p_2,p_3,p_4)}_{12}\nonumber\\
    =&e^{-il_1\cP}\Bigg[\left(4\kappa+4\kappa^{-1}-E_4\right)\Psi(l_1,l_2,l_3,l_4)-\Psi(l_1\pm1,l_2,l_3,l_4)\nonumber\\
    -&\Psi(l_1,l_2\pm1,l_3,l_4)-\Psi(l_1,l_2,l_3\pm1,l_4)-\Psi(l_1,l_2,l_3,l_4\pm1)\Bigg] \;, 
\end{align}
such that all the relative distances are positive integers $n,m,r>0$ and total momentum is given in \eqref{totalmomentum}. The notation $\pm1$ indicates that we have to include the term with $+1$ and $-1$ shifts of the position, as writing all the terms will clutter our equations. After substituting \eqref{four-magnon-wave}, we distinguish the plane wave contributions from the position-dependent corrections to keep track of the permutations of momenta,
\begin{align}
    \delta(n,m,r)=\sum_{\sigma\in\mathcal{S}_4}\delta_\sigma(n,m,r) e^{i n (p_{\sigma(2)}+p_{\sigma(3)}+p_{\sigma(4)})+i m(p_{\sigma(3)}+p_{\sigma(4)})+ir p_{\sigma(4)}} \;.
\end{align}
Then, to solve the non-interacting equations coming from the four-magnon eigenvalue problem, we require each $\delta_\sigma$ to vanish separately for any $n,m,r>0$. We explicitly give one of these expressions, as the rest can be obtained by permuting the indices of the momentum variables.
\begin{align}
    \delta_{1234}(n,m,r)=e^{i(3p_4 +2p_3 +p_2)} \Big( \EE D_{1234}^{n,m,r} -e^{-ip_4} D_{1234}^{n,m,r-1} -e^{ip_4} D_{1234}^{n,m,r+1}-e^{-ip_3} D_{1234}^{n,m-1,r+1} \nonumber\\
 -e^{ip_3} D_{1234}^{n,m+1,r-1}-e^{-ip_2} D_{1234}^{n-1,m+1,r} -e^{ip_2} D_{1234}^{n+1,m-1,r} -e^{-ip_1} D_{1234}^{n+1,m,r} -e^{ip_1} D_{1234}^{n-1,m,r} \Big)\label{non-int-rec}\;,
\end{align}
where we introduced a parameter which is related to the total energy as
\begin{align}
    \EE=4\kappa^{-1} +4\kappa -E_4=\sum_{i=1}^4 e^{ip_i}+e^{-ip_i}\;,\label{totalenergy}
\end{align}
and we denote the total momentum as
\begin{align}
    \PP=\sum_{i=1}^4 p_i\;.\label{totalmomentum}
\end{align}
By scaling the position-dependent corrections, we can acquire a form of the non-interacting equations that only depends on the parameters $\EE$ and the exponential of the total momentum.
\begin{align}
    \tilde D_\sigma^{n,m,r}=e^{-inp_{\sigma(1)}+im(p_{\sigma(3)}+p_{\sigma(4)})+irp_{\sigma(4)}}D_\sigma^{n,m,r}\label{scaledd}\;.
\end{align}
Under this transformation, we obtain,
\begin{align}
    \delta_{1234}(n,m,r)=e^{i(p_2 +2p_3 +3p_4)} \Big( \EE \tilde D_{1234}^{n,m,r} -\tilde D_{1234}^{n,m,r-1} -\tilde D_{1234}^{n,m,r+1}-\tilde D_{1234}^{n,m-1,r+1}\nonumber\\
 -\tilde D_{1234}^{n,m+1,r-1} -\invPP \tilde D_{1234}^{n-1,m+1,r} 
-\PP  \tilde D_{1234}^{n+1,m-1,r} -\tilde D_{1234}^{n+1,m,r} -\tilde D_{1234}^{n-1,m,r} \Big)\;.\label{non-int-recd}
\end{align}
It is crucial to keep in mind that there are other ways to scale position-dependent corrections which would still reduce the momentum-dependent coefficients of the eigenvalue problem to depend only on the total momentum and the total energy. As these other scalings are of the form $\tilde{E}^{n,m,r}_\sigma =e^{- \alpha im \cP} \tilde{D}^{n,m,r}_\sigma$, for $\alpha\in\mathbb R$, these scaled position-dependent corrections would change the position of the $\PP$ and $\invPP$ factors in the non-interacting eigenvalue equations but this would still give us a similar solution. A similar scaling transformation was used in the long-range Bethe ansatz for three-magnon states. In that case, there is only one way to reduce the coefficients of the non-interacting expression, which is to depend solely on the total momentum and total energy. For four-magnon states, it is not surprising to find distinct ways to perform this scaling, reflecting the increased complexity of the system. These scaling choices are reminiscent of the isometries of the three-dimensional lattice that encodes the position-dependent corrections for each permutation, $\sigma$. Furthermore, each scaling transforms the wave function into two distinct pieces. Under the scaling of the position-dependent correction given in \eqref{scaledd} the general structure of the wave function transforms to
\begin{multline}
    \quad\quad\ket{\Psi(p_1,p_2,p_3,p_4)}_{11}=\sum_{l_1<l_2<l_3<l_4}\sum_{\sigma\in\mathcal{S}_4}\Big(A_\sigma e^{i\Vec{p}_\sigma\cdot\Vec{l}}\nonumber\\+e^{ip_{\sigma(2)}+2ip_{\sigma(3)}+3ip_{\sigma(4)}}\tilde D_\sigma^{n,m,r}e^{il_2\cP}\Big)\ket{l_1,l_2,l_3,l_4}_{11}\;.
\end{multline}
The scaling $\tilde E_\sigma$ also transforms the wave function into a similar form. In particular, the scaling from $D_\sigma$ to $\tilde D_\sigma$ isolates all position-dependent corrections into a separate wave function in the center-of-mass limit, along with the scattering coefficient contributions, which remain in the CBA form. This approach significantly simplifies the long-range Bethe ansatz. Therefore, after presenting the eigenvalue equations, we consistently use the scaled position-dependent corrections $\tilde{D}_\sigma$ throughout the paper.\\\\
\textbf{Two-magnon Interaction Equations}\\
 When two of the magnons are adjacent and the remaining two are far apart, the eigenvalue problem, \eqref{eq:eig4} requires the \emph{left two-magnon interaction} equation to vanish,
\begin{align}
    \gamma_l(m,r)=&e^{-il_1\cP}\braket{l_1,l_1+1,l_3,l_4|\mathcal{H}-E_4|\Psi(p_1,p_2,p_3,p_4)}_{12}\nonumber\\
    =&e^{-il_1\cP}\Bigg[\left(2\kappa+4\kappa^{-1}-E_4\right)\Psi(l_1,l_1+1,l_3,l_4)-\Psi(l_1-1,l_1+1,l_3,l_4)\nonumber\\
    -&\Psi(l_1,l_1+2,l_3,l_4)-\Psi(l_1,l_1+1,l_3\pm1,l_4)-\Psi(l_1,l_1+1,l_3,l_4\pm1) \Bigg]\;,
\end{align}
for all $l_3-l_2-1=m>0$ and  $l_4-l_3-1=r>0$. Similarly, the \emph{middle two-magnon interaction} equation is given
\begin{align}
    \gamma_m(n,r)=&e^{-il_1\cP}\braket{l_1,l_2,l_2+1,l_4|\mathcal{H}-E_4|\Psi(p_1,p_2,p_3,p_4)}_{12}\nonumber\\
    =&e^{-il_1\cP}\Bigg[\left(4\kappa+2\kappa^{-1}-E_4\right)\Psi(l_1,l_2,l_2+1,l_4)-\Psi(l_1\pm1,l_2,l_2+1,l_4)\nonumber\\
    -&\Psi(l_1,l_2-1,l_2+1,l_4)-\Psi(l_1,l_2,l_2+2,l_4)-\Psi(l_1,l_2,l_2+1,l_4\pm1)\Bigg] \;,
\end{align}
for all $l_2-l_1-1=n>0$ and $l_4-l_3-1=r>0$, while the \emph{right two-magnon interaction} equation is given as,
\begin{align}
    \gamma_r(n,m)=&e^{-il_1\cP}\braket{l_1,l_2,l_3,l_3+1|\mathcal{H}-E_4|\Psi(p_1,p_2,p_3,p_4)}_{12}\nonumber\\
    =&e^{-il_1\cP}\Bigg[\left(2\kappa+4\kappa^{-1}-E_4\right)\Psi(l_1,l_2,l_3,l_3+1)-\Psi(l_1\pm1,l_2,l_3,l_3+1)\nonumber\\
    -&\Psi(l_1,l_2\pm1,l_3,l_3+1)-\Psi(l_1,l_2,l_3-1,l_3+1)-\Psi(l_1,l_2,l_3,l_3+2)\Bigg] \;,
\end{align}
for $l_2-l_1-1=n>0$ and $l_3-l_2-1=m>0$. Then, we decompose these expressions according to their plane wave factors and simplify the vanishing of the entire expression to the vanishing of each of the coefficients that accompany the plane wave contributions.

The left two-magnon interaction expression decomposes into its components as
\begin{align}
\gamma_l (m,r)&= \sum_{\substack{i,j\in\{1,2,3,4\}\\i<j}} \gamma_{l,ij}(m,r)e^{im(p_i+p_j)+irp_j} \;,
\end{align}
where each $\gamma_{l,ij}$ is obtained by permuting the indices of the following function,
\begin{align}
 \gamma_{l,34}(m,r) =&e^{i(p_2+2 p_3 +3 p_4)} \Big[e^{-ip_2}a(p_{2},p_{1},\kappa) A_{1234} + (\EE-2\kappa) D_{1234}^{0,m,r} \nonumber\\
 -&e^{i p_{2}} D_{1234}^{1,m-1,r} 
 -e^{-i p_{1}} D_{1234}^{1,m,r}-e^{i p_{3}} D_{1234}^{0,m+1,r-1}-e^{-i p_3} D_{1234}^{0,m-1,r+1} \nonumber\\
 -&e^{-i p_{4}} D_{1234}^{0,m,r-1} 
 -e^{i p_{4}} D_{1234}^{0,m,r+1}\Big] +[1\leftrightarrow2] \label{gamma-l-34}\;,
 \end{align}
here the function $a(p_{2},p_{1},\kappa)$ is given in \eqref{eq:defa} and we denote the sum over particular permutations by $1\leftrightarrow2$. For the middle two-magnon interaction, we have,
\begin{align}
\gamma_m (n,r)&= e^{i n \cP} \sum_{\substack{i,j\in\{1,2,3,4\}\\i< j}} \gamma_{m,ij}(n,r)e^{-inp_i+irp_j} \;,
\end{align}
where 
\begin{align}
\gamma_{m,14}(n,r) =&e^{i(p_{2}+2p_{3} +3p_{4})} \Big[ e^{-ip_{3}}a\left(p_{3},p_{2},\kappa^{-1}\right) A_{1234} + (\EE-2\kappa^{-1}) D_{1234}^{n,0,r}\nonumber\\
-&e^{-i p_{4}} D_{1234}^{n,0,r-1} 
-e^{i p_{4}} D_{1234}^{n,0,r+1} -e^{i p_{3}} D_{1234}^{n,1,r-1}-e^{i p_{1}} D_{1234}^{n-1,0,r} \nonumber\\
-&e^{-i p_{2}} D_{1234}^{n-1,1,r}
-e^{-i p_{1}} D_{1234}^{n+1,0,r} \label{gamma-m-14}
\Big]+\left[2\leftrightarrow3\right] \;.
\end{align}
Finally, for the right two-magnon interaction:
\begin{align}
\gamma_r (n,m)&= e^{i n \cP} \sum_{\substack{i,j\in\{1,2,3,4\}\\i< j}} \gamma_{r,ij}(m,r)e^{-inp_i+imp_j} \;,
\end{align}
where
\begin{align}
\gamma_{r,12}(n,m) =&e^{i(p_{2}+2p_{3} +3p_{4})} \Big[e^{-ip_{4}}a(p_{4},p_{3},\kappa) A_{1234} \nonumber\\
+& (\EE-2\kappa)D_{1234}^{n,m,0}
-e^{-ip_{3}} D_{1234}^{n,m-1,1} 
-e^{ip_{4}} D_{1234}^{n,m,1}-e^{ip_{2}} D_{1234}^{n+1,m-1,0}   \nonumber\\
-&e^{-ip_{2}} D_{1234}^{n-1,m+1,0} -e^{-ip_{1}} D_{1234}^{n+1,m,0} -e^{ip_{1}} D_{1234}^{n-1,m,0}\Big] + [3 \leftrightarrow 4] \;.\label{gamma-r-12}
\end{align}
In the next section, we will explore the symmetries of the four-magnon states. These symmetries enable us to relate the distinct two-magnon interaction equations through a map that combines parity and $\mathbb Z_2$ transformations.\\\\
\textbf{Two Two-magnon Interaction Equation}\\
Additionally, we are required to consider the case where the first two and the last two magnons are adjacent but these two interaction pairs are far away from each other. Then, the eigenvalue problem \eqref{eq:eig4} is projected to position space as,
\begin{align}
    \gamma_{rl}(m)=&e^{-il_1\cP}\braket{l_1,l_1+1,l_3,l_3+1|\mathcal{H}-E_4|\Psi(p_1,p_2,p_3,p_4}_{12}\;,\\
    &=e^{-il_1\cP}\Bigg[\left(4\kappa^{-1}-E_4\right)\Psi(l_1,l_1+1,l_3,l_3+1)-\Psi(l_1-1,l_1+1,l_3,l_3+1)\nonumber\\
    &-\Psi(l_1,l_1+2,l_3,l_3+1)-\Psi(l_1,l_1+1,l_3-1,l_3+1)-\Psi(l_1,l_1+1,l_3,l_3+2)\Bigg]
\end{align}
such that
\begin{align}
    \gamma_{rl}(m)= \sum_{\substack{i,j\in\{1,2,3,4\}\\i<j}}\gamma_{rl,ij}e^{im (p_i+p_j)}\;.
\end{align}
These expressions take the following form,
\begin{align}
\label{gamma-rl-34}
\gamma_{rl,34}(m)=& e^{i(p_2 +2p_3+ 3p_4)}\Big[ (e^{ip_1} +e^{-ip_2} +e^{ip_3} +e^{-ip_4} -4\kappa) A_{1234}\nonumber\\
+&(\EE-4\kappa) D_{1234}^{0,m,0} -e^{-ip_3} D_{1234}^{0,m-1,1} -e^{ip_4} D_{1234}^{0,m,1} -e^{-ip_1} D_{1234}^{1,m,0}  -e^{ip_2} D_{1234}^{1,m-1,0} \Big]\nonumber\\
&+[1\leftrightarrow2]+[3\leftrightarrow4]+[1\leftrightarrow2,3\leftrightarrow4] \; .
\end{align}
We observe that the particular case of interaction affects the sum over permutation labels, the $\kappa$ factors in the expression, and the combination of position variables. Therefore, each case gives us a different type of equation.
\\\\
\textbf{Three-magnon Interaction Equations}\\
We consider the cases where three magnons are next to each other and the fourth one is far away. In this setting, we obtain two distinct three-magnon interaction equations. First, \emph{the left three-magnon interaction} gives
\begin{align}
    \beta_l(r)=&e^{-il_1\cP}\braket{l_1,l_1+1,l_1+2,l_4|\mathcal{H}-E_4|\Psi(p_1,p_2,p_3,p_4)}_{12}\nonumber\\
    =&e^{-il_1\cP}\Bigg[\left(2\kappa+2\kappa^{-1}-E_4\right)\Psi(l_1,l_1+1,l_1+2,l_4)-\Psi(l_1,l_1+1,l_1+2,l_4\pm1)\nonumber\\
    -&\Psi(l_1-1,l_1+1,l_1+2,l_4)-\Psi(l_1,l_1+1,l_1+3,l_4)\Bigg] \;.
\end{align}
such that $l_4-l_1-3=r>0$. Then, the \emph{right three-magnon interaction} which takes the form,
\begin{align}
    \beta_r(n)=&e^{-il_1\cP}\braket{l_1,l_2,l_2+1,l_2+2|\mathcal{H}-E_4|\Psi(p_1,p_2,p_3,p_4)}_{12}\nonumber\\
    =&e^{-il_1\cP}\Bigg[\left(2\kappa+2\kappa^{-1}-E_4\right)\Psi(l_1,l_2,l_3,l_4)-\Psi(l_1,l_2,l_2+1,l_2+3)\nonumber\\
    -&\Psi(l_1\pm1,l_2,l_2+1,l_2+2)-\Psi(l_1,l_2-1,l_2+1,l_2+2)\Bigg]\;,
\end{align}
for $l_2-l_1-1=n>0$. These two expressions explicitly take the following form,
\begin{align}
\beta_{l}(r)= \sum_{i\in\{1,2,3,4\}}\;\beta_{l,i}(r) e^{i p_i r} \;,
\end{align}
where
\begin{align}
\label{eq:beta-l-4}
\beta_{l,4}(r) =& e^{i(p_j+ 2 p_k)+3ip_4} \Big[(e^{i p_1} +e^{i p_2} +e^{-i p_2} + e^{-i p_3} -2\kappa -2\kappa^{-1}) A_{1234}+(\EE-2\kappa-2\kappa^{-1}) D_{1234}^{0,0,r}\nonumber\\
 - &e^{-i p_1}D_{1234}^{1,0,r} -e^{i p_3}D_{1234}^{0,1,r-1} -e^{-i p_4} D_{1234}^{0,0,r-1} -e^{i p_4} D_{1234}^{0,0,r+1} \Big]+[1\leftrightarrow2\leftrightarrow3] \;,
\end{align}
such that we denote the permutation of three permutation indices by $1\leftrightarrow2\leftrightarrow3$, together with
\begin{align}
\beta_r(n) =e^{i n \cP} \sum_{i\in\{1,2,3,4\}}\;\beta_{r,i} e^{-in p_i } \;,
\end{align}
such that
\begin{align}
\label{eq:beta-r-1}
\beta_{r,1}(n) =& e^{i(p_2 +2p_3 +3p_4)} \Big[ (e^{ip_2} +e^{ip_3} +e^{-ip_3} +e^{-ip_4} -2\kappa -2\kappa^{-1}) A_{1234}+ (\EE-2\kappa-2\kappa^{-1}) D_{1234}^{n,0,0}\nonumber\\
 -&e^{-ip_2} D_{1234}^{n-1,1,0} -e^{ip_4} D_{1234}^{n,0,1} -e^{-ip_1} D_{1234}^{n+1,0,0} -e^{ip_1} D_{1234}^{n-1,0,0} \Big]+[2\leftrightarrow3\leftrightarrow4] \;,
\end{align}
these two configurations have the same contribution of $\kappa$ factors and scattering coefficients as in the case of the three-magnon problem. Later, we will show that under the $\mathbb Z_2$ transformation the form of the three-magnon interaction equations stays invariant and parity transforms the left three-magnon interaction equation to the right three-magnon interaction equation.\\\\
\textbf{Four-magnon Interaction Equation}\\
When all four of the magnons are next to each other we obtain the following expression,
\begin{align}
    \alpha=&e^{-il_1\cP }\braket{l_1,l_1+1,l_1+2,l_1+3|\mathcal{H}-E_4|\Psi(p_1,p_2,p_3,p_4)}_{12}\nonumber\\
    =&e^{-il_1\cP}\Big[\left(2\kappa^{-1}-E_4\right)\Psi(l_1,l_1+1,l_1+2,l_1+3)\nonumber\\
    -&\Psi(l_1-1,l_1+1,l_1+2,l_1+3)-\Psi(l_1,l_1+1,l_1+2,l_1+4)\Big] \;.
\end{align}
More explicitly it is given as
\begin{align}
    \alpha=&\sum_{\sigma\in\mathcal{S}_4} e^{i(p_{\sigma(2)} +2p_{\sigma(3)} +3p_{\sigma(4)})}\Bigg( \Big[e^{-ip_{\sigma(2)}}a(p_{\sigma(2)},p_{\sigma(1)},\kappa)+e^{-ip_{\sigma(3)}}a\left(p_{\sigma(3)},p_{\sigma(2)},\kappa^{-1} \right)\nonumber\\ 
    +&e^{-ip_{\sigma(4)}}a(p_{\sigma(4)},p_{\sigma(3)},\kappa)\Big] A_{\sigma}+(\EE-4\kappa-2\kappa^{-1}) D_{\sigma}^{0,0,0}-  e^{-i p_{\sigma(1)}} D_{\sigma}^{1,0,0} - e^{i p_{\sigma(4)}} D_{\sigma}^{0,0,1} \Bigg) \;.\label{fourint}
\end{align}
Before imposing these equations on the generating function to extract solutions, we first examine the symmetry properties of the four-magnon states and their relation to the three-magnon problem in order to gain intuition about the structure of the solutions.

\subsection{Parity and $\mathbb Z_2$}
\label{sec:parityz2}
The two discrete symmetries of the Hamiltonian \eqref{Hamiltonian}, $\mathbb Z_2$ and parity,\footnote{Since we are considering the four-magnon problem ($M=4$), the operator $\mathbb Z_2^M\mathbb{P}$ given in \eqref{eq:z2pcommute} reduces to the parity operator.} given in \eqref{eq:z2commute} and \eqref{eq:z2pcommute} play a crucial role in characterizing the spectrum of the model. Here we analyze the action of these two operators on the four-magnon long-range Bethe ansatz states and present the effect of these operators on the eigenvalue equations presented in the previous section.

As given in \eqref{eq:Z2map}, $\mathbb Z_2$ operator maps two distinct momentum states to each other,
\begin{align}
    \mathbb Z_2\ket{\Psi(p_1,p_2,p_3,p_4)}_{11}=\ket{\Psi(p_1,p_2,p_3,p_4)}_{22} \;.
\end{align}
This implies that the wave function of the state $\ket{\Psi}_{22}$ should be the $\kappa\to\kappa^{-1}$ version of the state $\ket{\Psi}_{11}$ and, more explicitly, we obtain the following relations on the scattering coefficients and the position-dependent corrections,
\begin{align}
    A_\sigma(p,\kappa^{-1})=\hat{A}_{\sigma}(p,\kappa)\;,\quad\quad D_\sigma^{n,m,r}(p,\kappa^{-1})=\hat{D}_\sigma^{n,m,r}(p,\kappa) \;,
\end{align}
where $A_\sigma(p,\kappa)$ and $D_\sigma^{n,m,r}(p,\kappa)$ are the coefficients associated with the state $\ket{\Psi}_{11}$ and $\hat{A}_\sigma(p,\kappa)$ and $\hat{D}_\sigma^{n,m,r}(p,\kappa)$ are the coefficients of state $\ket{\Psi}_{22}$. We observe that the transformation from $\kappa\to\kappa^{-1}$ leaves the form of the non-interacting equations \eqref{non-int-rec} and the three-magnon interaction equations, given in \eqref{eq:beta-l-4} and \eqref{eq:beta-r-1},  invariant
\begin{align}
    \delta(n,m,r)&\xrightarrow[]{\kappa\to\kappa^{-1}}\delta(n,m,r)_{\text{with}\;D_\sigma(p,\kappa^{-1})}\\
    \beta_{l(r)}(n)&\xrightarrow[]{\kappa\to\kappa^{-1}}\beta_{l(r)}(n)_{\text{with}\;D_\sigma(p,\kappa^{-1})}
\end{align}
which was also the case for the three-magnon problem. On the other hand, the coefficients of the two-magnon interaction equations, given in \eqref{gamma-l-34}, \eqref{gamma-m-14} and \eqref{gamma-r-12}, and the four-magnon interaction equation \eqref{fourint} depend on $\kappa$ in a way which is not symmetric under $\kappa\to\kappa^{-1}$, therefore they change under the $\mathbb Z_2$ transformation.

The action of parity cannot only reverse the order of the states by $l_i\to-l_i$. To follow the color contraction rules, additionally, we replace $Q_{12}$ with $Q_{21}$ while preserving the color indices of $\phi$'s in the reflection process. This is a particular definition of the parity, as preserving the color indices of $Q_{12}$ and $Q_{21}$ while changing the colors of $\phi$'s in the reflection process would mix with the action of the $\mathbb Z_2$ operator. Then, the action of parity on the momentum states is given
\begin{align}
    \mathbb{P}&\ket{\Psi(p_1,p_2,p_3,p_4)}_{11}=\sum_{l_1<l_2<l_3<l_4} \Psi(l_1,l_2,l_3,l_4)\ket{-l_4,-l_3,-l_2,-l_1}_{11}\\
    =&\sum_{l_1<l_2<l_3<l_4}\sum_{\sigma\in\mathcal{S}_4}\left(A_\sigma(\vec{p},\kappa)+D_\sigma^{l_2-l_1-1,l_3-l_2-1,l_4-l_3-1}(\vec{p},\kappa)\right)e^{i\Vec{p}_\sigma\cdot\Vec{l}}\ket{-l_4,-l_3,-l_2,-l_1}_{11}\label{eq:action-par}\;,
\end{align}
where $\Vec{p}_\sigma= (p_{\sigma(1)}, p_{\sigma(2)}, p_{\sigma(3)}, p_{\sigma(4)})$. If, for every element $\sigma=(ijkl) \in \mathcal{S}_4$, we introduced the reflected element of the symmetric group $\sigma^r=(lkji)$, and we invert the ordering of the position variables, $-l_{i}=x_{5-i}$, the parity transform of the four-magnon long-range Bethe ansatz \eqref{four-magnon-wave} is written as
\begin{align}
    \mathbb{P}&\ket{\Psi(p_1,p_2,p_3,p_4)}_{11}=\sum_{l_1<l_2<l_3<l_4} \Psi(l_1,l_2,l_3,l_4)\ket{-l_4,-l_3,-l_2,-l_1}_{11}\\
    =&\sum_{x_1<x_2<x_3<x_4} \sum_{\sigma\in\mathcal{S}_4}\left(A_\sigma(\vec{p},\kappa)+D_\sigma^{x_4-x_3-1,x_3-x_2-1,x_2-x_1-1}(\vec{p},\kappa)\right)e^{-i\Vec{p}_{\sigma^r}\cdot\Vec{x}}\ket{x_1,x_2,x_3,x_4}_{11}\;.
\end{align}
Then, we conclude that the four-magnon eigenstate is parity invariant
\begin{align}
    \mathbb{P}\ket{\Psi(p_1,p_2,p_3,p_4)}_{11}=e^{i\varphi(p,\kappa)}\ket{\Psi(-p_1,-p_2,-p_3,-p_4)}_{11}\;.\label{eq:parity-inv}
\end{align}
if the following conditions are satisfied by the scattering coefficients and the position-dependent corrections,
\begin{align}
    e^{i\varphi(\vec{p},\kappa)}A_{\sigma^r}(-\vec{p},\kappa)&=A_\sigma(\vec{p},\kappa)\;,\label{eq:parity-cond1}\\
    e^{i\varphi(p,\kappa)}D_{\sigma^r}^{n,m,r}(-\vec{p},\kappa)&=D_\sigma^{r,m,n}(\vec{p},\kappa)\label{eq:parity-cond2}\;.
\end{align}
Here, $e^{\varphi(\vec{p},\kappa)}$ is a phase factor that is defined with respect to the scattering coefficients and the position-dependent corrections. The action of parity can be summarized in terms of a map defined by the scattering coefficients and the position-dependent corrections,
\begin{align}
    \{p_1,p_2,p_3,p_4,\kappa,A_\sigma,D^{n,m,r}_\sigma\}\rightarrow\{-p_1,-p_2,-p_3,-p_4,\kappa,A_{\sigma^r},D^{r,m,n}_{\sigma^r}\} \;.\label{eq:parity-map}
\end{align}
This formulation of the parity transformation does not explicitly account for how the dependence of the scattering coefficients and position-dependent corrections on momentum variables changes when the momentum variables are flipped in sign, as it is given in \eqref{eq:parity-cond1} and \eqref{eq:parity-cond2}. Until we obtain the explicit solution for the position-dependent corrections, we use this transformation as a guiding principle, expecting the generating functions to stay invariant under it. This ensures that the overall solution satisfies \eqref{eq:parity-inv}, allowing us to determine the eigenvalues of the parity operator from the explicit form of the position-dependent corrections. From the parity transformation map, we observe that the scaled position-dependent corrections transform as follows:
\begin{align}
    \tilde D^{n,m,r}_\sigma\rightarrow e^{-i m\cP}\tilde D^{r,m,n}_{\sigma^r}\;.\label{eq:p-map2}
\end{align}
Furthermore, under the parity transformation, the left and right two-magnon interaction equations given in \eqref{gamma-l-34} and \eqref{gamma-r-12} transform to each other, as well as the left and right three-magnon interaction equations given in \eqref{eq:beta-l-4} and \eqref{eq:beta-r-1},
\begin{align}
    \gamma_{l,34}(m,r)&\leftrightarrow\gamma_{r,12}(r,m)\label{paritytwo}\\
    \beta_{l,4}(r)&\leftrightarrow\beta_{r,1}(r)\label{partiythree}
\end{align}
Additionally, the four-magnon interaction equation \eqref{fourint} preserves its form, while the two two-magnon interaction equations \eqref{gamma-rl-34} and the non-interacting equations \eqref{non-int-rec} get mapped to another equation from the same type with altered permutation indices,
\begin{align}
    \delta_\sigma(n,m,r)&\leftrightarrow\delta_{\sigma^r}(r,m,n)\label{parity-nonint}\\
    \gamma_{rl,34}(m)&\leftrightarrow\gamma_{rl,12}(m)\label{parity-rl}\\
    \alpha&\leftrightarrow\alpha\label{parity-four}
\end{align}
The relations between the scattering coefficients and the position-dependent corrections derived from the action of the $\mathbb Z_2$ and parity transformations will help us constrain the generating functions and obtain the special solution for the eigenvectors.
\subsection{Non-interacting Equations in Terms of Generating Functions}
After deriving the eigenvalue equations in position space, and highlighting the symmetries of the Hamiltonian underlying these expressions, we proceed to solve the eigenvalue problem. The initial step is to interpret these relations as recursion relations and reformulate the problem within the framework of generating functions. From the outset, it becomes evident that for the states of an open, infinite spin chain, the eigenvalue problem alone does not fully constrain the position-dependent corrections, leaving a significant number of degrees of freedom undetermined. Consequently, the generating functions will inherently encode this freedom. In the subsequent sections, we systematically reduce these additional degrees of freedom by leveraging the symmetries of the model and insights drawn from the two-magnon CBA eigenstates. Then, in Section \ref{sec:periodic} we further impose periodic boundary conditions.

We define a generating function of three complex variables that encodes the position-dependent corrections for each permutation in its expansion around the origin,
\begin{align}
    G_\sigma(x,y,z)=\sum_{n,m,r=0}^\infty D_\sigma^{n,m,r} x^ny^mz^r\label{eq:def-gen}\;.
\end{align}
We should keep in mind that the position-dependent corrections, $D_\sigma^{n,m,r}$ for all $\sigma\in\mathcal{S}_4$ are functions of momentum variables $(p_1,p_2,p_3,p_4)$ and $\kappa$. These position-dependent corrections come from the $\ket{\Psi}_{11}$ states, meaning that the $\mathbb Z_2$ conjugate states, $\ket{\Psi}_{22}$, would contain the $\kappa\to\kappa^{-1}$ transformed version of these $D_\sigma^{n,m,r}$. Although we focus on the solution of the former, the latter follows exactly the same procedure and can be obtained by $\kappa\to\kappa^{-1}$. Since the scaling of the position-dependent corrections \eqref{scaledd} simplifies the eigenvalue equations drastically, we use the generating function with the scaled position-dependent corrections. This one is related to the initial definition of the generating function \eqref{eq:def-gen} as
\begin{align}
    \tilde G_\sigma(x,y,z)=G_\sigma(e^{-ip_{\sigma(1)}}x,e^{ip_{\sigma(3)}+ip_{\sigma(4)}}y,e^{ip_{\sigma(4)}}z)\;.
\end{align}
Additionally, by combining the parity transformation on the wave function given in \eqref{eq:parity-map} with the parity transformation on the variables of the generating function,
\begin{align}
    \{x,y,z\}\leftrightarrow\{z,\PP y,x\} \; , \label{eq:parity-map-xyz}
\end{align}
where the rescaling of $y$ is a consequence of the observation \eqref{eq:p-map2}, the transformation of the generating function under the action of parity becomes,
\begin{align}
    \tilde G_\sigma(x,y,z)\rightarrow \tilde G_{\sigma^r}(z,y,x)\;.\label{eq:parity-gen}
\end{align}
This is consistent with the rest of our formalism. By applying the parity transformation at the level of the generating function, one can verify whether the solution respects parity invariance. Indeed, we demonstrate later that this is the case.

We start solving the eigenvalue problem in terms of generating functions starting from the non-interacting equations given in \eqref{non-int-recd}. We impose the vanishing of the non-interacting equations,
\begin{align}
    \delta_\sigma(n,m,r)=0\quad\text{for}\quad n,m,r>0\;,
\end{align}
and obtain the following form of the scaled generating function,
\begin{equation}
    \tilde{G}_{\sigma}(x,y,z)=\frac{F_\sigma(x,y,z)}{Q_4(x,y,z) } \;, \label{eq:gen-fun-def}
\end{equation}
where the denominator is given by the polynomial, 
\begin{align}
Q_4(x,y,z)=\EE x y z -\invPP  x^2 z -e^{i \cP} y^2 z - x y -y z  -xy^2 -xz^2 -x^2 y z -xyz^2\;,\label{eq:def-q4}
\end{align}
and the numerator takes the form,
\begin{align}
F_\sigma (x,y,z)=& \left(Q_4(x,y,z)+xz^2+xyz^2\right)\tilde{G}_\sigma(x ,y ,0) - (1+ y)xyz \partial_{z} \tilde{G}_\sigma(x ,y ,0) \nonumber \\
+& \left( Q_4(x,y,z)+\PP y^2z+xy^2  \right) \tilde{G}_\sigma(x ,0,z) - (z+\invPP  x)xyz \partial_{y} \tilde{G}_\sigma(x,0,z) \nonumber\\
+& \left( Q_4(x,y,z)+\invPP x^2z+x^2yz \right) \tilde{G}_\sigma(0,y ,z) - (1+\PP  y) x y z \partial_{x} \tilde{G}_\sigma(0,y ,z)\nonumber \\
-& (\EE x y z -\invPP  x^2 z - x y -y z -x^2 y z) \tilde{G}_\sigma(x,0,0) \nonumber\\
+& xyz \partial_{z} \tilde{G}_\sigma(x,0,0) + \invPP  x^2 yz \partial_{y} \tilde{G}_\sigma (x ,0,0) \nonumber\\
-& \left(\EE x y z -e^{i \cP} y^2 z - x y -y z  -xy^2\right) \tilde{G}_\sigma(0,y ,0) \nonumber\\
 +&(1+ y) xyz \partial_{z} \tilde{G}_\sigma(0,y,0) + (e^{i\cP} y+1) xyz \partial_{x} \tilde{G}_\sigma(0,y ,0) \nonumber\\
 -& \left(\EE x y z - x y -y z -xz^2 -xyz^2\right) \tilde{G}_\sigma(0,0,z) \nonumber\\
 +& x y z^2 \partial_{y} \tilde{G}_\sigma(0,0,z) + x y z \partial_{x} \tilde{G}_\sigma(0,0,z) + (\EE  xyz - xy - yz) \tilde{D}_\sigma^{0,0,0} 
 \nonumber \\
 -&x y z \left(\tilde{D}_\sigma^{0,0,1}+  \tilde{D}_\sigma^{1,0,0}\right) \;.\label{eq:nonint-gen}
\end{align}
We keep the contribution of $\tilde D^{0,0,0}$ for completeness but we will chose conventions where this vanishes. Here, we use $\partial_z \tilde G_\sigma(x,y,0)$ as shorthand notation for $\Big[\partial_zG_\sigma(x,y,z)\Big]_{z=0}$, and similarly for the other derivatives. After imposing the non-interacting equations on the initial form of the generating function \eqref{eq:def-gen}, we observe that the number of unknown position-dependent corrections is reduced to those encoded in six two-variable generating functions: $G_\sigma(x,y,0)$, $G_\sigma(x,0,z)$, $G_\sigma(0,y,z)$, $\partial_zG_\sigma(x,y,0)$, $\partial_yG_\sigma(x,0,z)$, and $\partial_xG_\sigma(0,y,z)$ in addition to nine single-variable generating functions such as $G_\sigma(x,0,0)$, $G_\sigma(0,y,0)$, $G_\sigma(0,0,z)$, among others. Because the numerator of the generating function consists of the sum of functions that depend at most on two of the three variables weighted with coefficients that are at most linear in the remaining variables (e.g. $\tilde{G}_\sigma(x,y,0)$ is multiplied by a polynomial that is at most linear in $z$, while $\partial_{x} \tilde{G}_\sigma(0,0,z)$ are multiplied by a polynomial that is at most linear in $x$ and $y$), it fulfills the differential equation 
\begin{align}
    \partial_x^2 \partial_y^2 \partial_z^2 \big( F_{\sigma}(x,y,z) \big) =0\;,
\end{align}
and interaction equations can also be interpreted as differential equations. Before deriving the relations between generating functions coming from the interaction equations we point out that a simple counting of the position-dependent corrections which are fixed by the non-interacting equations reveals the number of undetermined two-variable generating functions. It turns out that the undetermined position-dependent corrections counted in terms of two-variable generating functions should be four rather than six. This implies that the generating function given in \eqref{eq:nonint-gen} does not know about a set of non-interacting equations encoded by two of these two-variable generating functions. Furthermore, we observe that the generating function given in \eqref{eq:gen-fun-def} is not analytic around the origin, contradicting our initial assumption \eqref{eq:def-gen} and undermining the entire purpose of extracting position-dependent corrections from the generating function. These two observations lead us to eliminate the spurious two-variable generating functions by imposing analyticity at the origin, allowing us to obtain the final form of the generating function that incorporates the solution to all non-interacting equations, \eqref{non-int-rec}. At first glance, this may seem somewhat artificial; however, we find that there is a natural way to impose the analyticity constraints, and these additional constraints correspond to the non-interacting eigenvalue equations that are missing from \eqref{eq:nonint-gen}.

\subsubsection{Analyticity}
\label{sec:analy}
To complete the derivation of a generating function that includes the position-dependent corrections solving the non-interacting equations \eqref{non-int-rec}, we impose analyticity at the point $(x,y,z)=(0,0,0)$. More precisely, we examine the generating function defined in \eqref{eq:gen-fun-def} and eliminate the poles at the origin by solving a system of equations. The coefficients that accompany inverse powers in the expansion of the generating function are linear combinations of undetermined, position-dependent corrections. They form a system of equations that turns out to be a subset of the original non-interacting equations \eqref{non-int-rec}, which confirms the consistency of the approach and leads to the final form of the non-interacting generating function. Instead of canceling the poles one by one, we impose analyticity conditions altogether as follows. We start by identifying the curve defined by the vanishing of the polynomial in the denominator \eqref{eq:def-q4}. As a first step, we factorize the elliptic curve appearing in the denominator,
\begin{align}
    Q_4(x,y,z)=(z-z_+(x,y))(z-z_-(x,y))\;,\label{eq:Q4-zfac}
\end{align}
where
\begin{align}
    z_\pm(x,y)=&\frac{\EE x y-\left(\invPP x^2+y\right)(1+\PP y)}{2x (1+y)} \mp \sqrt{\left(\frac{\EE x y-\left(\invPP x^2+y\right)(1+\PP y)}{2x (1+y)}\right)^2-y}\;.\label{curvez}
\end{align}
Demanding analyticity of $\tilde{G}_\sigma(x,y,z)$ at the origin is equivalent to demanding the numerator of $\tilde{G}_\sigma(x,y,z)$, \eqref{eq:nonint-gen}, to vanish at the points where the curve $Q_4(x,y,z)$ given in \eqref{eq:def-q4} vanishes around the origin. In contrast to the three-magnon case analyzed in \cite{Bozkurt:2024tpz}, here both the $z_+$ curve and the $z_-$ curve contain the $(x,y,z)=(0,0,0)$ point, however, this point is a branch point for both of them. When we assume that $-\pi<\textit{Arg}(x)<\pi$, and similarly for $y,z$, and the total momentum, $\PP$, then $z_+$ admits the origin as a limit point and we can expand around origin.\footnote{Similarly, the $z_+$ curve is ill-defined for $x=0$ when we approach the origin from the outside of the range $-\pi<\textit{Arg}(x)<\pi$ for complex variables $x,y,z$ and $\PP$. Here we assume that we are in the positive cone of the three-dimensional complex space and $-\pi<\cP<\pi$. Note that similar analysis can be done for the rest of the complex space and we can choose appropriate curves to ensure analyticity as well. For simplicity here we present only one.} However, demanding analyticity at $z=z_+(x,y)$ alone,
\begin{equation}
    F_\sigma(x,y,z_+(x,y))=0\label{eq:first-analy} \;,
\end{equation}
is not enough to ensure analyticity around the origin. On the other hand, demanding that the numerator vanishes on the curve $z=z_-(x,y)$ together with the first analyticity condition given above would force all the position-dependent corrections, $D_\sigma^{n,m,r}$ to vanish for $n,m,r\geq1$ because these two conditions would cancel the entire denominator of the generating function, $Q_4(x,y,z)$ and when incorporating the interaction equations into our solution, this would lead to an inconsistency. 

A natural way to complete the analyticity conditions is to demand that the resulting generating function is invariant under parity transformation, \eqref{eq:parity-map} together with \eqref{eq:parity-map-xyz}. This condition leads us to consider,
\begin{align}
    z_\pm(x,y)\xrightarrow[]{\mathbb{P}} x_\pm(y,z)\;,
\end{align}
such that $x_\pm$ is coming from another factorization of the elliptic curve,
\begin{align}
    Q_4(x,y,z)=(x-x_+(y,z))(x-x_-(y,z))\;.\label{eq:Q4-xfac}
\end{align}
Exactly as in the case of $z_\pm$, the origin is a branch point of $x_\pm$, and only $x_+$ admits a well-defined limit around it.  We observe that together with \eqref{eq:first-analy}, imposing
\begin{equation}
F_\sigma(x_+(y,z),y,z)=0\label{eq:second-analy} \;,
\end{equation}
determines two of the two-variable generating functions and removes all poles near the origin, while still yielding a non-trivial denominator. The resulting generating function is also parity invariant. To achieve this, we first solve the condition \eqref{eq:first-analy} by fixing the $\partial_z\tilde G_\sigma(x,y,0)$ component of the generating function. Next, we solve the second condition \eqref{eq:second-analy} to obtain the two-variable function $\partial_x \tilde G_\sigma(0,y,z)$. Substituting these into the non-interacting ansatz gives the final form of the generating function.

In the following sections, we turn to the relations between the generating functions that arise from the interaction eigenvalue equations. When we combine the solutions of the interaction and non-interacting equations in Section \ref{sec:special}, we choose to first solve the interaction equations and then impose the analyticity conditions given in \eqref{eq:first-analy} and \eqref{eq:second-analy} on the numerator of the generating function, which already includes the solution to the interaction equations. This approach does not alter the final result, and it ensures the analyticity of the full generating function.

\subsection{Interaction Equations in Terms of Generating Functions}

The generating function \eqref{eq:gen-fun-def}, together with the analyticity conditions \eqref{eq:first-analy} and \eqref{eq:second-analy}, codifies the position-dependent corrections that solve the non-interacting equation \eqref{eq:nonint-gen}. As in the three-magnon case, the non-interacting equations fix almost all position-dependent corrections. The remaining undetermined position-dependent corrections lie on the boundaries of the three-dimensional lattice and they are encoded in the two-variable and one-variable generating functions that appear in \eqref{eq:nonint-gen}. Therefore, the interaction eigenvalue equations will be constraints on these two-variable generating functions and the nine single-variable generating functions.

As for the non-interacting equations, we first transform the recursion relations coming from the interaction equations to relations between generating functions, and then we solve these relations in the context of the special solution by imposing additional constraints in the next section. Note that the interaction eigenvalue equations for the minimum separation of the magnons,
\begin{equation}
    \alpha \; , \qquad \beta_l(1) \; , \qquad \beta_r(1) \; , \qquad \gamma_l(1,1) \; , \qquad \gamma_m(1,1) \; , \qquad \gamma_r(1,1) \; , \qquad \gamma_{lr}(1) \; ,\label{eq:first}
\end{equation}
play a key role throughout the solution. Therefore, it is worth noting that they appear as initial conditions for the relations of generating functions, which we present in this section. Furthermore, we explicitly solve these equations at the beginning of Section \ref{sec:special} under additional assumptions on the scattering coefficients and position-dependent corrections. This results in a remarkably simple form of the position-dependent corrections, which remains valid even for configurations with larger magnon separations as we construct the special solution.

\subsubsection{Two-Magnon Interaction in Terms of Generating Functions}

Let us start by considering the left two-magnon interaction equation, given in \eqref{gamma-l-34}. The generating function associated with this equation is given by the formal sum
\begin{equation}
    F^{\gamma_{l,34}}(y,z)=\sum_{n,m=0}^\infty \gamma_{l,34}(n+1,m+1) x^ny^m=0 \;.
\end{equation}
Using the definition \eqref{eq:def-gen} for $G_\sigma (x,y,z)$, this sum can be recast into
\begin{align}\label{eq:gen-two-l-A}
&F^{\gamma_{l,34}}(y,z) = e^{i (p_{2}+2p_3 + 3p_4)} 
\Big(\frac{e^{i(p_3+p_4)} y^2}{1-e^{i(p_3+p_4)}y} \frac{e^{i p_4}z^2}{1-e^{i p_4} z}  e^{-i p_{2}} \, a(p_{2} , p_{1}, \kappa ) A_{1234} \nonumber\\
 + &[ (\EE -2\kappa )yz - yz^2 - z^2 - y^2
- y] \tilde G_{1234}(0,y,z)- yz (1 +e^{i \cP} y) \partial_x \tilde G_{1234} (0,y,z)
\nonumber \\
-& [(\EE-2\kappa ) yz -y^2 -y] \tilde G_{1234} (0,y,0)+ y z (y +1) \partial_z \tilde G_{1234} (0,y,0)+ y z (1 +e^{i\cP} y) \partial_x \tilde G_{1234}(0,y,0) \nonumber\\
-& [ (\EE -2\kappa) yz -z^2 -yz^2 -y ] \tilde G_{1234}(0,0,z) +y z^2 \partial_y \tilde G_{1234}(0,0,z)+ yz \partial_x \tilde G_{1234}(0,0,z) \nonumber \\
-& yz \left[\tilde D_{1234}^{0,0,1} + \tilde D_{1234}^{1,0,0} \right] + [(\EE  - 2 \kappa) yz - y] \tilde D_{1234}^{0,0,0}\Big) + (1\leftrightarrow2) \; .
\end{align}
The generating functions associated with the other permutations can be obtained by appropriately permuting the momenta. Moreover, the same steps can be followed to obtain the generating functions associated with the middle two magnon interaction equations \eqref{gamma-m-14}, giving us
\begin{align}
&F^{\gamma_{m,14}}(x,z) = e^{i (p_{2}+2p_3 + 3p_4)} \Bigg( \frac{e^{-ip_1} x^2}{1-e^{-ip_1} x} \frac{e^{i p_4} z^2}{1-e^{i p_4}z} e^{-i p_{3}} \, a(p_{3} , p_{2}, \kappa^{-1} ) A_{1234}  \nonumber\\
+& [ (\EE -2\kappa^{-1})xz- x z^2 - x^2 z - x - z] \tilde G_{1234}(x,0,z) -x z(e^{-i \cP} x+ z) \partial_y \tilde G_{1234} (x,0,z) \nonumber\\
 -&  [ (\EE -2\kappa^{-1}) xz - x - x^2 z - z] \tilde G_{1234} (x,0,0) + x z \partial_z \tilde G_{1234}(x,0,0) - x^2 z e^{-i\cP} \partial_y \tilde G_{1234}(x,0,0) \nonumber\\
-&[ (\EE -2\kappa^{-1}) x z - x - x z^2 - z] \tilde G_{1234}(0,0,z) +x z \partial_x \tilde G_{1234}(0,0,z)
+x z^2 \partial_y \tilde G_{1234}(0,0,z)\nonumber\\
 +& [(\EE  -2\kappa^{-1}) xz - x - z] \tilde D_{1234}^{0,0,0} -xz \left[ \tilde D_{1234}^{0,0,1} + \tilde D_{1234}^{1,0,0}\right] \Bigg) +(2\leftrightarrow 3)\, , \label{eq:gen-two-m-A}
\end{align}
and the right two-magnon interaction equations, as given in \eqref{gamma-r-12},
\begin{align}
&F^{\gamma_{r,12}}(x,y) = e^{i (p_2+2p_3 + 3p_4)} \Bigg( \frac{e^{-ip_1}x^2}{1-e^{-ip_1} x} \frac{e^{i(p_3+p_4)} y^2}{1-e^{i(p_3+p_4)}y} e^{-i p_{4}} a(p_4 , p_{3},\kappa)  A_{1234} \nonumber \\
+ & [(\EE-2\kappa )x y -e^{i \cP} y^2 -e^{-i \cP} x^2 - x^2 y - y] \tilde G_{1234}(x,y,0) \nonumber \\
- & x y( y +1) \partial_z \tilde G_{1234}(x,y,0) 
- [(\EE-2\kappa) xy- x^2 y -e^{-i \cP} x^2 - y] \tilde G_{1234}(x,0,0) \nonumber \\
+& x y \partial_z \tilde G_{1234}(x,0,0) + x^2 y e^{-i\cP} \partial_y \tilde G_{1234} (x,0,0) 
- [ (\EE-2\kappa)xy - e^{i\cP} y^2 - y] \tilde G_{1234} (0,y,0) \nonumber \\
+& x y ( y+1 ) \partial_z \tilde G_{1234}(0,y,0) 
+ x y(e^{i \cP} y +1) \partial_x \tilde G_{1234}(0,y,0)- x y \left[ \tilde D_{1234}^{0,0,1} + \tilde D_{1234}^{1,0,0} \right] \nonumber \\
&+[(\EE  - 2 \kappa) - y] \tilde D_{1234}^{0,0,0} \Bigg) +( 3\leftrightarrow 4)  \, ,\label{eq:gen-two-r-A}
\end{align}
After obtaining the explicit expressions we apply the parity transformation given in \eqref{eq:parity-map} together with the appropriate transformation on the complex variables \eqref{eq:parity-map-xyz} and observe that the relation on the generating functions coming from the left two-magnon interaction equations \eqref{eq:gen-two-l-A} transform to the right two-magnon interaction one \eqref{eq:gen-two-r-A} while the relation for the middle two magnon interaction equations \eqref{eq:gen-two-m-A} stays invariant, as stated in \eqref{paritytwo}. Thus, we observe that the generating function relations are compatible with the parity relations. Additionally, $\mathbb Z_2$ transforms these expressions to the ones for the state $\ket{\Psi}_{22}$.

\subsubsection{Three-Magnon Interaction in Terms of Generating Functions}
We consider the relations coming from the three-magnon interaction equations, \eqref{eq:beta-l-4} and \eqref{eq:beta-r-1} and the two two-magnon interaction equations given in \eqref{gamma-rl-34}. Similarly to the two-magnon interaction case we obtain the relations between the generating functions by defining the formal sum. For the left three-magnon interaction equations, given in \eqref{eq:beta-l-4}, we define,
\begin{align}
    F^{\beta_{l,4}}(z)=\sum_{m=1}^\infty\beta_{l,4}(m+1)z^m=0\;,
\end{align}
together with the vanishing of shortest separation configuration $\beta(1)=0$, we obtain the solution of these eigenvalue expressions in terms of the generating functions
\begin{align}
F^{\beta_{l,4}}(z)=e^{i(p_2 +2p_3)} \Big(& ((\EE -2\kappa -2\kappa^{-1})z -1 - z^2) \tilde G_{1234}(0,0,z)- z \partial_x \tilde G_{1234}(0,0,z)\nonumber \\
& - z^2  \partial_y \tilde G_{1234}(0,0,z) 
-\tilde C_{1234}^{\beta_{l,4}}(z) \Big)+(1\leftrightarrow2\leftrightarrow3)\;,\label{eq:gen-beta-l}
\end{align}
where
\begin{align}
\tilde C_{1234}^{\beta_{l,4}}(z)=
&  - z \tilde D_{1234}^{0,0,1} - z \tilde D_{1234}^{1,0,0} 
+\frac{z^2}{1-e^{ip_4} z} \left( (\EE -2\kappa -2\kappa^{-1} ) \tilde D_{1234}^{0,0,1}- \tilde D_{1234}^{0,1,0} - \tilde D_{1234}^{0,0,2} - \tilde D_{1234}^{1,0,1} \right)\label{eq:const-beta-l}\;.
\end{align}
Here, we use $(1\leftrightarrow2\leftrightarrow3)$ to indicate the sum over all permutations of the indices $123$ while keeping the $4$th one fixed. The leftover position-dependent corrections, which are organized as $\tilde C(z)$, are related to a linear combination of the three-magnon interaction equation with minimum separation as anticipated. Furthermore, the relation between the rest of the one-variable generating functions coming from the three-magnon interaction equations is given as follows.

For the two two-magnon interaction equations \eqref{gamma-rl-34} we obtain,
\begin{align}
F^{\gamma_{lr,34}}(y)&=e^{i(p_2 +2p_3+ 3p_4)} \Big(  (\EE- 4\kappa ) \tilde G_{1234}(0,y,0) - (1 +\PP y) \partial_x \tilde G_{1234}(0,y,0)\nonumber \\
& - (1 +y) \partial_z \tilde G_{1234}(0,y,0) -\tilde C_{1234}^{\gamma_{lr}}(y) \Big)\label{eq:gen-beta-r}+(1\leftrightarrow2)+(3\leftrightarrow4)+(1\leftrightarrow2 , 3\leftrightarrow4)=0 \;,
\end{align}
where
\begin{align}
\tilde C_{1234}^{\gamma_{lr}}(y)&= -\tilde D_{1234}^{0,0,1} -\tilde D_{1234}^{1,0,0} - \frac{y}{1-e^{i(p_3+p_4)}y} \left(\tilde D_{1234}^{0,0,1} - (\EE- 4\kappa) \tilde D_{1234}^{0,1,0}+ \PP \tilde D_{1234}^{1,0,0} + \tilde D_{1234}^{0,1,1} +  \tilde D_{1234}^{1,1,0}\right)\;.\label{eq:const-beta-r}
\end{align}
Finally, for the three-magnon interaction equation from the right as given in \eqref{gamma-r-12}, we obtain the following relations between the generating functions,
\begin{align}
F^{\beta_{r,1}}(x)=e^{i( p_2 +2p_3 +3p_4)} &\Big( ((\EE -2\kappa^{-1} -2\kappa)x  -1 - x^2) \tilde G_{1234}(x,0,0)- x^2 \invPP \partial_y \tilde G_{1234}(x,0,0)\nonumber \\
&-x \partial_z \tilde G_{1234}(x,0,0)  -\tilde C_{1234}^{\beta_{r,1}}(x) \Big)+(1\leftrightarrow2\leftrightarrow3)=0\;,\label{eq:gen-gamma-rl}
\end{align}
where
\begin{align}
\tilde C_{1234}^{\beta_{r,1}}(x) =&- x \tilde D_{1234}^{0,0,1} -x \tilde D_{1234}^{1,0,0}\\
&-\frac{x^2}{1-e^{-ip_1} x} \left(\invPP \tilde D_{1234}^{0,1,0} - (\EE -2\kappa^{-1} -2\kappa) \tilde D_{1234}^{1,0,0} +  \tilde D_{1234}^{1,0,1} + \tilde D_{1234}^{2,0,0} \right)\;.\label{eq:const-gamma-rl}
\end{align}
As discussed in Section \ref{sec:parityz2}, under the action of $\mathbb Z_2$ the coefficients in front of the generating functions stay invariant. We conclude that the scattering coefficients and the position-dependent corrections of the $\ket{\Psi}_{22}$ state have to satisfy the same equations. On the other hand, under the parity transformation, the expression for the three-magnon from the left transforms to the three-magnon from the right, and the expression for the two two-magnon interaction stays invariant.

Here, we completed the derivation of the relations between the pieces of the generating function \eqref{eq:gen-fun-def} which are not fixed by the non-interacting equations. Combining all of these equations and obtaining the final form of the generating function is straightforward, as all of these relations are linear in terms of the one- and two-variable generating functions. However, there are infinitely many undetermined position-dependent corrections after combining all the pieces and the final form is not very illuminating. Therefore, instead of blindly trying to simplify our solution, we use this freedom to our advantage and impose additional conditions inspired by the CBA to obtain a simple solution.

\section{Special Solution for the Four-Magnon Problem}
\label{sec:special}
In this section, we present a special solution to the four-magnon eigenvalue problem. As the complexity of the system of linear equations increases from the three-magnons to the four-magnons, solving all the relations derived from the interaction equations leaves a large number of undetermined coefficients behind. Instead, we impose additional conditions, similar to those used in the three-magnon problem \cite{Bozkurt:2024tpz}, to obtain a solution with significantly fewer undetermined degrees of freedom.\footnote{We draw inspiration from dynamical Felder-like setups  \cite{Felder:2020tct} and choose these additional constraints to establish factorizability. Details of the reasoning are given in \cite{Bozkurt:2024tpz}.} However, even with these special conditions, not all position-dependent corrections are fully determined. We will utilize this remaining freedom in Section \ref{sec:smearing} to establish the connection between solutions with different numbers of magnons.

To obtain the special solution, we impose the factorization of scattering coefficients, consistent with the two-magnon solution \eqref{eq:two-wave1},
\begin{align}
    \frac{A_{jikl}}{A_{ijkl}}=\frac{A_{kjil}}{A_{kijl}}=\frac{A_{klji}}{A_{klij}}=S_\kappa(p_i,p_j)\;,\label{specialscat}
\end{align}
for any permutation, $A_\sigma$ such that $\sigma=(ijkl)\in\mathcal{S}_4$. Additionally, to enforce invariance under a subgroup of permutation group, we constrain the position-dependent corrections by a partial factorizability condition. For the middle two permutation indices of all position-dependent corrections, we impose the following relation,
\begin{align}
    \tilde D_{ikjl}^{n,m,r}=S_\kappa(p_j,p_k)\tilde D_{ijkl}^{n,m,r}\;,\label{partialfac}
\end{align}
for all $D_\sigma$, with $\sigma=(ijkl)\in \mathcal{S}_4$. Under these conditions, we first solve the eigenvalue equations with minimum separation as given in \eqref{eq:first} and obtain the position-dependent corrections for the lowest separation explicitly. Then, we move on and solve the rest of the eigenvalue equations in terms of generating functions. We only solve for the pieces of the generating function, \eqref{eq:nonint-gen} which are not fixed by the non-interacting equations and impose the analyticity condition only at the end to simplify the expressions for the generating functions. Nevertheless, the final form of the solution describes a true eigenstate which can be checked independently.

 Under the assumptions given in \eqref{specialscat} and \eqref{partialfac}, the eigenvalue equations simplify significantly. For the two-magnon interaction equations written in terms of generating functions in \eqref{eq:gen-two-l-A}, \eqref{eq:gen-two-m-A}, and \eqref{eq:gen-two-r-A}, the partial factorization property \eqref{partialfac} is sufficient to give a solution for two-variable generating functions in terms of the rest of the pieces of generating function without mixing different permutation labels. However, for the three-magnon interaction equations, given in \eqref{eq:gen-beta-l}, \eqref{eq:gen-gamma-rl}, and \eqref{eq:gen-beta-r}, the partial factorization alone is not enough to achieve the separation of permutation indices. Therefore, in that stage, we choose to enforce this by separating the terms with different permutation indices and solving those parts independently. We insist on seeking a solution for the position-dependent corrections that do not mix permutation indices because this would be a symmetric form of the solution in terms of the permutation group, $\mathcal{S}_4$. Indeed, we obtain the tower of YBEs for this particular solution which is symmetric under the permutation group.

\subsection{Special Solution for the Minimum Separation}
\label{sec:minsepspecial}

Among the eigenvalue equations arising from the four-magnon problem, the one corresponding to the minimum separation of magnons, given in \eqref{eq:first}, plays a distinctive role. As in the three-magnon case, these equations serve as initial conditions for solving a system of differential equations. We solve the minimum separation equations by making special choices designed to render the remaining equations as permutationally symmetric as possible. These equations effectively determine the behavior of all position-dependent corrections, as they appear in the generating function relations—such as \eqref{eq:const-beta-l} and \eqref{eq:const-beta-r}—in isolation from other contributions. Moreover, the role of these equations as initial values is not merely an analogy: the eigenvalue equations can be interpreted directly as difference equations and then transformed into differential equations via generating functions. Therefore, we will see that also in Section \ref{sec:specialspecial} when we are able to write a solution to the four-magnon problem based on the three-magnon solution we will do it by matching that solution to the the solution of the minimum separation equations, especially the coefficients $\tilde D_\sigma^{1,0,0}$ and $\tilde D_\sigma^{0,0,1}$, and this will be enough to fully fix the position-dependent corrections.

Starting with the eigenvalue equations at minimum separation, given in \eqref{eq:first}, we impose the special conditions from \eqref{specialscat} and \eqref{partialfac} to identify the position-dependent corrections within these expressions. In doing so, we set some of these corrections to zero, simplifying the solution. As a result, the solution we present is not unique; however, it retains key features of the spin chain model. These properties include the vanishing of position-dependent corrections in the orbifold point ($\kappa=1$), and the preservation of parity invariance, as specified in \eqref{eq:parity-cond1}, \eqref{eq:parity-cond2}, and \eqref{eq:parity-map}.

As a first step, we use the four-magnon interaction equation \eqref{fourint} to illustrate how the expressions can be simplified under the special conditions \eqref{specialscat} and \eqref{partialfac}, leading to a special solution. The four-magnon interaction equation \eqref{fourint} is given by:
\begin{equation}
    \alpha=\sum_{(ijkl)\in\mathcal{S}_4} e^{i(p_j +p_k +3p_l)}\omega(p_j,p_k;\kappa)\Big[\left(\kappa-\kappa^{-1}\right)A_{ijkl} +\left(\EE-4\kappa-2\kappa^{-1}\right)\tilde D_{ijkl}^{0,0,0}-  \tilde  D_{ijkl}^{1,0,0} - \tilde D_{ijkl}^{0,0,1} \Big]\;,\label{fourintspecial}
\end{equation}
such that 
\begin{align}
    \omega(p_j,p_k;\kappa)=\frac{\left(e^{ip_j}-e^{ip_k}\right)\left(1+e^{ip_j+ip_k}\right)}{1+e^{ip_j+ip_k}-2\kappa e^{ip_j}} \label{eq:omega}\;.
\end{align}
To obtain this form of  \eqref{fourint}, we make use of the following identity for the scattering coefficients which are factorized in terms of $S_\kappa$s,
\begin{align}
    \sum_{(ijkl)\in\mathcal{S}_4} e^{i(p_j +2p_k +3p_l)} (e^{-ip_j}a(p_j,p_i,\kappa)+e^{-ip_l}a(p_l,p_k,\kappa))A_{ijkl}=0\;,\label{identity1}
\end{align}
which vanishes because the sum can be re-arranged by using \eqref{specialscat} and the definition of the scattering coefficient $S_\kappa$, \eqref{eq:defscat}. Secondly, the partial factorization \eqref{partialfac} can be used to reduce the sum over all the permutations to a sum over only half of them. In this case, the summand, which includes the scattering coefficient, also simplifies and gives us an overall $\left(\kappa-\kappa^{-1}\right)$ factor,
\begin{align}
\sum_{i,l\in\{1,2,3,4\}}&e^{i(p_j+p_k +3p_l)}\left[a(p_k,p_j,\kappa^{-1})A_{ijkl}+a(p_j,p_k,\kappa^{-1})A_{ikjl}\right]\nonumber\\&=-\sum_{i,l\in\{1,2,3,4\}}e^{i(p_j+p_k +3p_l)}\frac{2\left(\kappa-\kappa^{-1}\right)\left(e^{ip_j}-e^{ip_k}\right)\left(1+e^{ip_j+ip_k}\right)}{1+e^{ip_j+ip_k}-2\kappa e^{ip_j}} A_{i\underline{jk}l}\;,\label{scatsum}
\end{align}
where the underlined indices indicate that, between $(jk)$ and $(kj)$, only the terms with $j<k$ are included. The $\left(\kappa-\kappa^{-1}\right)$ factor plays a crucial role, as it ensures that all position-dependent corrections vanish in the limit $\kappa\to1$, leaving only the CBA solution at the orbifold point. Similarly, for the rest of the four-magnon interaction equation, we make use of the partial factorizability property \eqref{partialfac} and obtain,
\begin{align}
    -\sum_{i,l\in\{1,2,3,4\}}&e^{i(p_j+ p_k +3p_l)}\left(e^{ip_k} +S_\kappa(p_j,p_k)e^{ip_j}\right)\left(\tilde D_{i\underline{jk}l}^{1,0,0}+\tilde D_{i\underline{jk}l}^{0,0,1}\right)\nonumber\\&=\sum_{i,l\in\{1,2,3,4\}}e^{i(p_j+ p_k +3p_l)}\frac{\left(e^{ip_j}-e^{ip_k}\right)\left(1+e^{ip_j+ip_k}\right)}{1+e^{ip_j+ip_k}-2\kappa e^{ip_j}} \left(\tilde D_{i\underline{jk}l}^{1,0,0}+\tilde D_{i\underline{jk}l}^{0,0,1}\right) \;.\label{dsum}
\end{align}
Therefore, we combine the final form of the sum over the permutations of scattering coefficients \eqref{scatsum} with the sum over the position-dependent corrections \eqref{dsum} and obtain \eqref{fourintspecial}. Finally, we conclude that, for any permutation $\sigma$, a simple relation between the scattering coefficients and the position-dependent corrections of the lowest separation makes the four-magnon interaction equation vanish,
\begin{align}
    \tilde  D_\sigma^{1,0,0}=\tilde D_\sigma^{0,0,1}=\left(\kappa -\kappa^{-1}\right)A_\sigma  \;, \label{eq:d101}
\end{align}
which is very similar to the special solution of the three magnon problem given in \cite{Bozkurt:2024tpz}. Note that taking $D^{0,0,0}=0$ is reasonable for open infinite spin chains, as we can always redefine the position-dependent corrections and the scattering coefficients in the long-range Bethe ansatz \eqref{four-magnon-wave} by shifting all of them with an expression that does not depend on position variables.

Moving our attention to solving the three-magnon interaction and two two-magnon interaction equations with minimum separation, imposing our assumptions, gives us the following solution,
\begin{align}
&\tilde D_{\sigma}^{0,0,2}= \left(\kappa-\kappa^{-1}\right)\left(2e^{ip_{\sigma(4)}}+\EE-2\kappa-2\kappa^{-1}\right)A_\sigma\;,\label{eq:d102}\\
&\tilde D_\sigma^{2,0,0}=\left(\kappa-\kappa^{-1}\right)\left(2e^{-ip_{\sigma(1)}}+\EE-2\kappa-2\kappa^{-1}\right)A_\sigma\;,\label{eq:d200} \\
&\tilde D_\sigma^{1,1,0}=-\PP \left(\kappa-\kappa^{-1}\right)A_\sigma\;,\quad\tilde D_\sigma^{0,1,1}= -\left(\kappa-\kappa^{-1}\right)A_\sigma\;,\label{eq:d121}
\end{align}
provided that we fix $\tilde D^{0,1,0}_{ijkl}=\tilde D^{1,0,1}_{ijkl}=0$. Finally, the solution of the minimum separation two-magnon interaction equations is given as
\begin{align}
&\tilde D_\sigma^{0,1,2}=-\left(2e^{ip_{\sigma(4)}}+2\EE-4\kappa-2\kappa^{-1}\right)\left(\kappa-\kappa^{-1}\right)A_\sigma \;,\\
&\tilde D_\sigma^{1,0,2}=\tilde D_\sigma^{2,0,1} =\left(e^{-ip_{\sigma(1)}+ip_{\sigma(4)}}+\frac{\invPP}{2}\left(1-\PP \right)^2\right)\left(\kappa-\kappa^{-1}\right)A_\sigma\;,\\
&\tilde D_\sigma^{2,1,0}=-\PP \left(2e^{-ip_{\sigma(1)}}+2\EE-4\kappa-2\kappa^{-1}\right)\left(\kappa-\kappa^{-1}\right)A_\sigma\;,
\end{align}
provided that we fix $\tilde D_{ijkl}^{0,2,0}=\tilde D_{ijkl}^{1,1,1}=0$. This completes the solution of the eigenvalue equations with minimum separation. An immediate question we can ask to these position-dependent corrections is whether or not they satisfy the conditions given in \eqref{eq:parity-cond1} and \eqref{eq:parity-cond2} for the overall wave function to be parity invariant. The answer is yes, but when we apply the parity transformation map \eqref{eq:parity-map} we should keep in mind that the scaled position-dependent corrections bring extra factors of $\PP$, as emphasized in \eqref{eq:p-map2}.

In solving the minimum separation equations, we made specific choices, such as setting $\tilde D_\sigma^{0,1,0} = 0$. Alternatively, one could choose $\tilde D_\sigma^{0,1,0} = (\kappa - \kappa^{-1})A_\sigma$, analogous to the position-dependent corrections in~\eqref{eq:d101}. This choice would shift the remaining coefficients by a factor of $(\kappa - \kappa^{-1})A_\sigma$. For example, instead of~\eqref{eq:d102}, we would obtain, $\tilde D_{\sigma}^{0,0,2} = (\kappa - \kappa^{-1})(2e^{ip_{\sigma(4)}} - 1 + \EE - 2\kappa - 2\kappa^{-1})A_\sigma$. At this point, we see no compelling reason to prefer one choice over the other, and therefore proceed with a representative option. A more natural prescription for fixing these position-dependent corrections may emerge by refining the solution to the eigenvalue equation with periodic boundary conditions, as discussed in Section~\ref{sec:periodic}, particularly for long spin chains. We leave this to future work.

\subsection{Special Solution for All Position-Dependent Corrections}\label{sec:specialGs}

The eigenvalue equations captured in terms of the generating functions which we derived in Section \ref{sec:fourmagnon}, always include a combination of the position-dependent corrections with minimum separation which we have found a special solution in Section \ref{sec:minsepspecial}. These pieces of the generating functions are given in \eqref{eq:const-beta-l} and \eqref{eq:const-beta-r} for three-magnon interaction equations and \eqref{eq:const-gamma-rl} for the two two-magnon interaction equations. In addition, the two-magnon interaction equations and the non-interacting expressions depend on $\tilde D_\sigma^{1,0,0}+\tilde D_\sigma^{0,0,1}$.  Therefore, the choices we made for the solution of the equations with minimum separation play the role of initial conditions. Then, we focus on fixing the undetermined pieces of the generating functions by substituting the special solution above and using the partial factorization of the position-dependent corrections, \eqref{partialfac} which can be rewritten in terms of the generating function as
\begin{align}
    \tilde G_{ikjl}(x,y,z)=S_\kappa(p_j,p_k)\tilde G_{ijkl}(x,y,z)\;.
\end{align}
This relation between generating functions reduces the number of distinct generating functions from twenty-four to twelve. Then, we calculate three two-variable generating functions for each permutation $\sigma\in\mathcal{S}_4$ by solving the two-magnon interaction equations given in \eqref{eq:gen-two-l-A}, \eqref{eq:gen-two-m-A}, and \eqref{eq:gen-two-r-A}.
The solution of the two-magnon interaction equations is given as follows, 
\begin{align}
    &\left(y(1-\EE z+z^2+2 \kappa  z+y)+z^2\right)\tilde G_\sigma(0,y,z)= -yz \left(1+\PP  y\right) \partial_x\tilde G_\sigma(0,y,z)-2yz\left(\kappa-\kappa^{-1}\right)A_\sigma\nonumber\\&\;\;+y(1-\EE z+y+2 \kappa  z)\tilde G_\sigma(0,y,0)+yz \left(1+\PP  y\right)\partial_x\tilde G_\sigma(0,y,0)+yz(1+y)\partial_z\tilde G_\sigma(0,y,0)\nonumber\\&\;\;+\left(y\left(1-\EE z+z^2+2 \kappa  z\right)+z^2\right)\tilde G_\sigma(0,0,z) +yz^2 \partial_y\tilde G_\sigma(0,0,z)+yz\partial_x\tilde G_\sigma(0,0,z)\label{eq:special-yz}\;.
\end{align}
We choose to express $\tilde G(0,y,z)$ in terms of $\partial_x \tilde G(0,y,z)$, so that $\tilde G(0,y,z)$ is thereby determined, while $\partial_x \tilde G(0,y,z)$ remains to be fixed. Here we omit the constraints coming from the analyticity condition \eqref{eq:first-analy} and \eqref{eq:second-analy} to simplify the expressions. At this stage, the solution of the generating function looks like it has poles around the origin, once we divide by the polynomial factor that appears on the left-hand side of \eqref{eq:special-yz}. These are again spurious poles coming from the two-magnon interaction equations that are not counted yet. Similar to the non-interacting equations we impose analyticity by fixing one of the single-variable generating functions on the curve, which is coming from the vanishing of the polynomial factor on the denominator of $\tilde G_\sigma(0,y,z)$, that passes from the origin. We can impose each of these analyticity conditions at the end when we ensemble everything. Similarly, we solve two more two-variable generating functions from the rest of the two-magnon interaction equations,
\begin{align}
    &\left(\kappa  (x z (-\EE+x+z)+x+z)+2 x z\right)\tilde G_\sigma(x,0,z)=-\kappa x z (\invPP x+ z) \partial_y G_\sigma(x,0,z)\nonumber\\&+\left(\kappa  (x z (-\EE+z)+x+z)+2 x z\right)\tilde G_\sigma(0,0,z)+\kappa x z^2\partial_y \tilde G_\sigma(0,0,z)+\kappa x z\partial_x \tilde G_\sigma(0,0,z)\nonumber\\&+\left(\kappa  (x z (-\EE+x)+x+z)+2 x z\right)\tilde G_\sigma(x,0,0)+\invPP\kappa x^2 z\partial_y \tilde G_\sigma(x,0,0)+\kappa x z\partial_z \tilde G_\sigma(x,0,0)\nonumber\\&-\frac{2xz\kappa\left(\kappa-\kappa^{-1}\right)A_\sigma\left(1-e^{-ip_{\sigma(1)}}x-e^{ip_{\sigma(4)}}z\right)}{\left(1-e^{-ip_{\sigma(1)}}x\right)\left(1-e^{ip_{\sigma(4)}}z\right)}\;.\label{eq:special-xz}
\end{align}
We highlight these factors because they will play a crucial role in the following sections. The rest of the two-magnon interaction equations are solved,
\begin{align}
    &\left((x^2 +\PP y)(\PP y+1)-\PP x y (\EE-2 \kappa )\right)\tilde G_\sigma(x,y,0)=-\PP x y (1+y)\partial_z\tilde G_\sigma(x,y,0)\nonumber\\&-2\PP xy(\kappa-\kappa^{-1})A_\sigma+\PP y\left(1+\PP y-\EE x+2 \kappa x\right)\tilde G_\sigma(0,y,0)\nonumber\\&+\PP xy \left((1+y)\partial_z \tilde G_\sigma(0,y,0)+ \left(1+\PP y\right)\partial_x \tilde G_\sigma(0,y,0)\right)
    \nonumber\\&+\left(x^2 -\PP x y (\EE-2 \kappa-x-x^{-1})\right)\tilde G_\sigma(x,0,0)+x^2y\partial_y \tilde G_\sigma(x,0,0)+\PP xy\partial_z \tilde G_\sigma(x,0,0)\;. \label{eq:special-xy}
\end{align}
Furthermore, we solve eigenvalue equations coming from the two two-magnon interaction equation \eqref{eq:gen-gamma-rl} as well as the three-magnon interaction equations \eqref{eq:gen-beta-l} and \eqref{eq:gen-beta-r} in terms of generating functions. In this case, the partial factorization property alone is not enough to get rid of the sum over permutation indices, therefore we choose to solve for each permutation separately to obtain the most symmetric solution. To demonstrate the strategy we focus on the left three-magnon interaction expression \eqref{eq:gen-beta-l},
\begin{align}
    F^{\beta_{l,a}}(z)=\sum_{(ijk)\in\mathcal{S}_3}F^{\beta_{l,a}}_{ijk}(z) \;,
\end{align}
such that $(ijka)\in\mathcal{S}_4$ and, instead of finding the general solution of $F^{\beta_{l,a}}(z)=0$, we find the solution of $F^{\beta_{l,a}}_{ijk}(z)=0$ and check if the resulting solution in terms of the one-variable generating function satisfies the partial factorization \eqref{partialfac}. In this case, we obtain,
\begin{align}
    & \tilde G_{\sigma}(0,0,z)=-\frac{ z \partial_x \tilde G_{\sigma}(0,0,z)+ z^2  \partial_y \tilde G_{\sigma}(0,0,z) 
-2z\left(\kappa-\kappa^{-1}\right)A_\sigma/\left(1-e^{ip_{\sigma(4)}}z\right)}{1-(\EE -2\kappa -2\kappa^{-1})z+z^2} \;.\label{eq:special-z}
\end{align}
Similarly, the solution of the three-magnon interaction from the right and the two-two magnon interaction equations give
\begin{align}
    \tilde G_{\sigma}(x,0,0)=-\frac{x \partial_z \tilde G_{\sigma}(x,0,0) + x^2 \invPP \partial_y \tilde G_{\sigma}(x,0,0)-2x\left(\kappa-\kappa^{-1}\right)A_\sigma/\left(1-e^{-ip_{\sigma(1)}}x\right)}{1-(\EE -2\kappa^{-1} -2\kappa)x + x^2} \;,\label{eq:special-x}
\end{align}
\begin{align}
    \tilde G_{\sigma}(0,y,0)=\frac{1}{\EE-4\kappa}\left((1 +y) \partial_z \tilde G_{\sigma}(0,y,0)+ (1 +\PP y) \partial_x \tilde G_{\sigma}(0,y,0) -2\left(\kappa-\kappa^{-1}\right)A_{\sigma}\right) \;.\label{eq:special-y}
\end{align}
At this point, we have solved all the interaction equations. The last step remaining is to impose the analyticity conditions \eqref{eq:first-analy} and \eqref{eq:second-analy} and substitute all the two-variable and one-variable generating functions into the generating function \eqref{eq:gen-fun-def}. Nevertheless, this final form of the special solution still has undetermined position-dependent corrections. Since the expressions from each of the pieces of the generating function are very involved we summarize the solution by color coding the pieces of the non-interacting generating function according to the stage we find the solution. When we bring everything together, we obtain
\begin{align}
    Q_4(x,y,z)G_\sigma(x,y,z)=& \left(Q_4(x,y,z)+xz^2+xyz^2\right)\textcolor{blue}{\tilde{G}_\sigma(x ,y ,0) }- (1+ y)xyz \textcolor{green}{\partial_{z} \tilde{G}_\sigma(x ,y ,0)} \nonumber \\
+& \left( Q_4(x,y,z)+\PP y^2z+xy^2  \right) \textcolor{blue}{\tilde{G}_\sigma(x ,0,z)} - (z+\invPP  x)xyz \underline{\partial_{y} \tilde{G}_\sigma(x,0,z)} \nonumber\\
+& \left( Q_4(x,y,z)+\invPP x^2z+x^2yz \right) \textcolor{blue}{\tilde{G}_\sigma(0,y ,z)} - (1+\PP  y) x y z \textcolor{green}{\partial_{x} \tilde{G}_\sigma(0,y ,z)}\nonumber \\
-& (\EE x y z -\invPP  x^2 z - x y -y z -x^2 y z) \textcolor{purple}{\tilde{G}_\sigma(x,0,0)} \nonumber\\
+& xyz \textcolor{orange}{\partial_{z} \tilde{G}_\sigma(x,0,0)} + \invPP  x^2 yz\underline{ \partial_{y} \tilde{G}_\sigma (x ,0,0)} \nonumber\\
-& \left(\EE x y z -e^{i \cP} y^2 z - x y -y z  -xy^2\right) \textcolor{purple}{\tilde{G}_\sigma(0,y ,0)} \nonumber\\
 +&(1+ y) xyz\textcolor{orange}{ \partial_{z} \tilde{G}_\sigma(0,y,0)} + (e^{i\cP} y+1) xyz \textcolor{orange}{\partial_{x} \tilde{G}_\sigma(0,y ,0)} \nonumber\\
 -& \left(\EE x y z - x y -y z -xz^2 -xyz^2\right) \textcolor{purple}{\tilde{G}_\sigma(0,0,z) }\nonumber\\
 +& x y z^2 \underline{\partial_{y} \tilde{G}_\sigma(0,0,z) }+ x y z \textcolor{orange}{\partial_{x} \tilde{G}_\sigma(0,0,z) } \nonumber \\
 -&x y z \left(\textcolor{red}{\tilde{D}_\sigma^{0,0,1}}+ \textcolor{red}{ \tilde{D}_\sigma^{1,0,0}}\right)\;.\label{eq:solG4}
\end{align}
For clarity, we color-code the components of the generating function based on how they are fixed. The \textcolor{blue}{blue} two-variable terms are determined in this section using the two-magnon interaction equations~\eqref{eq:special-yz}, \eqref{eq:special-xz}, and~\eqref{eq:special-xy}. The \textcolor{purple}{purple} one-variable terms are fixed by the three-magnon interaction equations~\eqref{eq:special-z}, \eqref{eq:special-x}, and~\eqref{eq:special-y}. The \textcolor{orange}{orange} derivative terms are fixed by imposing analyticity on the two-variable generating functions obtained from two-magnon interactions. This is done in a parity-invariant way and is explained in detail in Section~\ref{sec:specialspecial}. The \textcolor{green}{green} two-variable derivative terms are then fixed by the analyticity conditions~\eqref{eq:first-analy} and~\eqref{eq:second-analy}, as shown in Appendix~\ref{sec:cijkl}, equations~\eqref{solg3} and~\eqref{solf3}. The \textcolor{red}{red} position-dependent correction is determined from the minimum separation equations and given in~\eqref{eq:d101}. When solved in this order, the remaining undetermined coefficients appear only in the \underline{underlined} generating functions. These are fully fixed in Section~\ref{sec:specialspecial}, although in a different order than presented here.\\
\\\\
\textbf{Counting of Undetermined Coefficients}\\
\begin{table}[h]
    \centering
    \begin{tabular}{c|c||c|c}
         & Eq. Type & Two-variable& One-variable\\\hline
       Initial Freedom  & Non-Interacting& $6$& $9$\\ \hline
      Constraint  &  Analy. Non-Int. & $2$& $0$\\
       &  Two-Magnon Int.& $3$& $0$\\
        &  Two-M. Analy.+Parity& $0$& $4$\\
         &  Three-Magnon& $0$& $3$\\\hline
         Final Freedom& & $1$& $2$
    \end{tabular}
    \caption{We list the counting of the position-dependent corrections starting from implementing the non-interacting recursion relations to the generating function \eqref{eq:def-gen}. Non-interacting equations give the initial freedom as given in \eqref{eq:nonint-gen}. Then analyticity conditions on this form of the generating function \eqref{eq:first-analy} and \eqref{eq:second-analy} bring the first set of constraints. Two-magnon interacting equations solved with additional constraints result in \eqref{eq:special-yz}, \eqref{eq:special-xz} and \eqref{eq:special-xy}. Ensuring the analyticity of these expressions in a parity-invariant way brings four more constraints. Finally, solving the three-magnon interaction equations gives \eqref{eq:special-z}, \eqref{eq:special-x}, and \eqref{eq:special-y}. Here, we step-by-step count the constraints coming from the interaction equations and analyticity conditions. In the three-magnon problem, the generating function was a two-variable quantity and the special solution was leaving a one-variable function many freedom, in parallel four-magnon problem is a three-variable function, and after the special solution, we are left with a two-variable function.}
    \label{tab:count}
\end{table}\\
After imposing the recursion relations derived from the non-interacting equations \eqref{non-int-recd}, we find that there are six two-variable generating functions and nine one-variable generating functions that are not fixed, the latter being contained in the former. Imposing the analyticity condition reduces the number of undetermined two-variable generating functions to four, which agrees with the counting of the non-interacting equations as recurrence relations.

Then, the special solution fixes three out of four two-variable generating functions and leaves us with only one two-variable generating function, $\partial_y\tilde G(x,0,z)$ which we will use to connect the solution of the four-magnon problem with three-magnons. Note that the solution of analyticity conditions given in \eqref{eq:first-analy} and \eqref{eq:second-analy}, do not fix the one-variable generating functions which lie in the boundary of these two-variable ones, however the additional analyticity conditions coming from the special two-magnon solution above fixes three of the single-variable pieces, additionally imposing parity fixes one more. Then, incorporating the three-magnon interaction equations and two-two magnon interaction equations fix three one-variable generating functions and we are left with two one-variable generating functions as listed in Table \ref{tab:count}. Recall that the generating function of the three-magnon problem was a two-variable quantity and the special solution was leaving a one-variable function many freedom, in parallel four-magnon problem is a three-variable function and after the special solution, we are left with a two-variable function. And the two single-variable functions which remain undetermined are the boundaries of this one two-variable function.

After investigating the relation between the eigenvectors with different numbers of excitations we constrain this solution further and get rid of all the leftover freedom.
\\\\\textbf{Permutation Symmetry is Restored}\\
The main idea behind finding the special solution described in this section is to express the position-dependent corrections solely in terms of permutations of momentum variables, for fixed magnon separations (i.e., fixed values of $n, m, r$). In other words, we seek a single function that captures all such corrections, where the only difference between them is the ordering of the momentum variables according to the corresponding permutation label:
\begin{align}
    D_{1234}^{n,m,r} &\longleftrightarrow f(p_{1}, p_{2}, p_{3}, p_{4}), \\
    D_{2134}^{n,m,r} &\longleftrightarrow f(p_{2}, p_{1}, p_{3}, p_{4}).
\end{align}
To achieve this, we start with the generating functions which include the scaled position-dependent corrections. These form of generating function only include the factors of total energy, total momentum, and $\kappa$ which are invariant under the permutation of indices. This way all dependence on permutation of momenta are absorbed into the scaled position-dependent corrections. Then we construct generating functions that depend on a single permutation index, using initial conditions given in \eqref{eq:d101} and solutions to the two-variable components such as those in \eqref{eq:special-yz}. By either setting all undetermined parts of the generating function to zero or assuming they are analytic functions around the origin, which have the same permutational symmetry, we reduce the full set of twenty-four generating functions $\tilde G_\sigma(x, y, z)$, each encoding a distinct position-dependent correction, to a single unified generating function,
\begin{align}
    \tilde G_\sigma(x,y,z)\longrightarrow \tilde G(p_{\sigma(1)},p_{\sigma(2)},p_{\sigma(3)},p_{\sigma(4)},\kappa;x,y,z) \;.
\end{align}
This way by permuting the order of momentum variables that we input to the generating function we are able to obtain the position-dependent corrections for distinct permutation labels $\sigma$. This implies that the position-dependent corrections admit a unifying description among the permutation of the indices,
\begin{align}
    D_\sigma^{n,m,r}=(\kappa-\kappa^{-1})A_\sigma f(\Vec{p}_\sigma;n,m,r,\kappa)=\oint_B \frac{dx dy dz}{(2\pi i)^3}\frac{G(\Vec{p}_\sigma,\kappa;x,y,z)}{x^{n+1}y^{m+1}z^{r+1}}\;.\label{eq:fin-spe-sol}
\end{align}
To emphasize this point better we use the Yang operator formalism and analyze the coefficients that accompany the plane waves with respect to the permutation indices after fully constraining the solution.

\section{From $M+1$-Magnons to $M$-Magnons}
\label{sec:smearing}
In this section, we investigate the relation between eigenstates of the Hamiltonian with different numbers of excitations. Our goal is to develop a systematic method that starts from an $M+1$-magnon eigenstate and yields the corresponding $M$-magnon eigenstate. This method must operate directly on the CBA and its long-range generalization, while also accommodating the structure of our model—where the $M=1,2$ magnon problems are solved by the CBA, and the $M \geq 3$ magnon problems require the long-range Bethe ansatz.

In conventional integrable models solvable by the CBA, the factorization in terms of two-magnon scattering amplitudes provides strong evidence for underlying integrability. However, in the case of the Hamiltonian~\eqref{Hamiltonian}, only the $M = 1,2$ magnon sectors admit standard CBA solutions; for $M \geq 3$, position-dependent corrections must be incorporated. To bridge the gap between these cases, we introduce a limiting procedure that averages over all possible positions of a chosen magnon in an $(M+1)$-magnon eigenstate. Physically, this amounts to `smearing' one magnon, which we show yields the corresponding $M$-magnon eigenstate.

We demonstrate that, under suitable assumptions detailed below, this limit establishes a concrete relation between eigenstates with successive magnon numbers. As we will see, this approach not only clarifies the recursive structure of the wave functions but also allows us to constrain all previously undetermined position-dependent corrections in the solution constructed in Section~\ref{sec:special}.

\subsection{The Case of the Coordinate Bethe Ansatz}
\label{sec:CBAsmearing}

To explain the method, we begin by applying it to a generic three-magnon eigenstate expressed in standard CBA form. For an integrable model such as the Heisenberg XXX or XXZ spin chain, the two-magnon wave function takes the form
\begin{align}
\Psi(p_1,p_2,l_1,l_2)=A^{(2)}_{12} e^{ip_1l_1+ip_2l_2}+A^{(2)}_{21} e^{ip_2l_1+ip_1l_2}\;,
\end{align}
defined up to an overall normalization,
\begin{align}
    \ket{\Psi(p_1,p_2)}=C\sum_{l_1<l_2}\Psi(p_1,p_2,l_1,l_2) |l_1,l_2\rangle \;,
\end{align}
that can be chosen such that  $A^{(2)}_{12}=1$. Similarly, the three-magnon wave function takes the form,
\begin{align}
    \Psi(p_1,p_2,p_3,l_1,l_2,l_3)=\sum_{\sigma\in\mathcal{S}_3} A^{(3)}_{\sigma}e^{ip_{\sigma(1)}l_1+ip_{\sigma(2)}l_2+ip_{\sigma(3)}l_3} \; .
\end{align}
In integrable models, the three-magnon scattering coefficients $A^{(3)}_\alpha$ factorize into products of two-magnon coefficients  $A^{(2)}_\sigma$ as
\begin{equation}
    A^{(3)}_{\sigma}=\prod_{\alpha_i}A^{(2)}_{\alpha_i} \; ,\label{factor}
\end{equation}
where we have decomposed the permutation $\sigma$ into a product of transpositions $\sigma=\alpha_1 \alpha_2 \alpha_3$. Although this decomposition is not unique, the YBE guarantees consistency among the possible decompositions.

To smear the magnon at position $l_3$, we average the wave function over all possible positions it can occupy, normalized by the total number of positions:
\begin{align}
    \lim_{N\to\infty}\frac{1}{N}\sum_{m=0}^N \psi(p_1,p_2,p_3;n,m)&=\lim_{N\to\infty}\left(\sum_{\sigma\in\mathcal{S}_3}A^{(3)}_\sigma e^{i(n+1)(p_{\sigma(2)}+p_{\sigma(3)})+ip_{\sigma(3)}}\frac{1}{N}\sum_{m=0}^Ne^{ip_{\sigma(3)}m}\right)\;.\label{cbaNsum}
\end{align}
To `localize' the smeared magnon, we scale its momentum as $p_3=2\pi\frac{\bar{p}_3}{N}$. In this limit, the magnon becomes stationary at $\bar{p}_3=0$, and the summation yields:
\begin{align}
    \lim_{N\to\infty}\frac{1}{N}\sum_{m=0}^N &\Psi(p_1,p_2,p_3;n,m)=\frac{i}{2\pi}\left(A^{(3)}_{123} e^{i(n+1)p_{2}}+A^{(3)}_{213} e^{i(n+1)p_{1}}\right)\frac{1-e^{2\pi i\bar{p}_3}
    }{\bar{p}_{3}}\;.\label{cba3to2limN}
\end{align}
Here, the exponential factor 
$e^{i(n+2) p_{\sigma(3)}}$  appearing in front of the scattering coefficients $A^{(3)}_\sigma$ simplifies to unity when $\sigma(3)=3$  , and vanishes for
$\sigma(3)=1,2$ 
as we take $N\to \infty$ limit due to
\begin{align}
    \lim_{N\to\infty}\frac{1}{N}\frac{1-e^{ip_{\sigma(3)}N}}{1-e^{ip_{\sigma(3)}}}=0\;,\quad\quad \text{for}\quad \sigma(3)=1,2\;,
\end{align}
as neither $p_1$ nor $p_2$ is localized at zero. Thus, by subsequently taking the limit $\bar{p}_3 \to 0$ , we recover the two-magnon wave function (after removing the center-of-mass phase):
\begin{align}
    \lim_{\bar{p}_3\to0}\lim_{N\to\infty}\frac{1}{N}\sum_{m=0}^N &\tilde\Psi(p_1,p_2,p_3;n,m)=A^{(3)}_{123} e^{i(n+1)p_{2}}+A^{(3)}_{213} e^{i(n+1)p_{1}}\;.\label{cba3to2}
\end{align}
This demonstrates that the two-magnon eigenstate can be recovered from the three-magnon eigenstate via the smearing limit. Crucially, this construction relies on the integrability of the model, which ensures that the three-magnon scattering coefficients factorize according to~\eqref{factor}. 
For instance, assuming the normalization $A_{123}=1$, we obtain:
\begin{align}
    A^{(3)}_{123}=A^{(2)}_{12} =1 \;, \qquad A^{(3)}_{213}=A^{(2)}_{21} \;. 
\end{align}
We emphasize that this smearing procedure applies to any CBA solution with $M+1$ number of magnons (for $M\geq1)$), and yields the corresponding $M$-magnon solution. In particular, smearing a magnon in the two-magnon eigenstate recovers the corresponding one-magnon eigenstate, up to a normalization factor determined by the original wave function. In the following, we apply the same strategy to the long-range Bethe ansatz.

\subsection{From Three-Magnon to Two-Magnon in Long-Range Bethe Ansatz}

In this section, we test the relation between the two-magnon CBA solution and the three-magnon long-range Bethe ansatz solution of our Hamiltonian. We particularly focus on the special solution with factorized $S_\kappa$ scattering coefficients given in \cite{Bozkurt:2024tpz} for the states with $Q_{12}Q_{21}Q_{12}$ order of the magnons. We refer to the particular wave function coming from that special solution $\Psi(p_1,p_2,p_3;l_1,l_2,l_3)$. Recall that the long-range Bethe ansatz is given as
\begin{align}
    \Psi(p_1,p_2,p_3;&l_1,l_2,l_3)=\sum_{\sigma\in\mathcal{S}_3
    }\left(A_\sigma+D_\sigma^{l_2-l_1-1,l_3-l_2-1}\right)e^{i\Vec{p}_\sigma^{\,3}\cdot\Vec{l}^{\,3}}\\
    &=e^{il_1\sum_i^3p_i}\sum_{\sigma\in\mathcal{S}_3}\left(A_\sigma+D_\sigma^{n,m}\right)e^{i(n+1) (p_{\sigma(2)}+p_{\sigma(3)})+i(m+1)p_{\sigma(3)}} \;,
\end{align}
for simplicity, we replace the position of magnons with the separation between them, $n=l_2-l_1-1$ and $m=l_3-l_2-1$. Following the same steps given for the CBA, we sum over all possible positions that the magnon at $l_3$ can take
\begin{align}
    \lim_{N\to\infty} \frac{1}{N}\sum_{l_3=l_2+1}^N&\Psi(p_1,p_2,p_3,l_1,l_2,l_3)\nonumber\\&=e^{il_1\cPt}\lim_{N\to\infty} \sum_{\sigma\in\mathcal{S}_3}e^{i(n+1) (p_{\sigma(2)}+p_{\sigma(3)})}\frac{1}{N}\sum_{m=0}^N\left(A_\sigma+ D_{\sigma}^{n,m}\right)e^{i(m+1)p_{\sigma(3)}} \;.
\end{align}
After scaling one of the momentum variables as $p_3=\bar{p}_3/N$ and localizing the limit of the sum of the wave functions at $\bar{p}_3\to0$, we obtain,
\begin{align}
    \lim_{\bar{p}_3\to0}\lim_{N\to\infty} \frac{1}{N}\sum_{l_3=l_2+1}^N&\Psi(p_1,p_2,p_3,l_1,l_2,l_3)=\Psi_{11}(p_1,p_2,l_1,l_2)\nonumber\\&+\lim_{\bar{p}_3\to0}\lim_{N\to\infty}\frac{1}{N}e^{il_2\cPt} \sum_{\sigma\in\mathcal{S}_3}e^{i(p_{\sigma(3)}-p_{\sigma(1)})}\sum_{m=0}^N \tilde D_{\sigma}^{n,m} \;,
\end{align}
by using the result \eqref{cba3to2}, such that two-magnon wave function, $\Psi_{11}(p_1,p_2,l_1,l_2)$ is the two-magnon eigenstate of our model with $S_\kappa$ scattering coefficients given in \eqref{eq:two-wave1} and $\tilde D$ is the scaled position-dependent correction for three-magnons as defined in \eqref{eq:scaledDs}. Since the contribution of the scattering coefficients after smearing directly gives the wave function of the two-magnon CBA solution $\Psi_{11}(p_1,p_2,l_1,l_2)$, we expect the contribution with the position-dependent corrections to vanish. To show this, we strictly focus on that part of the computation. Crucially, we know the $N\to\infty$ limit for the sum of the position-dependent corrections can be written as
\begin{align}
     \lim_{N\to\infty}\sum_{m=0}^N \tilde D_{\sigma}^{n,m}=\sum_{m=0}^\infty \tilde D_{\sigma}^{n,m}=\frac{1}{n!}\frac{\partial^n \tilde G_\sigma(x,1)}{\partial x^n} \Bigg|_{x=0} \;,
\end{align}
such that $\tilde G_\sigma(x,y)=G_\sigma\left(e^{-ip_{\sigma(1)}}x,e^{ip_{\sigma(3)}}y\right)$ is the three-magnon generating function which encodes the $S_\kappa$ special solution of the three-magnon problem as it is given in Appendix B of \cite{Bozkurt:2024tpz} and it is summarized in Appendix \ref{sec:three} here. For each integer $n$, the coefficients $\frac{\partial^n \tilde G_\sigma(x,1)}{\partial x^n} \Bigg|_{x=0}$ are a function of total energy, $\EEt$ and total momentum, $\cPt$ as well as $\kappa$,
{\small\begin{align}
   \tilde G_\sigma(x,1)=\left[F_\sigma(x)+H_\sigma(x)\right]/\big[\kappa  e^{i p_3} \left(-\EEt\, e^{i\mathcal{P}_3} x+e^{2i\mathcal{P}_3}+e^{i\mathcal{P}_3} x^2+2 e^{i\mathcal{P}_3} x+e^{i\mathcal{P}_3}+x^2\right) \left(-\EEt\, \kappa  x+\kappa +\kappa  x^2+2 x\right)\nonumber\\\left(-\EEt\, \kappa  \rho_++\kappa +\kappa  \rho_+^2+2 \rho_+\right) (e^{i\mathcal{P}_3}-2 \kappa  \rho_++1) \left(e^{i p_1} S_\kappa(p_1,p_2)+e^{i p_2}\right) (S_\kappa(p_1,p_2) S_\kappa(p_2,p_3)+S_\kappa(p_1,p_3))\big]\;,
\end{align}}
such that the function $H_\sigma(x)$ includes only part of the generating function which is not fixed by the initial assumptions of the three-magnon special solution and we set it to zero previously,
\begin{align}
    H_\sigma(x)=&-\kappa  e^{i p_3} x^2 (1+e^{-i\mathcal{P}_3}x) \left(\rho_+ (2-\EEt\, \kappa )+\kappa +\kappa  \rho_+^2\right) \left(e^{i p_1} S_\kappa(p_1,p_2)+e^{i p_2}\right)\nonumber\\ &(S_\kappa(p_1,p_2) S_\kappa(p_2,p_3)+S_\kappa(p_1,p_3)) (e^{i\mathcal{P}_3} (\kappa -2 y)+\kappa  x) (e^{i\mathcal{P}_3} y-2 \kappa  \rho_++1)\partial_y G_\sigma(x,0)\nonumber\\ +&\kappa  e^{i p_3} \rho_+^2 (1+e^{-i\mathcal{P}_3}\rho_+) \left(x (2-\EEt\, \kappa )+\kappa +\kappa  x^2\right)\left(e^{i p_1} S_\kappa(p_1,p_2)+e^{i p_2}\right) \nonumber\\&(S_\kappa(p_1,p_2) S_\kappa(p_2,p_3)+S_\kappa(p_1,p_3)) (e^{i\mathcal{P}_3} y-2 \kappa  x+1) (e^{i\mathcal{P}_3}(\kappa -2 y)+\kappa  \rho_+)\partial_y G_\sigma(\rho_+,0)\;,
\end{align}
and $F_\sigma(x)$ represents the rest of the numerator of the three-magnon special solution of the generating function which is a polynomial in variable $x$ and its coefficients are parametrized by momentum variables and $\kappa$. It is explicitly given in the Mathematica file provided \texttt{fourmagnon.np}. Additionally, we adopt the shorthand notation for the expression,
\begin{align}
    \rho_+=\rho_+(1)=\frac{\sqrt{-e^{i\mathcal{P}_3} \left(e^{i\mathcal{P}_3} \left(-\EEt\,^2+4 \EEt\,+4 e^{i\mathcal{P}_3}+4\right)+4\right)}+(\EEt\,-2) e^{i\mathcal{P}_3}}{2 (e^{i\mathcal{P}_3}+1)}\;,
\end{align}
which is coming from the factorization of the three-magnon curve that we used to ensure the analyticity of the generating function. If we assume that $G_\sigma^{(0,1)}(x,0)$ is an analytic function around $x=0$ and does not have a pole at $x=\rho_+$ then  $\frac{\partial^n \tilde G_\sigma(x,1)}{\partial x^n} \Bigg|_{x=0}$ exists for each positive integer, and it is parametrized by nothing but momentum variables and $\kappa$. This is not affected by scaling one of the momentum variables as $p_3=\frac{\bar{p}_3}{N}$. Therefore we conclude that for any positive integer $n$
\begin{align}
    \lim_{N\to\infty}\sum_{m=0}^N \tilde D_{\sigma}^{n,m}=\textit{finite}
\end{align}
and since scaling the finite sum $\sum_{m=0}^N \tilde D_{\sigma}^{n,m}$ with a positive integer cannot cause a blow-up in the limit, 
\begin{align}
    \lim_{N\to\infty}\frac{1}{N}\sum_{m=0}^N \tilde D_{\sigma}^{n,m}=0\;.
\end{align}
Thus, when the third magnon is smeared, the effect of position-dependent corrections is washed off. Similarly, when we smear the first magnon instead of the last one, meaning the $Q_{12}$ magnon at $l_1$,  we should start from the special solution of the three-magnon problem with the scattering coefficients factorized in terms of $S_{1/\kappa}$ factors that were also computed in the Section $5.2$ of \cite{Bozkurt:2024tpz}. Then, the smearing process gives us the wave function of the eigenstate of the two-magnon problem with the magnons $Q_{21}Q_{12}$ and $S_{1/\kappa}$ as its scattering coefficient.

The case of the four-magnon long-range Bethe ansatz is more intricate. Unlike in the three-to-two magnon reduction, where the position-dependent corrections vanish in the smearing limit, here we require that they do not vanish. Instead, they must reduce consistently to the position-dependent corrections of the three-magnon problem. This will ensure compatibility of the long-range structure across different magnon sectors and provide a stringent consistency check of our results.

\subsection{From Four-Magnon to Three-Magnon in Long-Range Bethe Ansatz}

We now turn to the smearing procedure applied to the special solution of the four-magnon problem, as presented in Section~\ref{sec:special}. Here we derive the requirements of smearing a magnon in four-magnon eigenstate and recovering the corresponding special solution of the three-magnon problem with the correct factorized scattering coefficients. 
For all $M \geq 3$, the construction of eigenstates in our model necessarily involves position-dependent corrections. It is therefore essential to show that these corrections do not generically vanish under smearing—contrary to the behavior observed in the reduction from three-to-two magnons. 
Establishing this point further supports the existence of a family of $M$-magnon eigenstate solutions connected via smearing. This transition from the four- to the three-magnon state completes the web of relations among eigenstates in our model.

Moreover, we will demonstrate that these relations among eigenstates with differing numbers of excitations impose nontrivial constraints on the position-dependent corrections—constraints that are not manifest in either the general or special solutions of the eigenvalue problem alone. These additional conditions refine the structure of the allowed solutions and enhance our understanding of the solution space. Importantly, they lay the groundwork for a systematic extension of the long-range Bethe ansatz to an arbitrary number of magnons—a direction we pursue in forthcoming work~\cite{future}.

We now focus on the special solution of the four-magnon long-range Bethe ansatz introduced in Section~\ref{sec:special}. The corresponding wave function for an open infinite spin chain is conveniently expressed in terms of the  coordinate $l_1$ and the relative separations $(n,m,r)$ and takes the following form:
\begin{align}
\Psi&(p_1,p_2,p_3,p_4; l_1,l_2,l_3,l_4) \nonumber \\
&= e^{i l_1 \cP} \sum_{\sigma \in \mathcal{S}4} \left( A_\sigma + D_\sigma^{n,m,r} \right) e^{i(n+1) (p_{\sigma(2)} + p_{\sigma(3)} + p_{\sigma(4)}) + i(m+1)(p_{\sigma(3)} + p_{\sigma(4)}) + i(r+1)p_{\sigma(4)}} \;.
\end{align}

To implement the smearing of the fourth magnon, located at $l_4$, we sum over all possible values it can take while keeping the remaining coordinates fixed. This results in the following expression:
\begin{align}
\lim_{N \to \infty} \frac{1}{N} \sum_{l_4 = l_3 + 1}^N & \Psi(p_1,p_2,p_3,p_4; l_1,l_2,l_3,l_4) \nonumber \\
&= e^{i l_1 \cP} \lim_{N \to \infty} \frac{1}{N} \sum_{\sigma \in \mathcal{S}4} e^{i \vec{p} \cdot \vec{n}} \left( \sum_{r=0}^N A_\sigma \, e^{i(r+1)p_{\sigma(4)}} + \sum_{r=0}^N D_\sigma^{n,m,r} \, e^{i(r+1)p_{\sigma(4)}} \right) \;.
\end{align}
Here, $\Vec{p}\cdot\Vec{n}$ denotes the plane wave contributions which are not directly related to smearing, $\Vec{p}\cdot\Vec{n}=(n+1) (p_{\sigma(2)}+p_{\sigma(3)}+p_{\sigma(4)})+(m+1)(p_{\sigma(3)}+p_{\sigma(4)})$. Using the generalized version of the result in \eqref{cba3to2}, we extend the smearing procedure from the four-magnon wave function to the three-magnon wave function, where the scattering coefficients factorize in terms of $S_\kappa$. This allows us to derive the scattering coefficients for the three-magnon long-range Bethe ansatz special solution directly from those of the four-magnon case. To obtain the special solution for the three-magnon problem, we take the following limit 
\begin{align}
    \lim_{\bar{p}_4\to0}\lim_{N\to\infty} \frac{1}{N}\sum_{l_4=l_3+1}^N&\Psi_{11}(p_1,p_2,p_3,p_4,l_1,l_2,l_3,l_4)\stackrel{!}{=}\Psi_{12}(p_1,p_2,p_3,l_1,l_2,l_3)\;,\label{eq:lim4to3}
\end{align}
for $\bar{p}_4=p_4/N$. This implies a corresponding matching condition between the position-dependent corrections of the four- and three-magnon solutions:
\begin{align}
    \lim_{\bar{p}_4\to0}e^{il_1\sum_i^4p_i}\lim_{N\to\infty} \frac{1}{N}\sum_{\sigma\in\mathcal{S}_4}e^{i\Vec{p}\cdot\Vec{n}}\sum_{r=0}^N D_{\sigma}^{n,m,r}e^{i(r+1)p_{\sigma(4)}}=e^{il_1\sum_i^3p_i}\sum_{\sigma\in\mathcal{S}_3}e^{i\Vec{p}\cdot\Vec{n}}D_{\sigma}^{n,m}\;. \label{eq:matching4and3}
\end{align}
Since both the three- and four-magnon position-dependent corrections admit factorized representations—see equation~\eqref{eq:fin-spe-sol}—we expect this limit to hold separately for each permutation. Therefore, similarly to the three-magnon case given in the previous section, first we focus on the infinite sum of position-dependent corrections, and we express this in terms of the four-magnon generating function which encodes the special solution of the four-magnon eigenvalue problem as in \eqref{eq:solG4},
\begin{align}
    \lim_{N\to\infty}\sum_{r=0}^N \tilde D^{n,m,r}_\sigma=\sum_{r=0}^\infty \tilde D^{n,m,r}_\sigma=\frac{1}{n!\;m!}\frac{\partial^n\partial^m \tilde G^{(4)}_\sigma(x,y,1)}{\partial x^n\partial y^m}\Bigg|_{x,y=0}\;.\label{eq:coefGfour}
\end{align}
To avoid confusion, we label the generating functions $\tilde G^{(M)}$ to indicate the total number  $M$ of magnons in the corresponding spin chain.  A crucial observation is that taking the limit $z \to 1$ at the level of the generating function corresponds exactly to performing the infinite sum over the relative position $r$. This connection will play a key role in the forthcoming analysis.

Since analyticity around the origin was ensured in Section~\ref{sec:analy} by imposing the conditions~\eqref{eq:first-analy} and~\eqref{eq:second-analy}, the generating function encoding the four-magnon eigenvalue problem~\eqref{eq:eig4} admits a convergent Taylor expansion around~$z=0$. Consequently, if the limit  $z \to 1$ were nonsingular, the situation would mirror that of the three-magnon case described in the previous section, and the left-hand side of \eqref{eq:matching4and3} would vanish. However, this would result in a CBA-type three-magnon wave function that fails to solve the three-magnon eigenvalue problem.

Consequently, the four-magnon generating function must necessarily develop poles at $z = 1$ as the momentum variable localizes at $\bar{p} = 0$. More precisely, we expect the total smearing process to diverge in the absence of the $1/N$ prefactor,
\begin{align}
    \tilde G^{(4)}_\sigma(x,y,z)\to \infty\quad\quad\text{as}\quad\quad z\to1,\quad\bar{p}_4\to0\;.
\end{align}
We expect this divergence to occur in a controlled way: when we scale the partial sum by the factor $\frac{1}{N}$ and smear the magnon, this cancels the divergence and allows us to recover the correct three-magnon position-dependent coefficients. An asymptotic behavior consistent with this expectation is given by
\begin{align}
\tilde{G}^{(4)}_\sigma(x, y, z) \sim \frac{1}{1 - e^{i p_{\sigma(4)}} z}  \tilde{G}^{(3)}_\sigma(x, y) \; , \quad\quad\quad z \to 1 \; , \label{eq:4to3assum}
\end{align}
which produces the desired divergence for permutations with $\sigma(4) = 4$ when we localize $p_4 \to 0$. For permutations with $\sigma(4) \neq 4$, no divergence occurs as $z \to 1$ unless $p_{\sigma(4)} = 2 n \pi$ for some integer $n$.

This structure ensures that, during the smearing and localization process, a large subset of permutations is suppressed, leaving only those of the form $\sigma = (i j k 4)$, which correspond precisely to the permutations relevant for the three-magnon problem.

If we assume that the four-magnon generating function has the form given above, indeed
\begin{align}
    \lim_{N\to\infty}&\frac{1}{N}\sum_{r=0}^N \tilde D^{n,m,r}_\sigma=\lim_{N\to\infty}\frac{1}{N}\sum_{r=0}^N \frac{1}{n!m!r!}\frac{\partial^n\partial^m\partial^r \tilde G^{(4)}_\sigma(x,y,z)}{\partial x^n\partial y^m\partial z^r}\Bigg|_{{x,y,z=0}} \nonumber \\
&\sim\frac{1}{n!m!}\frac{\partial^n\partial^m \tilde G^{(3)}_\sigma(x,y)}{\partial x^n\partial y^m}\Bigg|_{x,y=0}\lim_{N\to\infty}\frac{1}{N}\sum_{r=0}^N \frac{1}{r!}\frac{\partial^r (z^r)}{\partial z^r}\Bigg|_{z=0}=\frac{1}{n!m!}\frac{\partial^n\partial^m \tilde G^{(3)}_\sigma(x,y)}{\partial x^n\partial y^m} \; .
\end{align}
In the remaining of this section, we analyze the special solution of the four-magnon eigenvalue problem and demonstrate, step by step, how the three-magnon position-dependent corrections emerge. We begin by examining the transition between the total energy and total momentum coefficients. Next, we show that the characteristic polynomial appearing in the denominator of the generating function—arising from the solution of the non-interacting equations \eqref{eq:def-q4}—reduces to its three-magnon counterpart. This allows us to identify the precise relations between the four-magnon and three-magnon special solutions which solve the non-interacting equations. After that, we refine relations between the solutions that also solve the two-magnon interaction equations. Finally, we show that the analyticity conditions and the solution of the lowest separation position-dependent corrections are compatible. For convenience, we summarize the relevant aspects of the special three-magnon solution coming from \cite{Bozkurt:2024tpz} in Appendix \ref{sec:three}.

Under the scaling $p_4=\frac{\bar{p}_4}{N}$, the total energy \eqref{totalenergy} and total momentum becomes
\begin{align}
    \EE&=\varepsilon_3+e^{i\frac{\bar{p}_4}{N}}+e^{-i\frac{\bar{p}_4}{N}} \; , \\
    \PP&=e^{i\mathcal{P}_3}e^{i\frac{\bar{p}_4}{N}} \; ,
\end{align}
such that the four-magnon total energy and total momentum reduce to their three-magnon subtotals in the limit $N\to\infty$. This is enough to show that the denominator of the four-magnon generating function reduces to the three-magnon expression as $N\to\infty$,
\begin{align}
    \lim_{N\to\infty} Q_4(x,y,1)= Q_3(x,y) \; , \label{eq:4to3denom}
\end{align}
where $Q_3(x,y)$ is given in \eqref{q3}. Then, to show that the special solution of the four-magnon problem has the form given in \eqref{eq:4to3assum}, first we point out that the pieces of the special solution coming from the middle two-magnon interaction equations as given in \eqref{eq:special-xz} and the three-magnon interaction equations from left as they are given in \eqref{eq:special-z}, explicitly contain factors of the form $\left(1-e^{ip_{\sigma(4)}}z\right)^{-1}$. In the limit $p_4=\bar{p}_4/N\to0$ as $N\to\infty$, these factors reduce precisely to the expected $(1-z)^{-1}$ form for permutations of the type $\sigma=(ijk4)$. After identifying evidence for a potential divergence in the four-magnon generating function, we explicitly construct the additional conditions necessary to recover, exactly, the three-magnon generating function that encodes the special solution of the three-magnon problem. We begin by recalling the form of the generating function that satisfies the four-magnon non-interacting equations, as given in \eqref{eq:gen-fun-def} and \eqref{eq:nonint-gen}. We then substitute the contributions from the special solution, specifically the expressions in \eqref{eq:special-xy}, \eqref{eq:special-xz}, and \eqref{eq:special-yz}, along with the remaining components in \eqref{eq:special-z}, \eqref{eq:special-x}, and \eqref{eq:special-y}. Finally, we impose the analyticity conditions given in \eqref{eq:first-analy} and \eqref{eq:second-analy} to arrive at the final form of the generating function, implicitly defined in \eqref{eq:solG4}.

To simplify the comparison between the four-magnon and three-magnon solutions, we focus on the structure of the generating function $G^{(4)}_\sigma(x,y,z)$. If we disregard the terms independent of the variable $z$ in \eqref{eq:nonint-gen}, we can write
\begin{align}
F^{(4)}_\sigma&(x,y,z)\supset \left( Q_4(x,y,z)+\PP y^2z+xy^2  \right) \tilde{G}_\sigma(x ,0,z) - (z+\invPP  x)xyz \partial_{y} \tilde{G}_\sigma(x,0,z) \nonumber\\
+& \left( Q_4(x,y,z)+\invPP x^2z+x^2yz \right) \tilde{G}_\sigma(0,y ,z) - (1+\PP  y) x y z \partial_{x} \tilde{G}_\sigma(0,y ,z)\nonumber \\
 -& \left(\EE x y z - x y -y z -xz^2 -xyz^2\right) \tilde{G}_\sigma(0,0,z)
 + x y z^2 \partial_{y} \tilde{G}_\sigma(0,0,z) + x y z \partial_{x} \tilde{G}_\sigma(0,0,z)\;.\label{eq:four-num-piece}
\end{align}
where $\supset$ indicates the terms contained in $F$ except for those independent of the variable $z$, and we have added a label to $F$ to indicate the number of magnons. We are not interested in the terms independent of the variable $z$ because they contribute only finite coefficients to the infinite sum over position-dependent terms, and thus will disappear in the $N\to \infty$ limit due to the $1/N$ factor. The terms that depend on the variable $z$ are potentially divergent in the limit $p_4=\bar{p}_4/N\to0$ and $z\to1$, and thus play a crucial role in recovering the position-dependent corrections of the three-magnon special solution.

Since we know the exact transformation of the denominator, as given in \eqref{eq:4to3denom}, we refine our expectation given in \eqref{eq:4to3assum} to
\begin{align}
    F^{(4)}_{ijk4}(x,y,z)\sim\left(1-e^{ip_4}z\right)^{-1}F^{(3)}_{ijk}(x,y) \; ,\label{refinedex}
\end{align}
for $(ijk)\in\mathcal{S}_3$ and we compare with the three-magnon generating function which includes the position-dependent corrections that solve the three-magnon non-interacting equations as given in \eqref{eq:generating-non-int3}. Although we recall the three-magnon special solution in Appendix \ref{sec:three}, to easily discuss the asymptotic relation given in \eqref{refinedex} we look at the numerator of the three-magnon generating function which solves the non-interacting equations,
\begin{align}
F^{(3)}_{ijk}&(x,y)=\left(Q_3(x,y)+\PPt y^2\left(1 +\invPPt x \right)\right)\tilde G^{(3)}_{ijk}(x,0)+\left(Q_3(x,y)+\invPPt x^2\left( 1 +\PPt y\right) \right)\tilde G^{(3)}_{ijk}(0,y) \nonumber \\
       &-xy\left(1+\invPPt x\right)\partial_y\tilde G^{(3)}_{ijk}(x,0)-xy\left(1+\PPt y\right)\partial_x\tilde G^{(3)}_{ijk}(0,y)+xy\left(\tilde D^{0,1}_{ijk}+ \tilde D^{1,0}_{ijk} \right)\label{eq:generating-non-int}\; .
   \end{align}
By using the transitions between the total energy and total momentum variables and the transition between the denominators of the generating function, we observe that the factors that accompany the distinct pieces of the generating function in the above expressions one-to-one match under the $\bar{p}_4\to0$ as $N\to\infty$ limit. Therefore, we can refine our expectation from the connection between the generating function even further than \eqref{refinedex} and get the following relations under the smearing limit,
\begin{align}
    &\tilde G^{(4)}_{ijk4}(x,0,z)\sim \left(1-e^{ip_4}z\right)^{-1}\tilde G^{(3)}_{ijk}(x,0)\label{refex1} \; , \\
    &\tilde G^{(4)}_{ijk4}(0,y,z)\sim \left(1-e^{ip_4}z\right)^{-1}\tilde G^{(3)}_{ijk}(0,y)\label{refex2} \; , \\
    &\partial_y\tilde G^{(4)}_{ijk4}(x,0,z)\sim \left(1-e^{ip_4}z\right)^{-1}\partial_y\tilde G^{(3)}_{ijk}(x,0)\label{refex3} \; , \\
    &\partial_x\tilde G^{(4)}_{ijk4}(0,y,z)\sim \left(1-e^{ip_4}z\right)^{-1}\partial_x\tilde G^{(3)}_{ijk}(0,y)\label{refex4} \; ,
\end{align}
furthermore, to be consistent with the asymptotic behavior given above, we expect,
\begin{align}
    -&\left(\EE x y z - x y -y z -xz^2 -xyz^2\right) \tilde{G}^{(4)}_{ijk4}(0,0,z)\nonumber\\&
 + x yz\left(\partial_{x} \tilde{G}^{(4)}_{ijk4}(0,0,z)+z \partial_{y} \tilde{G}^{(4)}_{ijk4}(0,0,z)\right)\sim\left(1-e^{ip_{4}}z\right)^{-1}2xy(\kappa-\kappa^{-1})A^{(3)}_{ijk}\;,\label{eq:cond1}
\end{align}
which we can deduce only by comparing the four-magnon generating function \eqref{eq:four-num-piece} to the three-magnon one \eqref{eq:generating-non-int3}. Intuitively, this implies the following relation between the single-variable generating functions coming from the four-magnon problem and the single position-dependent corrections of three-magnon problem,
\begin{align}
    \tilde G_{ijk4}^{(4)}(0,0,z)&\sim \textit{finite}\;,\label{eq:intz1}\\ \partial_x \tilde G_{ijk4}^{(4)}(0,0,z)&\sim \left(1-e^{ip_4}z\right)^{-1}D_\sigma^{1,0}\;,\label{eq:intz2}\\\partial_y \tilde G_{ijk4}^{(4)}(0,0,z)&\sim \left(1-e^{ip_4}z\right)^{-1}D_\sigma^{0,1}\;,\label{eq:intz3} 
\end{align}
as $z\to1$ such that $\tilde G_{ijk4}^{(4)}(0,0,z)$ does not blow up as $z\to1$ and once we combine it with $1/N$ we observe that under the smearing limit, this piece vanishes. This is in complete agreement with the fact that $\tilde D^{0,0}=0$ in our conventions. Up to this point, we have only considered the conditions imposed by the non-interacting equations. Now we take the special solution of the pieces of the generating function given in Section \ref{sec:specialGs} and explicitly show what happens to these pieces of the generating function under the smearing procedure to match the special solution of the pieces of the three-magnon special solution. We will show that obtaining the three-magnon special solution under smearing imposes additional constraints on the special solution of the four-magnon problem. First, we focus on the special solution given in \eqref{eq:special-yz} and \eqref{eq:special-xz} and isolate the pieces that may diverge under the smearing
\begin{align}
    &\tilde G^{(4)}_{ijk4}(0,y,z)\supset -\frac{yz \left(1+\PP  y\right) \partial_x\tilde G^{(4)}_{ijk4}(0,y,z)}{y(1-\EE z+z^2+2 \kappa  z+y)+z^2}
    \nonumber\\&\;\;+\frac{\left(y\left(1-\EE z+z^2+2 \kappa  z\right)+z^2\right)\tilde G^{(4)}_{ijk4}(0,0,z) +yz^2 \partial_y\tilde G^{(4)}_{ijk4}(0,0,z)+yz\partial_x\tilde G^{(4)}_{ijk4}(0,0,z)}{y(1-\EE z+z^2+2 \kappa  z+y)+z^2}\;.\label{subyz}
\end{align}
\begin{align}
    &\tilde G^{(4)}_{ijk4}(x,0,z)\supset-\frac{\kappa x z (\invPP x+ z) \partial_y G^{(4)}_{ijk4}(x,0,z)}{\kappa  (x z (-\EE+x+z)+x+z)+2 x z}-\frac{2xz\left(\kappa^2-1\right)A_{ijk4}^{(4)}\left(1-e^{-ip_{i}}x-e^{ip_{4}}z\right)h(x,z)}{\kappa  (x z (-\EE+x+z)+x+z)+2 x z}\nonumber\\&\;\;+\frac{\left(\kappa  (x z (-\EE+z)+x+z)+2 x z\right)\tilde G^{(4)}_{ijk4}(0,0,z)+\kappa x z^2\partial_y \tilde G^{(4)}_{ijk4}(0,0,z)+\kappa x z\partial_x \tilde G_{ijk4}^{(4)}(0,0,z)}{\kappa  (x z (-\EE+x+z)+x+z)+2 x z}\;,\label{subxz}
\end{align}
The condition \eqref{eq:cond1} is satisfied, by taking into account the special solution of the $\tilde G_\sigma(0,0,z)$ given in \eqref{eq:special-z} in the following way,
\begin{align}
    &\partial_x \tilde G^{(4)}_{ijk4}(0,0,z)+ z \partial_y \tilde G^{(4)}_{ijk4}(0,0,z) \sim \frac{2\left(\kappa-\kappa^{-1}\right)A^{(4)}_{ijk4}}{1-e^{ip_4}z}\label{smearingz}
\end{align}
which implies that in the smearing procedure $\tilde G_{\sigma}(0,0,z)\sim0$ as $z\to1$ and $N\to\infty$. The asymptotic behavior above is absolutely necessary to establish the connection between the four-magnon and three-magnon solutions, and therefore it is the first extra condition we derived between the pieces of the generating function in addition to the special solution. Note that this asymptotic relation suggests,
\begin{align}
    &\partial_x \tilde G^{(4)}_{ijk4}(0,0,z)+ z \partial_y \tilde G^{(4)}_{ijk4}(0,0,z)= \frac{2\left(\kappa-\kappa^{-1}\right)A^{(4)}_{ijk4}f_1(z)}{1-e^{ip_4}z}+f_2(z)\label{smearingzz}
\end{align}
such that as $z\to1$ these two unknown functions should behave as $f_1(z)\to1$ and $f_2(z)$ stays finite. Then, we use this to refine the special solution of two-magnon interaction equations under the smearing limit and obtain the following behavior,
\begin{align}
    &\tilde G^{(4)}_{ijk4}(0,y,z)\sim -\frac{\left(1+\PP  y\right) \partial_x\tilde G^{(4)}_{ijk4}(0,y,z)}{2\kappa-\EEt\,+y+y^{-1}}+\frac{2\left(\kappa-\kappa^{-1}\right)A_{ijkl}^{(4)}}{2\kappa-\EEt\,+y+y^{-1}}\;.\label{asym-yz}
\end{align}
Similarly, we obtain,
\begin{align}
    \tilde G^{(4)}_{ijk4}(x,0,z)\sim-&\frac{(1+\invPP x) \partial_y G^{(4)}_{ijk4}(x,0,z)}{ 2\kappa^{-1}-\EEt\,+x+x^{-1}}+\frac{2\left(\kappa-\kappa^{-1}\right)A^{(4)}_{ijk4}}{\left(2\kappa^{-1}-\EEt\,+x+x^{-1}\right)(1-e^{-ip_{i}}x)(1-e^{ip_{4}}z)}\;.\label{asym-xz}
\end{align}
Comparing these asymptotic expressions to the special solution of the three-magnon interaction equations given in \eqref{sol3y} and \eqref{sol3x}, we conclude that the initial expectation given in \eqref{refex3} should be refined as
\begin{align}
    &\partial_y\tilde G^{(4)}_{ijk4}(x,0,z)\sim \left(1-e^{ip_4}z\right)^{-1}\partial_y\tilde G^{(3)}_{ijk}(x,0)+\frac{2\left(\kappa-\kappa^{-1}\right)A^{(4)}_{ijk4}+(1-e^{-ip_{i}}x)c_{ijk}(x)}{\left(1+\invPPt x\right)(1-e^{-ip_{i}}x)(1-e^{ip_{4}}z)}\label{refex32}
\end{align}
Remarkably, once this piece of generating function behaves as given above, in addition to the behavior assumed in \eqref{smearingz}, we conclude by the two-magnon interaction equations that the rest of the two-variable generating functions should indeed behave as
\begin{align}
     &\tilde G^{(4)}_{ijk4}(x,0,z)\sim \left(1-e^{ip_4}z\right)^{-1}\tilde G^{(3)}_{ijk}(x,0)\label{refex12}\\
    &\tilde G^{(4)}_{ijk4}(0,y,z)\sim \left(1-e^{ip_4}z\right)^{-1}\tilde G^{(3)}_{ijk}(0,y)\label{refex22}\\
    &\partial_x\tilde G^{(4)}_{ijk4}(0,y,z)\sim \left(1-e^{ip_4}z\right)^{-1}\partial_x\tilde G^{(3)}_{ijk}(0,y)\label{refex42}
\end{align}
Since the three-magnon special solution does not fix the piece of the generating function $\partial_y\tilde G^{(3)}_{ijk}(x,0)$ and four-magnon special solution also does not fix $\partial_y\tilde G^{(4)}_{ijkl}(x,0,z)$, the extra factor that appears in the asymptotic relation between these pieces does not contradict with the initial expectation. 

The only thing left to complete the comparison and establish the limit given in \eqref{eq:lim4to3} is to show that the analyticity conditions are compatible and the solution of the lowest separation scattering coefficients is compatible. As we explained in Section \ref{sec:analy} and at the end of the special solution given in Section \ref{sec:specialGs}, we can impose the analyticity conditions \eqref{eq:first-analy} and \eqref{eq:second-analy} after inserting the solution of the two-magnon interaction equations to the generating function which includes the solution of non-interacting equations as in \eqref{eq:nonint-gen}. Additionally, the same strategy was followed to obtain the special solution of the three-magnon solution which was presented in Appendix B of \cite{Bozkurt:2024tpz}. Then, the comparison between the analyticity conditions boils down to point out that the first analyticity conditions given in \eqref{eq:first-analy} does not depend on the variable $z$. Therefore, even if we impose it to the generating function, under the smearing limit it will not have any effect on the result for smearing the magnon from the right. At this point, we are left with the second analyticity condition as given in \eqref{eq:second-analy}, where we observe that 
\begin{align}
   \lim_{N\to\infty} x^{(4)}_\pm(y,z)= \rho^{(3)}_\pm(y)
\end{align}
for $\rho^{(3)}_\pm(y)$ given in \eqref{eq:elpara}. This shows that the analyticity condition is consistent under the smearing limit and the relation we obtain about pieces of generating function by solving \eqref{eq:second-analy} will boil down to the relation coming from the analyticity condition of three-magnon solution. Finally, due to the factorized form of the scattering coefficients, we immediately see
\begin{align}
    A^{(4)}_{ijk4}=A^{(3)}_{ijk}\label{incond1}
\end{align}
which implies that
\begin{align}
    D_{ijk4}^{1,0,0}=D_{ijk4}^{0,0,1}=D_{ijk}^{1,0}=D_{ijk}^{0,1}\;.\label{incond2}
\end{align}
Even inside the set of position-dependent corrections that are involved in minimum separation equations, these two pairs of position-dependent corrections play a distinguished role. Since the pair, $\tilde D_\sigma^{1,0,0}$ and $\tilde D_\sigma^{0,0,1}$ are the only position-dependent corrections involved in four-magnon interaction equations and the pair  $\tilde D_\sigma^{1,0}$ and $\tilde D_\sigma^{0,1}$ are involved in three-magnon interaction equations, we are able to solve them in the most symmetric way possible while setting the initial values of the eigenvalue problem. Since they are also the only pair independently appearing in the generating function which solves the non-interacting equations such as \eqref{eq:nonint-gen} their symmetric solution plays a crucial role in the solution of the rest of the position-dependent corrections. Therefore it makes sense to have such a connection between the four and three-magnon solutions, more precisely it makes sense to be able to connect a four-magnon solution to another special three-magnon solution and equations \eqref{incond1} and \eqref{incond2} are the necessary conditions which ensure that the initial conditions of these two eigenvalue problems match. It is certainly not a sufficient condition for smearing limit to work. For the rest of the position-dependent corrections, we can only write a direct relation under the smearing limit, once the conditions given above are satisfied,
\begin{align}
    \lim_{\bar{p}_4\to0}\lim_{N\to\infty}&\frac{1}{N}\sum_{r=0}^N \tilde D^{n,m,r}_{ijk4}=\tilde D_{ijk}^{n,m}
\end{align}
and 
\begin{align}
    \lim_{\bar{p}_4\to0}\lim_{N\to\infty}&\frac{1}{N}\sum_{r=0}^N \tilde D^{n,m,r}_{ijkl}=0\quad\quad\text{for}\quad\quad l\neq 4
\end{align}
which leads us to conclude \eqref{eq:lim4to3}.

Additionally, we can perform the smearing limit from the four-magnon solution to the three-magnon solution from the left as well. In this case, the limit would take the following form,
\begin{align}
    \lim_{\bar{p}_1\to0}\lim_{N\to\infty} \frac{1}{N}\sum_{l_1=-N}^{l_2-1}&\Psi_{11}(p_1,p_2,p_3,p_4,l_1,l_2,l_3,l_4)=\Psi_{21}(p_2,p_3,p_4,l_2,l_3,l_4)\;,\label{eq:lim4to3left}
\end{align}
which would correspond to the asymptotic behaviour,
\begin{align}
\tilde{G}^{(4)}_\sigma(x, y, z) \sim \frac{1}{1 - e^{-i p_{\sigma(1)}} x}  \tilde{H}^{(3)}_\sigma(y,z) \; , \quad\quad\quad x \to 1 \; , \label{eq:4to3assumleft}
\end{align}
for the solution of the three-magnon problem for the configuration $Q_{21}Q_{12}Q_{21}$ given in Appendix \ref{sec:three}. This limit would correspond to looking at asymptotics of the generating function as $x\to1$ together with $N\to\infty$ and we would pick up the poles of the type $\left(1-e^{ip_{\sigma(1)}}x\right)^{-1}$ as it can be seen from the special solution given in Section \ref{sec:special}. Note that the three-magnon solution we obtain under the smearing limit would correspond to an eigenstate with overall color indices, $\ket{\Psi}_{21}$ and the position states would be the $\mathbb Z_2$ conjugates of the ones we obtain for smearing limit from the right in this section. However, scattering coefficients would still factorize in terms of $S_\kappa$ since we started with the four-magnon solution which only has $S_\kappa$s. Nevertheless, we know that this is another special solution which is the $\mathbb Z_2$ conjugate of the solution given in Section $(5.2)$ of \cite{Bozkurt:2024tpz}.

Moreover, we could also perform a `double' smearing limit by taking the separation between the second and the third magnon to infinity,
\begin{align}
    \lim_{\bar{p}_4\to0}\lim_{\bar{p}_3\to0}\lim_{N\to\infty} \frac{1}{N}\sum_{l_3=l_2+1}^{N}&\Psi_{11}(p_1,p_2,p_3,p_4,l_1,l_2,l_3,l_3+1)=\Psi_{11}(p_1,p_2,l_1,l_2)\;.\label{eq:lim4to3middle}
\end{align}
Then, we are left with the two-magnon solution since we do not have any poles of type $(1-y)^{-1}$ in the special solution of the special solution as $y\to1$. This observation makes the $y\to1$ limit very similar to the case of smearing the three-magnon to the two-magnon solution and the position-dependent corrections vanish under this limit and we are left with the two-magnon solution.

\section{Further Constraining the Four-Magnon Solution}
\label{sec:specialspecial}
In this section, inspired by the smearing limit, we further constrain the solution we constructed in Section~\ref{sec:special} using the special three-magnon solution obtained in Section 5 of \cite{Bozkurt:2024tpz}. The final form of the four-magnon solution we obtain here, both solves the eigenvalue equations obtained in Section~\ref{sec:fourmagnon} and reduces to the three-magnon solution under the smearing procedure, but we do not claim this solution to be unique because it depends on the particular solution we chose for minimum separation equations given in Section \ref{sec:minsepspecial}.

In Section~\ref{sec:special}, we followed a specific order to solve the eigenvalue equations and analyticity conditions. However, for the purposes of this section, it is more convenient to change this order. This allows us to construct a solution that both satisfies the relations between generating functions—derived in Section~\ref{sec:special}—and explicitly depends on the three-magnon solution. Our strategy begins with the two-magnon interaction equations, starting from~\eqref{eq:special-y}. We incorporate the asymptotic behavior introduced in~\eqref{eq:lim4to3middle}, and use the remaining freedom from Section~\ref{sec:special} to fix certain single-variable functions. We then verify that the resulting solution agrees with the one in Section~\ref{sec:minsepspecial}, at leading orders in the Taylor expansion.

Next, we express the two-variable generating functions in terms of the single-variable generating functions obtained from the three-magnon solution. In this construction, we ensure that all interaction equations of the four-magnon problem are satisfied:~\eqref{eq:special-z}, \eqref{eq:special-x}, \eqref{eq:special-y}, as well as~\eqref{eq:special-yz}, \eqref{eq:special-xz}, and~\eqref{eq:special-xy}. We also impose analyticity on the two-magnon interaction equations to determine the remaining single-variable generating functions. Finally, we complete the solution by imposing analyticity conditions on the full three-variable generating function.

Let us start by considering the smearing of two magnons in a four-magnon eigenstate. From  \eqref{eq:lim4to3middle}, we know that the generating functions do not have any poles at $y\to 1$. Also, the initial conditions discussed in section \ref{sec:minsepspecial} are compatible with setting $\tilde D^{0,1,0}_{ijkl}=0$. Thus, we have the freedom to fix the single-variable generating function,
\begin{align}
    G_{ijkl}^{(4)}(0,y,0)=0\;.
\end{align}
Then, the two two-magnon interaction equation \eqref{eq:special-y} implies,
\begin{align}
    (1 +y) \partial_z \tilde G^{(4)}_{ijkl}(0,y,0)+ (1 +\PP y) \partial_x \tilde G^{(4)}_{ijkl}(0,y,0)=2\left(\kappa-\kappa^{-1}\right)A_{ijkl} \;,
\end{align}
which can be solved by taking
\begin{align}
    \partial_z \tilde G_{ijkl}^{(4)}(0,y,0)=\frac{\left(\kappa-\kappa^{-1}\right)A_{ijkl}}{1+y}\;,\quad \partial_x \tilde G_{ijkl}^{(4)}(0,y,0)=\frac{\left(\kappa-\kappa^{-1}\right)A_{ijkl}}{1+\PP y}\;.
\end{align}
This is a simple and parity invariant (under the maps \eqref{eq:parity-map} and \eqref{eq:parity-map-xyz}) solution of the single-variable generating functions which depend on $y$. In addition, if we expand these expressions around the origin, we get
\begin{align}
    \partial_z \tilde G^{(4)}_{{ijkl}}(0,y,0)=\tilde D_{ijkl}^{0,0,1} +\tilde D_{ijkl}^{0,1,1} y+\mathcal{O}(y^2)=(\kappa-\kappa^{-1})A_{ijkl}-(\kappa-\kappa^{-1})A_{ijkl}\; y+ \mathcal{O}(y^2) \;,\\
    \partial_x \tilde G^{(4)}_{ijkl}(0,y,0)=\tilde D_{ijkl}^{1,0,0} +\tilde D_{ijkl}^{1,1,0}\; y+\mathcal{O}(y^2)=(\kappa-\kappa^{-1})A_{ijkl} -\PP(\kappa-\kappa^{-1})A_{ijkl}\; y+ \mathcal{O}(y^2) \;,
\end{align}
which excatly match the position-dependent coefficients computed using the minimum separation equations, see equations \eqref{eq:d101} and \eqref{eq:d121}. We use this solution and combine it with the one-variable generating functions given in \eqref{eq:special-z} and \eqref{eq:special-x} and consider refining the solution of the generating functions $\tilde G_{ijkl}(0,y,z)$ and $\partial_x\tilde G_{ijkl}(0,y,z)$. From the asymptotic behavior of these generating functions under the smearing limit, given in \eqref{refex22} and \eqref{refex42}, we expect to have
\begin{align}
    &\tilde G^{(4)}_{ijkl}(0,y,z)=\frac{\tilde G^{(3)}_{ijk}(0,y)f_1(y)}{\left(1-e^{ip_{l}}z\right)f_2(y,z)}+f_3(y,z) \;, \\
    &\partial_x\tilde G^{(4)}_{ijkl}(0,y,z)=\frac{\partial_x\tilde G^{(3)}_{ijk}(0,y)}{\left(1-e^{ip_{l}}z\right)}+f_4(y,z)  \;.
\end{align}
Using the solution of the single-variable generating functions obtained in the previous step, we can substitute this ansatz into the special solution of the two-magnon interaction equation given in \eqref{eq:special-yz} and observe,
\begin{align}
\frac{\tilde G^{(3)}_{ijk}(0,y)f_1(y)}{\left(1-e^{ip_{l}}z\right)f_2(y,z)}&+f_3(y,z)\nonumber\\
=&-\frac{yz \left(1+\PP  y\right) }{\left(y(1-\EE z+z^2+2 \kappa  z+y)+z^2\right)} \left(\frac{ \partial_x\tilde G^{(3)}_{ijk}(0,y)}{\left(1-e^{ip_{l}}z\right)} +f_4(y,z)\right)\nonumber\\
&+\frac{z\left(1-e^{ip_{l}}z\right)\left(z-2\kappa^{-1}y\right)\tilde G^{(4)}_{ijkl}(0,0,z) +2yz (\kappa -\kappa^{-1}) A_{ijkl}}{\left(1-e^{ip_{l}}z\right)\left(y(1-\EE z+z^2+2 \kappa  z+y)+z^2\right)}\label{intertwoleft}\;,
\end{align}
where we reduce the one-variable dependence on $z$ to $\tilde G^{(4)}_{ijkl}(0,0,z)$ by using \eqref{eq:special-z}. The fact that the pieces of the three-magnon generating function $G^{(3)}$ have to satisfy the two-magnon eigenvalue equations of the three-magnon problem, given in \eqref{eq:two-body-rec12} ensures that \eqref{intertwoleft} is satisfied by taking
\begin{align}
    f_1(y)&=2\kappa-\EEt\,+y+y^{-1} \;,\label{f1} \\
    f_2(y,z)&=z+ z^{-1} +yz^{-1}+z y^{-1} +2 \kappa-\EE \label{f2}\;, \\
    f_3(y,z)&=\frac{\left(z y^{-1}-2\kappa^{-1}\right)\tilde G^{(4)}_{ijkl}(0,0,z)+\tilde f_3(y,z) }{z+ z^{-1} +yz^{-1}+z y^{-1} +2 \kappa-\EE} \label{f3}\;,
\end{align}
also,
\begin{align}
    f_4(y,z)=-\frac{\tilde f_3(y,z)}{1+\PP y}\;.\label{f6}
\end{align}
Note that for both two-variable generating functions given above, there is a common two-variable function, $\tilde f_3(y,z)$ which cancels inside the two-magnon interaction equation \eqref{intertwoleft}. Therefore, this function does not play a role in two-magnon interaction equations, but it contributes to the non-interacting generating function \eqref{eq:nonint-gen}. We will use this extra freedom at the end of this section to fix the analyticity of the non-interacting generating function as given in \eqref{eq:first-analy} and \eqref{eq:second-analy} in Appendix \ref{sec:cijkl}. To explicitly show the effect of the expressions given in equations \eqref{f1}-\eqref{f6}, we plug these solutions into \eqref{intertwoleft} and observe that thanks to the three-magnon solution which satisfies \eqref{eq:two-body-rec12}, the two-magnon interaction equations of four-magnon problem are also satisfied. This implies that the two-variable generating functions can be written in terms of the initial conditions and the three-magnon solution,
\begin{align}
    \tilde G^{(4)}_{ijkl}(0,y,z)=&\frac{\tilde G^{(3)}_{ijk}(0,y)yz\left(2\kappa-\EEt\,+y+y^{-1}\right)}{\left(1-e^{ip_{l}}z\right)\left(y(1-\EE z+z^2+2 \kappa  z+y)+z^2\right)}\nonumber\\
    &\quad\quad\quad+\frac{z\left(z-2\kappa^{-1}y\right)\tilde G^{(4)}_{ijkl}(0,0,z)+\tilde f_3(y,z) }{y(1-\EE z+z^2+2 \kappa  z+y)+z^2} \; ,\\
    \partial_x\tilde G^{(4)}_{ijkl}(0,y,z)=&\frac{\partial_x\tilde G^{(3)}_{ijk}(0,y)}{1-e^{ip_{l}}z} -\frac{\tilde f_3(y,z)}{ 1+\PP y}\;,
\end{align}
which is a consistent solution of the four-magnon eigenvalue equations. At this point we can impose analyticity condition on $\tilde G^{(4)}_{ijkl}(0,y,z)$ by demanding that the numerator of this two-variable function vanish at the curve, coming from one of the two factors in $f_2(y,z)=0$.\footnote{How to impose analyticity condition on two-variable generating functions also detailly explained in Part I, \cite{Bozkurt:2024tpz}.} This way we ensure that the two-variable generating function $\tilde G^{(4)}_{ijkl}(0,y,z)$ is analytic and we can expand it around the origin to extract the position-dependent corrections. This step should not be confused by the analyticity of the whole generating function $\tilde G^{(4)}_{ijkl}(x,y,z)$ as discussed in Section \ref{sec:analy} which we will impose at Appendix \ref{sec:cijkl}. We demonstrate how to impose the analyticity condition for a two-variable function in the following way. The curve, we obtain from $f_2(y,z)=0$ that passes from the origin is given as
\begin{align}
    y_+(z)=\frac{1}{2} \left(\EE z-z^2-2 \kappa  z-1+\sqrt{\left(-\EE z+z^2+2 \kappa  z+1\right)^2-4 z^2}\right)\;,\label{curve1}
\end{align}
then we ensure analyticity by solving 
\begin{align}
    \Big[ \left(y(1-\EE z+z^2+2 \kappa  z+y)+z^2\right)\tilde G^{(4)}_{ijkl}(0,y,z)\Big]_{y\to y_+(z)}=0\;,
\end{align}
which gives us
\begin{align}
    \tilde G_{ijkl}(0,0,z)=-\frac{ y_+(z) G_{ijk}^{(3)}(0,y_+(z)) \left(-\EEt+y_+(z)+2 \kappa +y_+(z)^{-1}\right)-\left(1-e^{ip_{l}}z\right)\tilde f_3(y_+(z),z)}{\left(1-z e^{i p_l}\right) (z-2\kappa^{-1} y_+(z))}\label{eq:solz2}\;.
\end{align}
In this case, the asymptotic relations given in \eqref{eq:intz1}-\eqref{eq:intz3} are satisfied because, as $z\to1$ and $N\to\infty$, the $f_2(y,z)$ given in \eqref{f2} reduces to $\left(2\kappa-\EEt\,+y+y^{-1}\right)$. Therefore, although the $\left(1-z e^{i p_l}\right)$ on the denominator goes to zero under the smearing limit, the numerator of the final expression \eqref{eq:solz2} have a zero in the smearing limit and these two factors cancel, and we are left with a finite quantity, as predicted in \eqref{eq:intz1}.

Furthermore, we can now repeat this process for the smearing from the left, given in \eqref{eq:lim4to3left}. In that case, smearing a magnon from left starting from the state $Q_{12}Q_{21}Q_{12}Q_{21}$ results in the three-magnon special solution with $S_\kappa$ scattering coefficients with the ordering $Q_{21}Q_{12}Q_{21}$. This solution is given in Section \ref{sec:q21} and it has the same form as the solution given in Section \ref{sec:q12} of the Appendix apart from the swapped constant functions between two-magnon interaction equations as given in \eqref{eq:confunc211} and \eqref{eq:confunc212}. Repeating the same steps as before, we find
\begin{align}
    &\tilde G^{(4)}_{ijkl}(x,y,0)=\frac{\tilde H^{(3)}_{jkl}(y,0)\PP x y\left(2\kappa -\EEt +y^{-1} +y\right)}{\left(1-e^{-ip_{i}}x\right)g(x,y)}+\frac{x\left(x-2\PP\kappa^{-1}y \right)\tilde G^{(4)}_{ijkl}(x,0,0)+\tilde g_3(x,y)}{g(x,y)}\;,\\
    &\partial_z\tilde G^{(4)}_{ijkl}(x,y,0)= \frac{(1+\invPP y) \partial_z\tilde H^{(3)}_{jkl}(y,0)}{(1+y) \left(1-e^{-ip_{i}}x\right)}-\frac{\tilde g_3(x,y)}{\PP xy(1+y)} \;,
\end{align}
where
\begin{align}
    g(x,y)=(x^2 +\PP y)(\PP y+1)-\PP x y (\EE-2 \kappa )\;.
\end{align}
Additionally, we use $\tilde G^{(4)}_{ijkl}(x,0,0)$ to insert the unfixed single-variable equations and demand analyticity of the two-variable generating function $\tilde G^{(4)}_{ijkl}(x,y,0)$. For the curve, $\tilde y_+(x)$ that passes from the origin defined by $g(x,y)=0$,
\begin{align}
    \tilde y_+(x)=\invPP y_+(x)\label{curve2}
\end{align}
as it is given in \eqref{curve1}, we obtain,
\begin{align}
    \tilde G^{(4)}_{ijkl}(x,0,0)=-\frac{\tilde H^{(3)}_{jkl}(\tilde y_+(x),0)\PP \tilde y_+(x)\left(2\kappa -\EEt +\tilde y_+(x)^{-1} +\tilde y_+(x)\right)-\left(1-e^{-ip_{i}}x\right)\tilde g_3(x,y_+(x))}{\left(1-e^{-ip_{i}}x\right)\left(x-2\PP\kappa^{-1}\tilde y_+(x) \right)}\label{eq:solx2}\;.
\end{align}

Similar to the previous case, we leave $\tilde g_3(x,y)$ free to use it to ensure the analyticity of the non-interacting generating function. The final step is to repeat this process for $G_{ijkl}^{(4)}(x,0,z)$. Since this piece plays a role in both smearing from the left and smearing from the right, we use an ansatz that would accommodate both of these cases and we start with 
\begin{align}
    &\tilde G^{(4)}_{ijkl}(x,0,z)=\frac{\tilde G^{(3)}_{ijk}(x,0)h_1(x)}{\left(1-e^{ip_{l}}z\right)h_2(x,z)}+\frac{\tilde H^{(3)}_{jkl}(0,z)h_1(z)}{\left(1-e^{-ip_{i}}x\right)h_2(x,z)}+h_3(x,z)\label{fin1}\;,\\
    &\partial_y\tilde G^{(4)}_{ijkl}(x,0,z)=\frac{(1+\invPP x)\partial_y\tilde G^{(3)}_{ijk}(x,0)}{\left(1-e^{ip_{l}}z\right) (z+\invPP x)}+\frac{(1+\PP z)\partial_y\tilde H^{(3)}_{jkl}(0,z)}{\left(1-e^{-ip_{i}}x\right) (z+\invPP x)}+h_4(x,z)\label{fin2}\;.
\end{align}
Imposing what we know about three-magnon interaction equations reduces the solution of the two-magnon special solution to a combination of two, two-magnon interaction equations of the three-magnon problem. We conclude that we can fix the undetermined expressions in \eqref{fin1} and \eqref{fin2} to the following forms
\begin{align}
    h_1(x)=2\kappa^{-1} -\EEt  + x + x^{-1} \label{eq:h1}\;,
\end{align}
\begin{align}
   h_2(x,z)=x+z+x^{-1}+z^{-1}+2\kappa ^{-1} -\EE \label{eq:h2}\;,
\end{align}
\begin{align}
h_3(x,z)&=\frac{x^{-1}-2\kappa }{h_2(x,z)}\tilde{G}^{(4)}_{ijkl}(0,0,z) +\frac{z^{-1}-2\kappa }{h_2(x,z)}\tilde{G}^{(4)}_{ijkl}(x,0,0) \;,
\end{align}
where both $\tilde{G}^{(4)}_{ijkl}(x,0,0)$ and $\tilde{G}^{(4)}_{ijkl}(0,0,z)$ are given in terms of the initial conditions $\tilde D^{1,0}$ and $\tilde D^{0,1}$ of the three magnon problem in \eqref{eq:solz2} and \eqref{eq:solx2} and analyticity conditions. Consequently, we obtain
\begin{align}
h_4(x,z)&=-\frac{ c_{ijk}(x)\left(1-e^{-i p_i}x\right)+ c_{jkl}(z) \left(1- e^{i p_l} z\right)-2 A_{ijkl} \left(\kappa -\kappa^{-1}\right)}{(1-e^{-ip_{i}}x)(1-e^{ip_{l}}z) (z+\invPP x)}\;,
\end{align}
and the function $c_{jkl}(z)$ is defined in \eqref{cijkx}. As in the previous cases, the analyticity of the two-magnon interacting equations is satisfied by fixing the leftover freedom with respect to the curve $h_1(x,z)=0$ and $\invPP x+z=0$ that passes from the origin. Since previously we used all the single-variable generating functions to fix the analytic structure of two-magnon interaction from left and right, now we use the undetermined pieces of the three-magnon solution (in our conventions they are $\partial_y\tilde G^{(3)}_{ijk}(x,0)$ and $\partial_y\tilde H^{(3)}_{jkl}(0,z)$). After this step, all the pieces of the generating functions are fixed by the three-magnon solution and the only thing left to do is to ensure the analyticity of non-interacting equations in which we bring all parts together \eqref{eq:generating-non-int}.

We can assemble the four-magnon solution before imposing the non-interacting analyticity conditions as\footnote{Notice that we do not need to compute the value of the one-variable generating functions $\partial_y \tilde G^{(4)}_{ijkl}(x,0,0)$, $\partial_z \tilde G^{(4)}_{ijkl}(x,0,0)$, $\partial_x \tilde G^{(4)}_{ijkl}(0,0,z)$ and $\partial_y \tilde G^{(4)}_{ijkl}(0,0,z)$, as they can be expressed in terms of $\tilde G^{(4)}_{ijkl}(x,0,0)$ and $\tilde G^{(4)}_{ijkl}(0,0,z)$ using equations \eqref{eq:special-z} and \eqref{eq:special-x}.}
\begin{equation}
   \begin{split}
    \tilde G_{ijkl}^{(4)}(x,y,z)&=\frac{A^{(4)}_{ijkl}}{A^{(3)}_{ijk}}\frac{r_1\,\tilde G_{ijk}^{(3)}(x,0)+r_2\,\tilde G_{ijk}^{(3)}(0,y)+r_3\,\partial_y\tilde G_{ijk}^{(3)}(x,0)+r_4\,\partial_x\tilde G_{ijk}^{(3)}(0,y)+x y z\, c_{ijk}(x)}{Q_4(x,y,z)(1-e^{ip_{l}}z)}\\
    &+\frac{A^{(4)}_{ijkl}}{A^{(3)}_{jkl}}\frac{r_5\,\tilde H_{jkl}^{(3)}(y,0)+r_6\,\tilde H_{jkl}^{(3)}(0,z)+r_7\,\partial_z\tilde H_{jkl}^{(3)}(y,0)+r_8\,\partial_y\tilde H_{jkl}^{(3)}(0,z)+x y z\, c_{jkl}(z)}{Q_4(x,y,z)(1-e^{-ip_{i}}x)}\\
    &+\frac{2x y z A^{(4)}_{ijkl} \left(\kappa-\kappa^{-1}\right) \left(1- e^{i p_l}z-e^{-i p_i}x\right)}{Q_4(x,y,z)(1-e^{ip_{l}}z)(1-e^{-ip_{i}}x)}+\frac{C_{ijkl}(x,y,z)}{Q_4(x,y,z)}\;.\label{finalform}\end{split}
\end{equation}
In this expression for the four-magnon generating function, the functions
$\tilde{G}_\sigma$ and $\tilde{H}_\sigma$  are the generating functions of the three magnon long-range corrections for the configurations
$Q_{12}Q_{21}Q_{12}$ and $Q_{21}Q_{12}Q_{21}$ respectively,
related to each other by $\mathbb{Z}_2$ plus parity. Moreover, $c_\sigma (x)$ is given in
\eqref{cijkx}
and is also a function of only three magnon data.
The expression  \eqref{finalform} is naturally divided into four parts according to their poles:
\begin{itemize}
    \item  The part with $(1-e^{ip_{l}}z)Q_4(x,y,z)$ in the denominator corresponds to the three magnon configuration $Q_{12}Q_{21}Q_{12}$ which captures the asymptotic behavior \eqref{eq:4to3assum} (the smearing limit from the right).

    \item  The part with $(1-e^{-ip_{i}}x) Q_4(x,y,z)$ in the denominator corresponds to the configuration $Q_{21}Q_{12}Q_{21}$ which gives us the asymptotic behavior    \eqref{eq:4to3assumleft} (the smearing limit from the left).

    \item  The part with $(1-e^{-ip_{i}}x)(1-e^{ip_{l}}z)Q_4(x,y,z)$ has a very simple numerator given directly with factorized scattering coefficients. The numerator of this part is precisely such that  upon taking the limits $z\to 1$ or $x\to 1$ it combines with the  $Q_{12}Q_{21}Q_{12}$ or  $Q_{21}Q_{12}Q_{21}$ parts respectively.

    \item Lastly, the part with only $Q_4(x,y,z)$ disappears upon taking either the $z\to 1$ or $x\to 1$ limits, nonetheless is needed to satisfy the four magnon eigenvalue equations. This part differes from the ones above as it is given in terms of the three magnon pieces but evaluated on specific curves \eqref{curvez}, \eqref{curve1}, and \eqref{curve2}. The coefficient  $C_{ijkl}(x,y,z)$ is given in equation \eqref{cijkl} of the Appendix where we also spell out the evaluation procedure.
\end{itemize}

Lastly,
the factors $r_i$ refer to the rational functions which are a combination of simple polynomial factors computed throughout this section,
\begin{align}
    r_1&=\frac{\left( Q_4(x,y,z)+\PP y^2z+xy^2  \right)\left( 2\kappa^{-1} -\EEt  + x + x^{-1} \right)}{x+z+x^{-1}+z^{-1}+2\kappa ^{-1} -\EE} \;, \\
     r_2&=\frac{\left( Q_4(x,y,z)+\invPP x^2z+x^2yz \right)\left( 2\kappa -\EEt+y+y^{-1} \right)}{z+ z^{-1} +yz^{-1}+z y^{-1} +2 \kappa-\EE } \;, \\
    r_5&=\frac{ \PP xy\left( Q_4(x,y,z)+xz^2 + xyz^2 \right)\left( 2\kappa -\EEt+y+y^{-1} \right)}{(x^2 +\PP y)(\PP y+1)-\PP x y (\EE-2 \kappa )} \;, \\
    r_6&=\frac{\left( Q_4(x,y,z)+\PP y^2z+xy^2  \right)\left( 2\kappa^{-1} -\EEt  + z + z^{-1} \right)}{x+z+x^{-1}+z^{-1}+2\kappa ^{-1} -\EE} \;, \\
    r_3&=-xyz(1+\invPP x)\;,\quad\quad\quad\quad r_4=-xyz (1+\PP y)  \;, \\
    r_7&=-xyz (1+ \invPP y )\;,\quad\quad\quad\quad r_8=-xyz (1+\PP z) \;.
\end{align}
Then we can expand the generating function given above and obtain the solution of the four-magnon position-dependent corrections. After imposing analyticity condition as described in Appendix \ref{sec:cijkl} we extract the position-dependent corrections $\tilde D^{n,m,r}$ up to $n+m+r=8$ and checked that they satisfy the eigenvalue equations given in Section \ref{sec:fourmagnon} independently. Although the connection between the four-magnon solution and the three-magnon solution is rather implicit we are able to determine each four-magnon position-dependent correction from the three-magnon solution and define the common function that gives all position-dependent corrections,
\begin{align}
    D_\sigma^{n,m,r}=(\kappa-\kappa^{-1})A_\sigma f(\Vec{p}_\sigma;n,m,r,\kappa)=\oint_B \frac{dx dy dz}{(2\pi i)^3}\frac{G_\sigma^{(4)}(x,y,z)}{x^{n+1}y^{m+1}z^{r+1}}\;.\label{eq:finalf}
\end{align}
This final form of the solution still possesses the permutational symmetry highlighted at the end of Section \ref{sec:special} and to point this out further we use the Yang operator formalism in the following section before completing our analysis on the four-magnon solution for open infinite states. 

\section{Yang Operator Formalism}
\label{sec:yang}
Yang operators provide a means to quantify the degree of permutational symmetry in a model \cite{YangPhysRev.168.1920,arutyunov2019elements}. They are rooted in quantum mechanical frameworks that allow for a decomposition of wave functions into plane waves. Since plane waves are symmetric under the transformation $\vec{p}_\sigma \cdot \vec{l} \leftrightarrow \vec{p} \cdot \vec{l}_{\sigma^{-1}}$, wave functions that admit such a decomposition can be tested for underlying permutational symmetry by comparing their behavior across different regimes—for example, different orderings of excitations in position space. In our analysis, we express the scattering coefficients in a plane wave basis and use the Yang operator to capture the effects of permuting momenta and positions. We investigate whether the Yang operator admits a consistent definition for arbitrary configurations of position coordinates and momenta. Furthermore, we explore possible relationships among the diagonal and off-diagonal entries of the operator within a fixed representation. Although the standard method applies to position-independent scattering coefficients, we make use of the generalized framework developed in \cite{Bozkurt:2024tpz} to incorporate position dependence.

\subsection{Yang operators of the CBA}

In the coordinate representation, the CBA uses the assumptions that at large separation, the state is well described by a plane wave together with the non-diffractive scattering condition coming from integrability to write the wave function for $N$ excitations as the weighted sum of plane waves,
\begin{align}
    \Psi(l_1,\cdots,l_N)=\sum_{\sigma\in\mathcal{S}_N}\mathcal{A}_{\sigma}(p_1,\cdots,p_N)e^{i\Vec{p}_{\sigma}\cdot\Vec{l}}\;,\label{eq:integrable-wave}
\end{align}
with the restriction $l_1<\cdots<l_N$, which we call the \emph{fundamental sector}.

As any other ordering is related to the fundamental sector by a permutation $\tau\in\mathcal{S}_N$ of the positions, we can write the solution in the kinematical domain $l_{\tau(1)}<\cdots<l_{\tau(N)}$ as
\begin{align}
    \Psi(l_1,\cdots,l_N|\tau)=\sum_{\sigma\in\mathcal{S}_N}\mathcal{A}_{\tau|\sigma}(p_1,\cdots,p_N)e^{i\Vec{p}_{\sigma}\cdot\Vec{l}_{\tau}}\;,\label{eq:integrable-wave-gen}
\end{align}
where the argument $\tau$ appears on the wave function to indicate the kinematical domain in which the expression is valid. The complex weights $\mathcal{A}_{\tau|\sigma}$ thus form an $N! \times N!$ matrix that depends on the momenta. We encode these coefficients for a fixed permutation of the momenta, $\sigma$, within any kinematic domain labeled by $\tau$ into a column vector formed by the coefficients accompanying the plane wave contributions,
\begin{align}
    \Phi(\sigma)=\{\mathcal{A}_{\tau|\sigma},\quad \forall \tau\in\mathcal{S}_N\} \;.
\end{align}

The \emph{Yang operator} is defined as the map that relates two vectors with different permutations of momenta,
\begin{align}
    \Phi(\alpha_j\sigma)=Y_j(\alpha_j)\Phi(\sigma)\;. \label{eq:YangDef}
\end{align}
Here, $\alpha_j\in\mathcal{S}_N$ is the transposition that exchanges the $j$th element with $j+1$th. This gives the weights $\mathcal{A}_{\tau|\sigma}$ a permutational structure and relates them to scattering coefficients. In fact, for an integrable system that admits a non-diffracting scattering, the coefficients of the wave function satisfy the following relations in terms of transmission, $A$, and reflection, $B$, coefficients
\begin{align}
    \mathcal{A}_{\tau|\alpha_j\sigma}&=A(p_{\tau(j)},p_{\tau(j+1)})\mathcal{A}_{\tau|\sigma} \;,\\
    \mathcal{A}_{\alpha_j\tau|\alpha_j\sigma}&=B(p_{\tau(j)},p_{\tau(j+1)})\mathcal{A}_{\tau|\sigma} \;.
\end{align}
In this case, the Yang operators include the reflection coefficients on the diagonal and the transmission coefficients on the off-diagonal,
\begin{align}
    Y_j(\alpha_j)=A(p_{\tau(j)},p_{\tau(j+1)})\mathbf{1}+B(p_{\tau(j)},p_{\tau(j+1)})\pi(\alpha_j)\;,
\end{align}
where $\mathbf{1}$ is the identity matrix for $N!$-dimensional vector space, and the $\pi(\alpha_j)$ is an $N!$-dimensional faithful representation of the transposition $\alpha_j$. From the fact that transpositions fulfill $\alpha_j^2=\mathbf{1}$ and $\alpha_j \alpha_{j+1} \alpha_j=\alpha_{j+1} \alpha_j \alpha_{j+1}$, we get that the Yang operators must fulfill
\begin{equation}
    Y_j (p_1, p_2) Y_j (p_2,p_1)=\mathbf{1}\;,
\end{equation}
and the YBE
\begin{align}
    Y_j(p_2,p_3)Y_{j+1}(p_1,p_3)Y_j(p_1,p_2)=Y_{j+1}(p_1,p_2)Y_j(p_1,p_3)Y_{j+1}(p_2,p_3)\;.\label{eq:yang-ybe}
\end{align}

\subsection{The Yang Operators of the Long Range Ansatz}
\label{sec:yangXZ}
In this work, we apply a generalized version of the Yang operators to both the scattering coefficients and the position-dependent corrections that arise within the framework of the long-range Bethe ansatz. The asymptotic behavior of these position-dependent corrections remains subtle: as the magnons become more widely separated in position space, the corrections do not exhibit a simple decay with distance.\footnote{This behavior is also linked to the fact that under open or infinite boundary conditions, the momenta remain unconstrained. Consequently, for generic momenta, one does not observe a straightforward suppression of the corrections with increasing separation; instead, the dependence retains a more intricate structure.} It is important to emphasize that the spin chain model still admits a plane wave description of magnons, as the one-magnon eigenstates are expressed in terms of single plane waves \eqref{eq:one-mag12}. However, there are two key distinctions between the Yang operators employed in this work and those appearing in the conventional CBA framework.

The first and most obvious difference is the fact that the Hilbert space does not have the structure of a tensor product due to the constraints imposed by the color contraction rules. For the wave function $|\Psi\rangle_{11}$ we are forced to have $Q_{12}$ at $l_1$ and $l_3$ and $Q_{21}$ at $l_2$ and $l_4$, and vice versa for $|\Psi\rangle_{22}$. This reduces the allowed kinematic domains and, by extension, the dimensionality of the column vectors $\Phi(\sigma)$. For the three-magnon problem, $\Phi(\sigma)$ is two-dimensional instead of six-dimensional, as the only allowed kinematical domains are the ones that arise from exchanging the first and last magnons in $Q_{ij}Q_{ji}Q_{ij}$,\footnote{Moreover the kinematic domains for the  configurations $Q_{12}Q_{21}Q_{12}$ and $Q_{21}Q_{12}Q_{21}$ are separated because they include different number of $Q_{12}$ and $Q_{21}$.}
\begin{align}
    l_1<l_2<l_3\quad\quad\text{and}\quad\quad l_3<l_2<l_1 \;.
\end{align}
Similarly, we would have expected to have kinematic domains $\vec{l}_\tau=(l_{\tau(1)},l_{\tau(2)},l_{\tau(3)},l_{\tau(4)})$ for each $\tau\in\mathcal{S}_4$ for the four-magnon problem, but the color contractions only allow for $\tau\in\mathcal{S}_2\times \mathcal{S}_2$. Specifically, only the following four kinematic domains are allowed
\begin{align}
    l_1<l_2<l_3<l_4\;,\quad  l_1<l_4<l_3<l_2\;,\quad  l_3<l_2<l_1<l_4 \;,\quad l_3<l_4<l_1<l_2 \;.\label{posperm}
\end{align}
The second difference is the presence of position-dependent corrections. In the presence of these corrections, the permutation labels for momentum and position variables take the following roles
\begin{equation}
   \mathcal{A}_{\tau|\sigma}(\Vec{p};\Vec{l}) =A_{\sigma}(\Vec{p})+ D^{n_\tau,m_\tau,r_\tau}_{\sigma} (\Vec{p}) \;,
\end{equation}
where $n_\tau=l_{\tau(2)}-l_{\tau(1)}-1$, $m_\tau=l_{\tau(3)}-l_{\tau(2)}-1$ and $r_\tau=l_{\tau(4)}-l_{\tau(3)}-1$. Therefore, the column vector $\Phi (\sigma)$ takes the form,
\begin{equation}
    \Phi^{n,m,r} (\sigma_0)=\begin{bmatrix}
        \mathcal{A}_{1234|\sigma_0} \\
        \mathcal{A}_{1432|\sigma_1} \\
        \mathcal{A}_{3214|\sigma_2}  \\
        \mathcal{A}_{3412|\sigma_3} 
    \end{bmatrix}=\begin{bmatrix}
        A_{\sigma_0} (\Vec{p}) +D^{n,m,r}_{\sigma_0} (\Vec{p}) \\
        A_{\sigma_1} (\Vec{p}) +D^{r+m+n+2,-2-r,-2-m}_{\sigma_1} (\Vec{p}) \\
        A_{\sigma_2} (\Vec{p}) +D^{-2-m,-2-n,r+m+n+2}_{\sigma_2} (\Vec{p}) \\
        A_{\sigma_3} (\Vec{p}) +D^{r,-4-r-m-n,n}_{\sigma_3} (\Vec{p})
    \end{bmatrix} \;,\label{phimat}
\end{equation}
such that for the given permutation $\sigma_0=(ijkl)$ the rest of the labels are defined as \begin{align}
    \sigma_1=(ilkj)\;,\quad \sigma_2=(kjil)\;,\quad \sigma_3=(klij)\;,\label{permutations}
\end{align}
related to the kinematic domains given in \eqref{posperm}. The Yang operator $Y_1$, which is associated to the transposition $\alpha_1=(12)$ and relates the vector $\Phi((1234))$ to the vector $\Phi((2134))$ as
\begin{align}
    \Phi^{n,m,r}((2134))=Y^{n,m,r}_{1}(\vec{p}) \Phi^{n,m,r}((1234)) \;,
\end{align}
can be computed for the solution discussed in the previous section by substituting the explicit expressions of the column vectors $\Phi$ for the long-range Bethe ansatz and using \eqref{eq:finalf},
\begin{align}
    \begin{bmatrix}
        S_\kappa(p_1,p_2)\left(1+(\kappa-\kappa^{-1}) f(\Vec{p}_{2134},n,m,r,\kappa)\right)\\
        S_\kappa(p_1,p_2)S_\kappa(p_1,p_3)S_\kappa(p_1,p_4)S_\kappa(p_3,p_4)\left(1+(\kappa-\kappa^{-1}) f(\Vec{p}_{2431},n_{\tau_1},m_{\tau_1},r_{\tau_1},\kappa)\right)\\
        S_\kappa(p_1,p_3)S_\kappa(p_2,p_3)\left(1+(\kappa-\kappa^{-1}) f(\Vec{p}_{3124},n_{\tau_2},m_{\tau_2},r_{\tau_2},\kappa)\right)\\
        S_\kappa(p_1,p_2) S_\kappa(p_1,p_3) S_\kappa(p_2,p_3) S_\kappa(p_1,p_4) S_\kappa(p_2,p_4) \left(1+(\kappa-\kappa^{-1}) f(\Vec{p}_{3421},n_{\tau_3},m_{\tau_3},r_{\tau_3},\kappa) \right)
    \end{bmatrix}\nonumber\\=Y_{1}^{n,m,r}(\Vec{p}) \begin{bmatrix}
        \left(1+(\kappa-\kappa^{-1}) f(\Vec{p},n,m,r,\kappa)\right)\\
        S_\kappa(p_2,p_3)S_\kappa(p_2,p_4)S_\kappa(p_3,p_4)\left(1+(\kappa-\kappa^{-1}) f(\Vec{p}_{1432},n_{\tau_1},m_{\tau_1},r_{\tau_1},\kappa)\right)\\
        S_\kappa(p_1,p_2)S_\kappa(p_1,p_3)S_\kappa(p_2,p_3)\left(1+(\kappa-\kappa^{-1}) f(\Vec{p}_{3214},n_{\tau_2},m_{\tau_2},r_{\tau_2},\kappa)\right)\\
        S_\kappa(p_1,p_3)S_\kappa(p_2,p_3)S_\kappa(p_1,p_4)S_\kappa(p_2,p_4)\left(1+(\kappa-\kappa^{-1}) f(\Vec{p}_{3412},n_{\tau_3},m_{\tau_3},r_{\tau_3},\kappa)\right)
    \end{bmatrix}
\end{align}
where the momentum and position variables with no labels refers to identity permutation $(1234)$ and the permutations labels on the position variables, $\tau_i$ refers to the different kinematic domains listed in the same order in \eqref{phimat}. In this case the Yang operator is given as
\begin{align}
    Y_{1}^{n,m,r}(\Vec{p}) =S_\kappa(p_1,p_2)\cdot \text{diag}& \left( \frac{1+(\kappa-\kappa^{-1}) f(\Vec{p}_{2134},\Vec{l})}{1+(\kappa-\kappa^{-1}) f(\Vec{p},\Vec{l})},\right. \nonumber \\
    &\frac{S_\kappa (p_1 ,p_3) S_\kappa (p_1 ,p_4)}{S_\kappa (p_2 ,p_3) S_\kappa (p_2 ,p_4)} \frac{1+(\kappa-\kappa^{-1}) f(\Vec{p}_{2431},\Vec{l}_{\tau_1})}{1+(\kappa-\kappa^{-1}) f(\Vec{p}_{1432}),\Vec{l}_{\tau_1}))} , \nonumber \\
    &\left. \frac{1}{S_\kappa(p_1,p_2)^{2}}\frac{1+(\kappa-\kappa^{-1}) f(\Vec{p}_{3124},\Vec{l}_{\tau_2})}{1+(\kappa-\kappa^{-1}) f(\Vec{p}_{3214}),\Vec{l}_{\tau_2}))} , \frac{1+(\kappa-\kappa^{-1}) f(\Vec{p}_{3421},\Vec{l}_{\tau_3})}{1+(\kappa-\kappa^{-1}) f(\Vec{p}_{3412}),\Vec{l}_{\tau_3}))} \right) \;.\label{yang12}
\end{align}
Note that apart from an overall factor of $S_\kappa(p_1,p_2)$ the Yang operator includes additional factors of scattering coefficients which are related to the relative position of the permutation indices that are being permuted. In the case of three-magnon as given in \eqref{eq:yangskappa} we only had $S_\kappa$ and $S_\kappa^{-1}$, however in the case of four-magnon these factors are more subtle. We can repeat this computation for all possible permutations and we conclude that the Yang operator that related the permutation $\sigma$ with the permutation $\alpha_j \sigma$ has to have the form.
\begin{align}
    Y_{j}^{n,m,r}(\Vec{p}_{\sigma_0})=\text{diag}\left(\frac{A_{\alpha_j\sigma_0}}{A_{\sigma_0}} \frac{1+(\kappa-\kappa^{-1}) f(\Vec{p}_{\alpha_{j}\sigma_0},\Vec{l})}{1+(\kappa-\kappa^{-1}) f(\Vec{p}_{\sigma_0},\Vec{l})} ,\frac{A_{\alpha_j\sigma_1}}{A_{\sigma_1}} \frac{1+(\kappa-\kappa^{-1}) f(\Vec{p}_{\alpha_{j}\sigma_1},\Vec{l_{\tau_1}})}{1+(\kappa-\kappa^{-1}) f(\Vec{p}_{\sigma_1},\Vec{l_{\tau_1}})} ,\cdots\right) \;,
\label{fouryang}\end{align}
such that starting from $\sigma_0=(ijkl)$ the rest of the permutations read as \eqref{permutations}. The transposition $\alpha_n$ is defined as $n$th and $n+1$th integers that are given inside index $\sigma_0=(ijkl)$. Then it acts on any permutation by exchanging the positions of those two integers.\footnote{In the example \eqref{yang12}, $\sigma_0=(1234)$ and Yang operator $Y_1(\vec{p})$ has the index $1$. Therefore, the corresponding transposition is $\alpha_1=(12)$. Then Yang operator acts on $\Phi(\sigma_0)$ and exchanges the permutation indices $1$ with $2$ in each row. This is achieved by having the diagonal factors given in \eqref{yang12}.}  It is straightforward to check that this Yang operator fulfills
\begin{equation}
    Y_{j}^{n,m,r}(\Vec{p}_{\alpha_j}) Y_{j}^{n,m,r}(\Vec{p}) =\mathbf{1} \;,
\end{equation}
and the modified\footnote{
Here, the Yang–Baxter equation involving only the scattering coefficients $S_\kappa$ is modified by the presence of position-dependent coefficients in two ways: first, by the introduction of $\kappa$ dependence, and second, by position dependence. When we take $\kappa\to1$ or $n,m,r\to0$ the modified YBE reduces to YBE with only $S_\kappa$s.} YBE reads,
\begin{align}
    Y_{j}^{n,m,r}(\Vec{p}_{\alpha_j\alpha_{j+1}})Y_{j+1}^{n,m,r}(\Vec{p}_{\alpha_j})Y_{j}^{n,m,r}(\Vec{p})=Y_{j+1}^{n,m,r}(\vec{p}_{\alpha_{j+1}\alpha_j})Y_{j}^{n,m,r}(\Vec{p}_{\alpha_{j+1}})Y_{j+1}^{n,m,r}(\Vec{p}) \;.
\end{align}
Indeed, this modified form of the Yang--Baxter equation holds for all choices of \( n \), \( m \), and \( r \), thereby giving rise to an infinite tower of modified Yang--Baxter equations, analogous to the case of the three-magnon solution. The Yang operators corresponding to the special three-magnon solution are given in~\eqref{eq:yangskappa}, and they satisfy the associated three-magnon modified Yang--Baxter equations as presented in~\eqref{eq:towerofYBE}. Furthermore, since the explicit relation between the three- and four-magnon solutions is established in Section~\ref{sec:specialspecial}, the Yang operators for the four-magnon problem are completely determined by the special three-magnon solution.

In light of Sections~\ref{sec:smearing} and~\ref{sec:specialspecial}, we observe that once the relative position of the magnon located at either end of the spin chain is fixed, the remaining three magnons exhibit behavior identical to that of a purely three-magnon state. This correspondence is explicitly visible in the final form of the generating function given in~\eqref{finalform}. Since we do not have control over the large-separation behavior, as emphasized at the beginning of this section, this observation is meaningful either when one of the relative distances is fixed, or when an average is taken over all possible relative configurations, as carried out in Section~\ref{sec:smearing}. Consequently, the Yang operators for the four-magnon problem are directly related to those of the three-magnon problem.

In \cite{Bozkurt:2024tpz}, we interpreted the form of the Yang operators \eqref{eq:yangskappa} heuristically as a combination of scattering coefficients and coassociators. Drawing inspiration from the elliptic quantum groups which admit R-matrices that satisfy DYBE, coassociators govern the co-associativity of the co-product,
\begin{align}
    \varphi_{ijk}:(v_i\tilde\otimes v_j)\tilde\otimes v_k\rightarrow v_i\tilde\otimes( v_j\tilde\otimes v_k)\label{eq:coass}\;.
\end{align}
We believe that our solution at the level of a CBA-like solution exhibits this behavior. To demonstrate the analogy, we think of a scattering operator that acts as
\begin{align}
    S_{12}(p,\lambda)((p_1\otimes p_2)\otimes p_3)=S_\kappa(p_1,p_2)((p_2\otimes p_1)\otimes p_3)\label{eq:scatdem1}\;,
\end{align}
the modified YBE should take the following form,
\begin{align}
    \varphi_{321}\;S_{23}(p,\lambda)\;\varphi_{231}^{-1}\;S_{13}(p,\lambda)\;\varphi_{213}\;S_{12}(p,\lambda)=S_{12}(p,\lambda)\;\varphi_{312}\;S_{13}(p,\lambda)\;\varphi_{132}^{-1}\;S_{23}(p,\lambda)\;\varphi_{123}\label{quasiYBE}\;,
\end{align}
acting on a configuration $((p_1\otimes p_2)\otimes p_3)$ from left to right. With respect to this analogy, the form of the four-magnon Yang operator given in \eqref{fouryang} should include the contributions of scattering coefficients and four-site coassociators $\varphi_{ijkl}$ which is a combination of $\varphi_{ijk}\otimes\mathbf{1}$ or $\mathbf{1}\otimes\varphi_{jkl}$. It would be very interesting to make this interpretation concrete and have a definition of scattering operators (equivalently R-matrix) and coassociators in our context.

Finally, the existence of a Yang operator for the four-magnon problem demonstrates that contrary to the naive expectation that color contractions would break permutational symmetry, such symmetry remains present in our solution. This tower of modified YBEs is implicitly related to the one obtained in the three-magnon problem because the generating function is fully determined by the three-magnon solutions and analyticity condition over them as given in \eqref{finalform}.

\section{Periodic Boundary Conditions}
\label{sec:periodic}
In this section, we introduce periodic boundary conditions into the spin chain model. The spectrum of these finite-length spin chains corresponds to the single-trace operators of the interpolating theory. From the gauge theory perspective, it is known that operators with generalized traces—introduced in \cite{Gadde:2010zi}—belong to either the untwisted or twisted sectors, which correspond to the $\mathbb{Z}_2$-even and $\mathbb{Z}_2$-odd states of the spin chain, respectively. We define the single-trace operators of the interpolating theory by relating them to their counterparts in $\mathcal{N}=4$ SYM. Using the orbifold projection dictionary for the scalar fields provided in \eqref{eq:fields}, along with the projection matrix $\gamma$ defined in \eqref{eq:gamma}, the twisted and untwisted operators map to the single-trace operators of $\mathcal{N}=4$ SYM as follows:
\begin{align}
    \text{Tr}\left(Z\cdots Z XZ\cdots Z X\right)&\rightarrow\text{Tr}\left(\phi\cdots\phi Q\phi\cdots \phi Q\right)\;,\\
    \text{Tr}\left(\gamma Z\cdots Z XZ\cdots Z X\right)&\rightarrow\text{Tr}\left(\gamma\phi\cdots\phi Q\phi\cdots \phi Q\right)\;.
\end{align}
Furthermore, a generalized single-trace operator can be decomposed into a sum of two distinct traces, depending on how the representations of the gauge group factors $SU(N)_1 \times SU(N)_2$ are contracted. In particular, the overall trace can be expressed as:
\begin{align}
    \text{Tr}\left(\phi\cdots\phi Q\phi\cdots \phi Q\right)=\text{tr}_1\left(\phi_1\cdots\phi_1 Q_{12}\phi_2\cdots \phi_2 Q_{21}\right)+\text{tr}_2\left(\phi_2\cdots\phi_2 Q_{21}\phi_1\cdots \phi_1 Q_{12}\right)\label{eq:untwisted-ex}\;.
\end{align}
This operator belongs to the untwisted sector, where $\text{tr}_i$ denotes the trace over the $SU(N)_i$ gauge group. Similarly,
\begin{align}
    \text{Tr}\left(\gamma\phi\cdots\phi Q\phi\cdots \phi Q\right)=\text{tr}_1\left(\phi_1\cdots\phi_1 Q_{12}\phi_2\cdots \phi_2 Q_{21}\right)-\text{tr}_2\left(\phi_2\cdots\phi_2 Q_{21}\phi_1\cdots \phi_1 Q_{12}\right) \; , \label{eq:twisted-ex}
\end{align}
is an operator from the twisted sector and equivalently it corresponds to an $\mathbb Z_2$-odd operator.

To compute the eigenvalues of these states in the spin chain picture, first, we add the periodic boundary conditions to the Hamiltonian \eqref{eq:total-hamiltonian},
\begin{align}
    \mathcal{H}=\sum_{i=1}^L\mathcal{H}_{i,i+1}\;,\quad\quad \mathcal{H}_{L,L+1}=\mathcal{H}_{L,1}\label{eq:ham-per}\;.
\end{align}
Due to the color contraction rules that govern the states, the Hilbert space is restricted to spin chains containing an even number of $Q_{12}$ and $Q_{21}$ excitations. In the following subsections, we present solutions for the eigenstates of the Hamiltonian \eqref{eq:ham-per} in the two-magnon case—previously studied in \cite{Gadde:2010zi}—and in the four-magnon case, which we examined in earlier sections. To implement periodic boundary conditions, we consider finite spin chain states of the form:
\begin{align}
    \textbf{(}Q_{12}(l_1)Q_{21}(l_2)\textbf{)}_{11}=\textbf{(}\phi_1\cdots\phi_1Q_{12}\phi_2\cdots\phi_2Q_{21}\phi_1\cdots\phi_1\textbf{)}_{11}\;, \nonumber\\\textbf{(}Q_{21}(l_1)Q_{12}(l_2)\textbf{)}_{22}=\textbf{(}\phi_2\cdots\phi_2Q_{21}\phi_1\cdots\phi_1Q_{12}\phi_2\cdots\phi_2\textbf{)}_{22}\;,
\end{align}
for $1\leq l_1<l_2\leq L$ such that by using parenthesis $ \textbf{(}\cdots\textbf{)}_{ii}$ we distinguish between the closed finite states and the open infinite states, denoted by $\ket{\cdots}_{ii}
$ and keep track of the corresponding trace $\text{tr}_i$ by using the overall color labels. Similarly, for the four magnons, we have the states in position space,
\begin{align}
    \textbf{(}Q_{12}(l_1)Q_{21}(l_2)Q_{12}(l_3)Q_{21}(l_4)\textbf{)}_{11}\;, \quad\text{and}\quad\textbf{(}Q_{21}(l_1)Q_{12}(l_2)Q_{21}(l_3)Q_{12}(l_4)\textbf{)}_{22}\;,
\end{align}
for $1\leq l_1<l_2<l_3<l_4\leq L$. In addition to the finite length, the periodic boundary conditions imply
\begin{align}
    \textbf{(}Q_{ij}(l_1)Q_{ji}(l_2)\textbf{)}_{ii}&\equiv\textbf{(}Q_{ji}(l_2)Q_{ij}(l_1+L)\textbf{)}_{jj}  \; ,  \label{eq:per-state-2}\\
    \textbf{(}Q_{ij}(l_1)Q_{ji}(l_2)Q_{ij}(l_3)Q_{ji}(l_4)\textbf{)}_{ii}&\equiv\textbf{(}Q_{ji}(l_2)Q_{ij}(l_3)Q_{ji}(l_4)Q_{ij}(l_1+L)\textbf{)}_{jj}  \; ,  \label{eq:per-state-4}
\end{align}
for $i,j\in\{1,2\}$, which holds both for twisted and untwisted states. Note that the periodicity condition transforms a state with overall color structure $\textbf{(} \cdots \textbf{)}_{11}$ into $\textbf{(} \cdots \textbf{)}_{22}$, and vice versa. Moreover, when the beginning and end of a closed finite-position state do not match in terms of $\phi$ and $Q$ fields, the action of the Hamiltonian \eqref{eq:ham-per} induces a transformation of the overall color structure. For example:
\begin{align}
    \mathcal{H}_{L1}\textbf{(}Q_{12}\cdots Q_{21}\cdots\phi_1\textbf{)}_{11}=\kappa^{-1}\textbf{(}Q_{12}\cdots Q_{21}\cdots\phi_1\textbf{)}_{11}-\textbf{(}\phi_2\cdots Q_{21}\cdots Q_{12}\textbf{)}_{22}\label{eq:ex-boundary} \;,
\end{align}
where $\mathcal{H}_{L1}$ is the Hamiltonian density \eqref{Hamiltonian} acting on the last and the first elements of the state. As a result, the momentum eigenstates must incorporate both color structures. Since the states are given in terms of closed chains, we can sum over states with distinct color indices—an operation that is not possible for open infinite states. The momentum eigenstate for length $L$ state with $2M$ magnons is defined as
\begin{align}
    \ket{\Psi(\Vec{p})}_{L,\pm}=\sum_{l_1<\cdots<l_{2M}} \psi_{11}&(\Vec{p},\Vec{l}) \textbf{(}Q_{12}(l_1)\cdots Q_{21}(l_{2M})\textbf{)}_{11}\nonumber\\& +T_\pm(\vec{p},\kappa)\sum_{l_1<\cdots<l_{2M}} \psi_{22}(\Vec{p},\Vec{l}) \textbf{(}Q_{21}(l_1)\cdots Q_{12}(l_{2M})\textbf{)}_{22}  \; , 
\end{align}
such that $T_\pm(\vec{p},\kappa)$ is the relative normalization factor between the $\mathbb Z_2$ conjugate wave functions $\psi_{11}$ and $\psi_{22}$ which depends on $2M$ momentum variables and $\kappa$ and it proves to be essential to the solution. This factor may vary depending on if the state is untwisted $\ket{\Psi(\Vec{p})}_{L,+}$ or twisted $\ket{\Psi(\Vec{p})}_{L,-}$. These distinct states act as follows under the action of the $\mathbb Z_2$ under a particular normalization,
\begin{align}
    \mathbb Z_2 \ket{\Psi(\Vec{p})}_{L,\pm}=\pm\ket{\Psi(\Vec{p})}_{L,\pm}\;.
\end{align}
For the two-magnon case, the wave function is expressed in the CBA form, incorporating the scattering coefficients given in equations \eqref{eq:defscat} and \eqref{eq:defscat2}. The Bethe equations are then solved to determine the two momenta, and thus the spectrum. In contrast, for the four-magnon case, we adopt a different approach due to the more involved form of the wave function. Similarly to the wave functions of open infinite states \eqref{four-magnon-wave}, we start with a long-range Bethe ansatz,
\begin{align}
    \psi_{ii}(n,m,r,s)&=\sum_{\sigma\in\mathcal{S}_4}\left(A_\sigma+D_\sigma^{n,m,r,s}\right)e^{i\Vec{p}_\sigma\cdot\Vec{l}} \;, \label{closed-wave}
\end{align}
such that $n=l_2-l_1-1$, $m=l_3-l_2-1$, $r=l_4-l_3-1$ and the distance between the first and the last magnon is denoted by an additional positive integer variable, $s=l_1+L-l_4-1$. Also in \eqref{closed-wave}, we omit the momentum variables for simplicity and exchange the usual notation of the wave function in terms of $l_i\in\{0,1,\cdots,L-1\}$ to distances between magnons to make the connection between the distinct position-dependent corrections more apparent. Note that this way we also omit the difference between the configurations $\{l_1,l_2,l_3,l_4\}$ and $\{l_1+a,l_2+a,l_3+a,l_4+a\}$ for $a\in\mathbb Z$ as this distinction disappears under the vanishing of total momentum condition $\cP=0$. In contrast to the open infinite long-range Bethe ansatz given in \eqref{four-magnon-wave}, the position-dependent corrections for the closed spin chains are finitely many, and they are enumerated by four integers instead of three. However, this additional variable is fixed in terms of the other three variables and total length,
\begin{align}
   s=L-4-n-m-r\;.
\end{align}
Here, we analyze the position-dependent corrections for closed spin chains separately from those of the open infinite chains, whose special solutions were presented in Section \ref{sec:special} and Section \ref{sec:specialspecial}. Despite their differences, we adopt the same strategy for closed and open boundary conditions: deriving the position-dependent corrections while ensuring that the solution respects the permutation symmetry of the wave function. Along the way, we highlight the connections we observe between the eigenvalue problems of closed finite and open infinite spin chains.

Although the analytic form of the position-dependent corrections obtained after imposing periodicity differs significantly from the open infinite case \eqref{eq:fin-spe-sol}, it retains key structural features—such as the overall $(\kappa - \kappa^{-1}) A_\sigma$ factor and the partial factorization property \eqref{partialfac}. Notably, as the length of the spin chain increases, the similarities between the closed finite and open infinite position-dependent corrections become more pronounced. This observation suggests a deeper interplay between the open infinite and closed finite solutions, which we aim to further elucidate through the development of an appropriate rapidity parametrization. Establishing such a parametrization is expected to clarify the role of boundary conditions in models governed by a long-range Bethe ansatz.

A direct connection between the open infinite and closed finite solutions has not yet been established within the framework of the long-range Bethe ansatz. Nevertheless, the shared structural features of their position-dependent corrections suggest that a deeper relationship remains to be uncovered. In the level of position-dependent corrections we expect to have a relation between the open-infinite and closed-finite coefficients as
\begin{align}
    \tilde D^{n,m,r,s}\propto c_1 \tilde D^{n,m,r}+ c_2 \tilde D^{m,r,n}+ c_3 \tilde D^{r,n,m}+\cdots
\end{align}
such that $c_i$ are coefficients in terms of momenta and $\kappa$. Once this map is understood, with an appropriate truncation procedure, the generating function of the closed finite chain, currently a polynomial in four complex variables, should be related to the open infinite generating function via the transformation,
\begin{align} t^{L-4} G\left(\frac{x}{t},\frac{y}{t},\frac{z}{t}\right) \longrightarrow G(x,y,z,t)\;,\end{align} 
such that
\begin{align}
    G_\sigma(x,y,z,t)=\sum_{\substack{n,m,r=0\\n+m+r<L-4}}^{L-4}D_\sigma^{n,m,r,L-4-n-m-r}x^ny^mz^rt^{L-4-n-m-r}\;.
\end{align}
However, truncating the sum over relative distances alone does not yield a meaningful solution for (un)twisted boundary conditions without introducing a suitable boundary term, the precise form of which remains to be discovered.

In the following subsections we consider the eigenvalue problem of two- and four-magnons separately and present our results. We solve the four-magnon problem for short length, $L=4,5,6$ and obtain the corresponding position-dependent corrections.

\subsection{Finite Length Two-Magnon Solution}
\label{sec:twoper}
In this section, we impose periodic boundary conditions on the two-magnon problem. The CBA solution was originally studied in \cite{Gadde:2010zi}; here, we combine the two distinct CBA solutions corresponding to the two possible orderings of the magnons, as given in \eqref{eq:two-magnonCBA}, since the periodic boundary conditions mix these two states \eqref{eq:ex-boundary}. To fully account for this mixing, we include a relative normalization factor, as the overall normalization of the wave functions depends on the parameter $\kappa$. We then derive the Bethe equations from the periodicity condition, which also explicitly depends on $\kappa$. After solving the Bethe equations, we compare the resulting eigenvalues and eigenstates with brute-force diagonalization of the spin chain Hamiltonian for $L =4, 5, 6$, and find complete agreement.

To impose periodic boundary conditions, we consider the momentum state for a generic two-magnon spin chain,
\begin{align}
    \ket{\Psi(p_1,p_2)}_{L,\pm}=\textbf{(}\psi(p_1,p_2)\textbf{)}_{11}+ T_\pm(p_1,p_2,\kappa)\textbf{(}\psi(p_1,p_2)\textbf{)}_{22}\;.\label{eq:closeddefp}
\end{align}
In this case, the periodic boundary conditions on wave functions imply,
\begin{align}
    \psi_{11}(p_1,p_2;l_2,l_1+L)=T_\pm(p_1,p_2,\kappa)\psi_{22}(p_1,p_2;l_1,l_2) \label{eq:per2ma}\;,
\end{align}
which leads to the following Bethe equations,
\begin{align}
    e^{-ip_1L}&= T_\pm(p_1,p_2,\kappa)S_{1/\kappa}(p_1,p_2)\label{eq:bethe1}\;,\\
    S_\kappa(p_1,p_2)&= T_\pm(p_1,p_2,\kappa)e^{ip_2L}\label{eq:bethe2}\;.
\end{align}
Due to the color structure of the spin chain states, the scattering coefficients in the Bethe equations exhibit distinct dependencies on the $\kappa$ factor. This difference distinguishes \eqref{eq:bethe1} and \eqref{eq:bethe2} from the Bethe equations of the Heisenberg XXZ model, which is characterized by an anisotropy parameter $(\kappa+\kappa^{-1})/2$ while sharing the same dispersion relation as the present model. The $\kappa$-structure of the scattering coefficients implies that the Bethe equations determine the momentum variables along with a nontrivial relative normalization factor ($T\neq1$), while still satisfying the vanishing total momentum condition,
\begin{align}
    p_1+p_2=2\pi m\;.\label{eq:totalmom0}
\end{align}
As we want both momenta $p_1$ and $p_2$ to be in the first Brillouin zone $p\in (-\pi,\pi]$, we set $m=0$ from now on and focus on the case $p=p_1=-p_2$. We observe that for $L>2$ and $E_2\neq0$, we can obtain the solution of the momentum variables analytically for any length $L$. However, the $E_4=0$ cases require special attention, since for $L=2$ it corresponds to a $Q$-vacuum state (a bound state derived from the $\phi$-vacuum), and for $L>2$ we obtain a descendant of the vacuum state.  To validate our results, we compare the spectrum of the two-magnon states with numerical diagonalization of the Hamiltonian \eqref{Hamiltonian} for short closed spin chains ($L=4,5,6$), finding perfect agreement up to an overall normalization factor.
\subsubsection{Solving the Two-Magnon Bethe Equations}
For the spin chains with length $L>2$,  the Bethe equations \eqref{eq:bethe1} and \eqref{eq:bethe2}, together with the total momentum condition \eqref{eq:totalmom0}, provide the following solution for the relative normalization factor,
\begin{align}
    T_\pm(p,\kappa)\equiv T_\pm(p,-p,\kappa)=\pm\frac{\kappa- e^{ip}}{1-\kappa e^{ip}}\label{eq:relnor2}
\end{align}
which obeys the property $T(p,\kappa)=\left(T(p,\kappa^{-1})\right)^{-1}$ and, for the untwisted states, the remaining Bethe equation takes the following form,
\begin{align}
     e^{-ipL}&=e^{-i p}
\end{align}
and this equation is solved for
\begin{align}
    p&=\frac{2n\pi}{L-1}\;, \label{L2momenta1}
\end{align}
whereas for the twisted states the remaining Bethe equation takes the form,
\begin{align}
     e^{-ipL}&=e^{-i p+i\pi}\;,
\end{align}
and it is solved for
\begin{align}
    p&=\frac{(2n+1)\pi}{L-1}\;,\label{L2momenta2}
\end{align}
for $n\in \mathbb{Z}$ with $\frac{L-1}{2}>n>0$ since the states $\ket{\Psi(p,-p)}_{L,\pm}$ and $\ket{\Psi(-p,p)}_{L,\pm}$ are the same up to a proportionality factor. Note that the solution $p=0$ and $p=\pi$ make the wave function vanish, therefore we do not consider them as valid solutions. 
\begin{figure}
    \centering
    \includegraphics[width=0.6\linewidth]{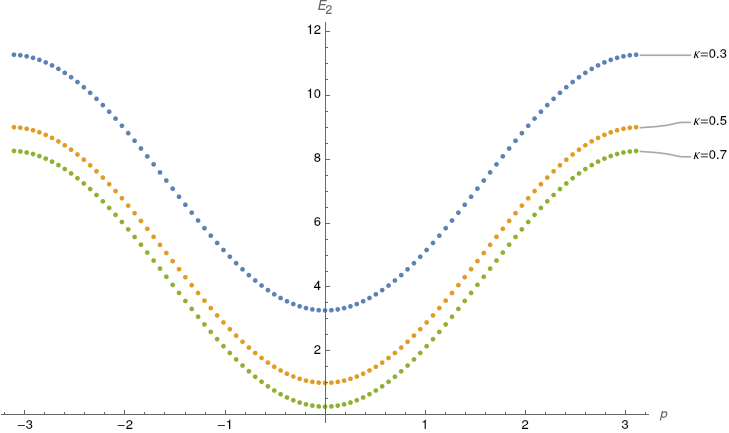}
    \caption{For length, $L=100$ spectrum of the untwisted states for $\kappa=0.3,0.5,0.7$ where $E_2$ denotes the energy eigenvalues.}
    \label{fig:eigdist}
\end{figure}

At this point, we should perform a counting of states to check if we are finding the complete spectrum. On the one hand, if we ignore for a moment the color contractions, the counting of states is the same as for the two copies of XXZ spin chain, which is $\lfloor \frac{L}{2} \rfloor$ for the case of two excitations and zero total momentum for each copy. Taking into account the color contractions, we naively would guess that we have $2\lfloor \frac{L}{2} \rfloor$ states. However, for even $L$ we have the state with total momentum $p_1+p_2=0$ created from $\textbf{(}\phi_1^{\frac{L-2}{2}} Q_{12} \phi_2^{\frac{L-2}{2}} Q_{21}\textbf{)}_{11}$, which maps into itself under $\mathbb{Z}_2$. This implies that we have to subtract one state from the naive counting for even values of $L$. Taking this into account, we have $L-1$ states both for even and odd values of $L$. On the other hand, for even values of $L$ both \eqref{L2momenta1} and \eqref{L2momenta2} have $\frac{L-2}{2}$ solutions. In contrast, for odd values of $L$ the equation \eqref{L2momenta1} has $\frac{L-3}{2}$ and the equation \eqref{L2momenta2} has $\frac{L-1}{2}$. Thus, in total, we have $L-2$ solutions both for even and odd values of $L$. This means that we are missing exactly one solution for every value of $L$ for untwisted and twisted states. Apart from these two states, the distribution of eigenvalues for each of these eigenstates for $L=100$ is given in Figure \ref{fig:eigdist}.

\subsubsection{Vacuum Solution and Descendants}
\label{sec:l2}
Notice that, among the momenta \eqref{L2momenta1} and \eqref{L2momenta2}, we are missing the state that corresponds to the $Q-$vacuum of the model for $L=2$. This means that the missing solution is a descendant of the $Q-$vacuum (or the $Q-$vacuum itself for $L=2$). Therefore, we first use the vanishing of the total momentum, $p_1=-p_2$ and plug this relation into the total energy to find the solution of the momentum variables,
\begin{align}
    E_2(p,-p)=2\kappa+2\kappa^{-1}-2 e^{i p}-2e^{-i p}=0\;.
\end{align}
This equation has two solutions,
\begin{align}
    p=\pm i\log(\kappa) \;, \label{eq:momzeroeig}
\end{align}
which corresponds to a bound state of the two magnons, as it gives
\begin{align}
    S_{1/\kappa}(i\log(\kappa),-i\log(\kappa))=0\;.
\end{align}
In this case, the previous form of the relative normalization coming from the Bethe equations is not valid, therefore we directly use the periodicity relation \eqref{eq:per2ma} to define the relative normalization such that fixing $e^{ip}=\kappa$ gives
\begin{align}
    T(p,\kappa)=\frac{\psi_{11}(p,-p,2,L+1)}{\psi_{22}(p,-p,1,2)}= \frac{1}{\kappa^{L}-\kappa^{L-2}}(1-e^{ip}/\kappa)\;.\label{eq:relnor22}
\end{align}
If we consider this zero as a divergence of $\psi_{22}(p,-p,1,2)$, we can extract the following two solutions, which correspond to the $\mathbb Z_2$ even and $\mathbb Z_2$ odd states
\begin{align}
    \ket{\Psi(-i\log\left(\kappa\right),i\log\left(\kappa\right)}_{L=2,+}=\textbf{(}Q_{12}(1)Q_{21}(2)\textbf{)}_{11}+\textbf{(}Q_{21}(1)Q_{12}(2)\textbf{)}_{22} \;,
\end{align}
and
\begin{align}
    \ket{\Psi(-i\log\left(\kappa\right),i\log\left(\kappa\right)}_{L=2,-}=\textbf{(}Q_{12}(1)Q_{21}(2)\textbf{)}_{11}-\textbf{(}Q_{21}(1)Q_{12}(2)\textbf{)}_{22}\;.
\end{align}
The twisted $L = 2$ state, as given above, does not correspond to a non-trivial single-trace operator because the trace over the two components cancels out. Specifically,
$ \text{tr}_1(Q_{12} Q_{21}) - \text{tr}_2(Q_{21} Q_{12}) = 0.
$ For $L=2$, it is trivial to check that indeed the action of the Hamiltonian on these two states gives zero eigenvalue. Furthermore, the single-trace operator which corresponds to the $\mathbb Z_2$ odd state does not exist because under the tracing the two distinct states cancel each other for $L>2$. An example of a less trivial check is the $L=4$ solution. For untwisted sector is given as
\begin{align}
    \ket{\Psi(-p,p)}_{4,+}=&\kappa ^2 \textbf{(}Q_{12}(0)Q_{21}(1)\textbf{)}_{11}+\kappa ^2 \textbf{(}Q_{12}(1)Q_{21}(2)\textbf{)}_{11}+\kappa ^2 \textbf{(}Q_{12}(2)Q_{21}(3)\textbf{)}_{11}\nonumber\\&+\kappa  \textbf{(}Q_{12}(0)Q_{21}(2)\textbf{)}_{11}+\kappa  \textbf{(}Q_{12}(1)Q_{21}(3)\textbf{)}_{11}+\textbf{(}Q_{12}(0)Q_{21}(3)\textbf{)}_{11}\nonumber\\
    &+\textbf{(}Q_{21}(0)Q_{12}(1)\textbf{)}_{22}+\textbf{(}Q_{21}(1)Q_{12}(2)\textbf{)}_{22}+ \textbf{(}Q_{21}(2)Q_{12}(3)\textbf{)}_{22}\nonumber\\&+\kappa  \textbf{(}Q_{21}(0)Q_{12}(2)\textbf{)}_{22}+\kappa  \textbf{(}Q_{21}(1)Q_{12}(3)\textbf{)}_{22}+\kappa^2\textbf{(}Q_{21}(0)Q_{12}(3)\textbf{)}_{22}\;.
\end{align}
This solution for the zero-energy eigenstate, along with its $L=5,6$ analogs, agrees with the numerical diagonalization of the Hamiltonian \eqref{eq:ham-per}. Therefore, the momentum $p=i\log (\kappa)$, together with the momenta \eqref{L2momenta1} and \eqref{L2momenta2}, give us the full spectrum for $M=2$.
\subsection{Finite Length Four-Magnon Solution}
\label{sec:fourper}
We now turn to the momentum state for a generic four-magnon spin chain with long-range Bethe ansatz, and impose periodic boundary conditions,
\begin{align}
    \ket{\Psi(p_1,p_2,p_3,p_4)}_{L,\pm}=\textbf{(}\Psi(p_1,p_2,p_3,p_4)\textbf{)}_{11}+ T_\pm(p,\kappa)\textbf{(}\Psi(p_1,p_2,p_3,p_4)\textbf{)}_{22}\;.\label{eq:closeddefp4}
\end{align}
In this case, the periodic boundary conditions implies,
\begin{align}
    \psi_{11}(n,m,r,s)=T_\pm(p,\kappa)\psi_{22}(m,r,s,n)\;,\label{eq:per4ma}
\end{align}
such that the wave functions are defined in \eqref{closed-wave} and the relative normalization factor initially is a function of the four momenta and $\kappa$. Due to the contributions from the position-dependent corrections, only omitting plane wave corrections and obtaining Bethe-like equations do not capture the solution of the periodic boundary conditions. The underlying reason is that the position-dependent corrections imply a subtle modification of the plane wave basis, as we discussed using the Yang operators in Section \ref{sec:yang}. Consequently, simply omitting the plane wave factors without altering the position-dependent corrections and reducing the periodicity relations to Bethe-like equations does not yield a solution. We plan to report on a more systematic approach for solving the periodic boundary conditions once the appropriate coordinate transformations and the rapidity structure of the model are fully understood.

Now, we focus on the periodicity conditions on the level of wave functions. By using the general property of the relative normalization factor, $T(p,\kappa)=\left(T(p,\kappa^{-1})\right)^{-1}$ in addition to \eqref{eq:per4ma}, we conclude that the model is 2L periodic both in untwisted and twisted sectors,
\begin{align}
    \psi_{ii}(n,m,r,s)=\psi_{ii}(r,s,n,m)\;,\label{eq:per4ma2}
\end{align}
for $i\in\{1,2\}$, which will play an important role in our solution, since it gives a relation between the wave functions with the same color indices. To impose 
the periodic boundary conditions, we follow an experimental route and discover how to solve the eigenvalue problem for the long-range Bethe ansatz while keeping the permutationally symmetric form of the solution. We focus on short chains ($L=4,5,6$) and we compute the action of the Hamiltonian on the momentum states, $\left(\mathcal{H}-E_4\right)\ket{\Psi(p_1,p_2,p_3,p_4)}_{L}$. Since the action of the Hamiltonian mixes the $\mathbb Z_2$ conjugate states as demonstrated in \eqref{eq:ex-boundary}, we distinguish between the eigenvalue equations coming from $\left(\mathcal{H}-E_4\right)\ket{\Psi(p_1,p_2,p_3,p_4)}_{L}$ which involve both of the $\mathbb Z_2$ conjugate contributions and those which do not, by referring to the former as \emph{boundary eigenvalue equations} and to the latter as \emph{bulk eigenvalue equations}. We observe that, by combining the bulk eigenvalue equations with the eigenvalue equations for open infinite states as presented in Section \ref{sec:eigvalueeqs} we are able to determine the position-dependent corrections analytically as a function of momenta and $\kappa$ for each distinct separation of magnons.\footnote{The fact that we can fully fix the form of position-dependent corrections once we impose boundary conditions without leftover freedom was a claim of \cite{Bozkurt:2024tpz} and here we confirm that this is the case.} To make use of the eigenvalue equations coming from the open infinite case we transform those position-dependent corrections of the open infinite states to closed ones by
\begin{align}
    \tilde D^{n,m,r}_\sigma\rightarrow \tilde D^{n,m,r,L-4-n-m-r}_\sigma \;.\label{eq:opentoclose}
\end{align}
Then, by solving the combined set of bulk equations and the open-infinite equations for short separation using the procedure we demonstrated in Section \ref{sec:special}, we obtain a solution for the position-dependent corrections. Then, we compute the relative normalization factor from one of the boundary eigenvalue equations and impose all periodicity relations as given in \eqref{eq:per4ma2} to ensure that all boundary eigenvalue equations are solved. Imposing these relations for all wave functions, we numerically solve for momentum variables for fixed values of $\kappa$ and obtain eigenstates of the closed Hamiltonian \eqref{eq:ham-per}. To improve the numerical accuracy of our solution we fix one of the wave functions (for example $\psi_{ii}(0,0,0,L-4)$ to have only constant position-dependent corrections which is very similar to having a CBA type wave function. This improves the numerical accuracy of the solution significantly. Finally, we check our solution against brute force diagonalization of the Hamiltonian for $L=5,6$. Up to an overall normalization factor, we obtain agreement.

\subsubsection*{$L=4$}
As in the case of zero-energy two-magnon states given in Section \ref{sec:l2}, the four-magnon states for $L=4$ correspond to the $Q$-vacuum of the model. Unlike the rest of the four-magnon states with finite length, this state does not include non-trivial position-dependent corrections since four magnons can only be next to each other.\footnote{In principle, this state contains the position-dependent corrections denoted as $D^{0,0,0}$ but they do not play any crucial role and they can be absorbed into the definition of scattering coefficients.} Therefore, we obtain the solution by using the periodicity relation \eqref{eq:per4ma}, and choose the relative normalization to be defined as,
\begin{align}
T_\pm(p,\kappa)=\pm\frac{\psi(0,0,0,1)_{11}}{\psi(0,0,0,1)_{22}} \;.
\end{align}
The relative normalization factor given for the descendants of the vacuum in the two-magnon problem, \eqref{eq:relnor22} also has the same description. Here, the wave function in the numerator is coming from the state $(l_1,l_2,l_3,l_4)=(1,2,3,4)$ whereas the wave function in the denominator is coming from $(l_1,l_2,l_3,l_4)=(0,1,2,3)$. Nevertheless, under the vanishing total momentum, these are the same wave functions up to $\kappa\to\kappa^{-1}$. In addition to the vanishing total momentum condition, we impose the vanishing of the total energy \eqref{en4}
\begin{align} 
E_4(p_1,p_2,p_3,-p_1-p_2-p_3) = 0\; .
\end{align} 
Moreover, the $L=4$ state corresponds to a bound state solution from the perspective of the $\phi$ vacuum. Therefore, we select two distinct scattering coefficients and solve for the momentum configurations that cause them to vanish: 
\begin{align} S_\kappa(p_1,p_2) = 0\; , \quad\quad\quad S_{1/\kappa}(p_2,p_3) = 0\; . 
\end{align}
Note that these choices are not unique, but the other solutions are related to each other by permutation of momentum indices. Under these conditions, we obtain the eigenvectors,
\begin{align}
    \ket{\Psi(p_1,p_2,p_3,p_4)}_{L=4,\pm}=\mathcal{N}(\kappa)\Big[\textbf{(}Q_{12}Q_{21}Q_{12}Q_{21}\textbf{)}_{11}\pm \textbf{(}Q_{21}Q_{12}Q_{21}Q_{12}\textbf{)}_{22}\Big] \; ,
\end{align}
for the momentum variables,
\begin{align}
    p_1= -i \log \left(\frac{3 \kappa}{\kappa ^2+2 }\right)\;,\quad p_2= i \log \left(\frac{\kappa(1+2 \kappa ^2) }{\kappa ^2+2}\right)\;,\quad p_3= i \log \left(\frac{3 \kappa }{2 \kappa ^2+1}\right)\;.
\end{align}
and the overall normalization,
\begin{align}
    \mathcal{N}(\kappa)=-\frac{2 \left(\kappa ^2+2\right) \left(4 \kappa ^4+9 \kappa ^2+5\right)}{9 \kappa ^4}\;.
\end{align}
As in the two-magnon case, the $\mathbb Z_2$-odd state vanishes when we map the spin chain states to the single-trace operators. In the following sections, we continue our search for the eigenstates in the long-range Bethe ansatz formalism and omit the overall factors of the normalization, as they are not relevant to our discussion.
\subsubsection*{$L=5$}
The momentum states for the $L=5$ spin chains take the form,
\begin{align}
    \ket{\Psi(p_1,p_2,p_3,p_4)}_{L=5,\pm}=\textbf{(}\Psi(p_1,p_2,p_3,p_4)\textbf{)}_{11}+ T_\pm(p_1,p_2,p_3,p_4,\kappa)\textbf{(}\Psi(p_1,p_2,p_3,p_4)\textbf{)}_{22} \; .
\end{align}
As there are five distinct position states for each color index, we can write
\begin{align}
    \textbf{(}\Psi(p_1,p_2,p_3,p_4)\textbf{)}_{11}=\sum_{\substack{n,m,r=0\\m+n+r\;\leq1}}^1\psi_{11}(n,m,r,1-n-m-r)\textbf{(}\phi(i)\textbf{)}_{11}\;.
\end{align}
Here, we use the short-hand notation, $\textbf{(}\phi(j)\textbf{)}_{11}$ to refer to the length five position state which is characterized by the $\phi_i$ state in the $j$th position, for example, $\textbf{(}\phi(2)\textbf{)}_{11}=\textbf{(}Q_{12}\phi_2 Q_{21}Q_{12}Q_{21}\textbf{)}_{11}$. The eigenvalue problem for each of these position-space states takes a distinct form, which we label by $\alpha_i$ and $\tilde \alpha_i$
\begin{align}
    \left(\mathcal{H}-E_4\right)\ket{\Psi(p_1,p_2,p_3,p_4)}_{L=5}=\sum_{i=1}^5 \alpha_i \textbf{(}\phi(i)\textbf{)}_{11}+\sum_{i=1}^5 \tilde\alpha_i \textbf{(}\phi(i)\textbf{)}_{22}\label{eq:eigL5} \;.
\end{align}
Here, the distinct eigenvalue expressions correspond to permuted versions of the four-magnon interaction equation. This can be seen by taking the open infinite four-magnon interaction equation \eqref{fourint} and applying the transformation \eqref{eq:opentoclose}. After this transformation, the resulting expressions can be rewritten in terms of the wave functions of the $L=5$ closed-chain states,
\begin{align}
    \alpha^c=\left(2\kappa^{-1}-E_4\right)\psi_{11}(0,0,0,1)-\psi_{11}(1,0,0,0)-\psi_{11}(0,0,1,0)\;.\label{alphaclosed}
\end{align}
Then, by applying the periodicity condition given in \eqref{eq:per4ma}, we reproduce all the eigenvalue equations that arise in \eqref{eq:eigL5}. As a consequence of the $\mathbb Z_2$ symmetry, the eigenvalue equations transform to their conjugates $\alpha_i\leftrightarrow\tilde\alpha_i$. Explicitly, these expressions are given by
\begin{align}
    &\alpha_1=\left(2\kappa^{-1}-E_4\right)\psi_{11}(0,0,0,1)-\psi_{11}(1,0,0,0)-T_\pm(p,\kappa)\psi_{22}(0,0,0,1)=0 \;,\label{alpha1}\\
    &\alpha_2=\left(2\kappa-E_4\right)\psi_{11}(1,0,0,0)-\psi_{11}(0,0,0,1)-\psi_{11}(0,1,0,0)=0 \;,\\
    &\alpha_3=\left(2\kappa^{-1}-E_4\right)\psi_{11}(0,1,0,0)-\psi_{11}(1,0,0,0)-\psi_{11}(0,0,1,0)=0 \;,\\
    &\alpha_4=\left(2\kappa-E_4\right)\psi_{11}(0,0,1,0)-\psi_{11}(0,1,0,0)-\psi_{11}(0,0,0,1)=0 \;,\\
    &\alpha_5=\left(2\kappa^{-1}-E_4\right)\psi_{11}(0,0,0,1)-\psi_{11}(0,0,0,1)-T_\pm(p,\kappa)\psi_{22}(0,0,0,1)=0 \;.\label{alpha5}
\end{align}
As an intermediate step, we check whether taking the four distinct wave functions $\psi_{11}(n,m,r,s)$ as unknowns and solve the bulk equations $\{\alpha_2,\alpha_3,\alpha_4\}$ is sufficient to obtain the correct solution. For $E_4=0$ state we obtain
\begin{align}
    \left(\frac{\psi_{11}(1,0,0,0)}{\psi_{11}(0,0,0,1)},\frac{\psi_{11}(0,1,0,0)}{\psi_{11}(0,0,0,1)},\frac{\psi_{11}(0,0,1,0)}{\psi_{11}(0,0,0,1)}\right)=\left(\kappa^{-1},1,\kappa^{-1}\right)\;,\label{eq:expl5}
\end{align}
which agrees with the result obtained from brute-force diagonalization. Since in this check we only solve for the bulk equations, this provides a nontrivial check of our method. The same approach produces the correct result for $E_4=2\kappa+2\kappa^{-1}$ as well.

Remarkably, if we also include the four-magnon interaction equation from the open infinite case, as given in \eqref{alphaclosed}, we find that the tentative solution obtained in \eqref{eq:expl5} ensures the vanishing of the open infinite expression as well. This observation suggests a deeper connection between the open infinite case and the closed periodic boundary conditions, which we will uncover step by step in the following sections.

Then, by imposing the factorization of the scattering coefficients, \eqref{specialscat} and 2L periodicity relation \eqref{eq:per4ma2} we bring all the bulk eigenvalue equations ($\alpha_2,\alpha_3,\alpha_4,\tilde\alpha_2,\tilde\alpha_3,\tilde\alpha_4$) into a similar form to the special four-magnon interaction equation \eqref{fourintspecial}, for example\footnote{Note that to obtain this particular form of the four-magnon interaction equations $\alpha_i$ we need to use the 2L periodicity equations \eqref{eq:per4ma2} and replace the wave functions with their permuted versions if necessary.}
\begin{align}
    \alpha_2=\sum_{ijkl\in\mathcal{S}_4}e^{i (2p_{j}+ 2 p_{k}+ 4p_{l})}\Bigg(\frac{e^{-ip_{j}-ip_{k}}\left(\kappa^{-1}-\kappa\right)\left(e^{ip_{j}}-e^{ip_{k}}\right)\left(1+e^{ip_{j}+ip_{k}}\right)^2}{a(p_{j},p_{k},\kappa)}A_{ijkl}\nonumber\\+\left(2\kappa-E_4-e^{-ip_i}a(p_i,p_l,\kappa^{-1})\right) e^{ip_{k}}D_{ijkl}^{1,0,0,0}-e^{-ip_{j}+ip_{k}}D_{ijkl}^{0,1,0,0}\nonumber\\-e^{i p_{i}+p_{k}-i\mathcal{P}_4} D_{ijkl}^{0,0,0,1}+e^{-i (p_{j}+p_k)}a(p_k,p_j,\kappa^{-1})D_{ijkl}^{0,0,1,0}\Bigg) \;, \label{alpha2last}
\end{align}
such that scattering coefficients are always accompanied by a factor $\kappa-\kappa^{-1}$. This ensures that the position-dependent corrections vanish at the orbifold point. We also observe that the factor $\left(e^{ip_{j}}-e^{ip_{k}}\right)\left(1+e^{ip_{j}+ip_{k}}\right)/a(p_{j},p_{k},\kappa)$ is reminiscent of a solution of position-dependent corrections which satisfies partial factorization \eqref{partialfac}. Additionally, we can always solve the linear equations for each permutation label $\sigma$ separately and obtain a solution of the position-dependent corrections with a common description for each permutation. The analytical solution of these position-dependent corrections is given in the Mathematica file attached to our paper \texttt{fourmagnon.np} since the expressions are highly complicated and do not offer additional insight than the features described here.

Next, we obtain the solution of the relative normalization factor from the boundary eigenvalue equation $\alpha_1$ given in \eqref{alpha1},
\begin{align}
    T_\pm(p,\kappa)=\frac{\left(2\kappa^{-1}-E_4\right)\psi_{11}(0,0,0,1)-\psi_{11}(1,0,0,0)}{\psi_{22}(0,0,0,1)} \;.\label{alpha1reln}
\end{align}
While we obtain a different solution from the boundary eigenvalue equation $\alpha_5$, given in \eqref{alpha5}, both results reduce to the same expression under the 2L periodicity conditions. The distinction between the relative normalization factor for the $\mathbb Z_2$ even and odd states comes into play when we fix the eigenvalue to specific values.\footnote{In the case of $L=5$, zero energy ($E_4=0$) gives the $\mathbb Z_2$ even state and $E_4=2\kappa+2\kappa^{-1}$ gives the $\mathbb Z_2$ odd state.} Therefore by imposing
\begin{align}
    \psi_{11}(0,1,0,0)=\psi_{11}(0,0,0,1)\;,\quad\quad\psi_{11}(1,0,0,0)=\psi_{11}(0,0,1,0) \;,\label{eq:percond}
\end{align}
together with vanishing total momentum and total energy we can solve for the momentum variables and obtain the spectrum of the model. However, with \eqref{eq:percond}, we could only match the result of the brute-force diagonalization by $2$ digits.\footnote{Note that we compute the relative error by
\begin{align}
\text{Relative Error} &= \left| \frac{x_{\text{numerical}} - x_{\text{analytical}}}{x_{\text{analytical}}} \right| \; ,
\end{align}
then one can compute the number of significant digits as
\begin{align}
    \text{Correct Digits} &\approx -\log_{10} \left( \text{Relative Error} \right) \; .
\end{align}
} To improve the accuracy, we leave one of the equations $\{\alpha^c,\alpha_2,\alpha_3,\alpha_4\}$ out. This leaves us with an undetermined position-dependent correction which we choose to be $D^{0,0,0,1}_\sigma$ without loss of generality. Then we assume that for any permutation $\sigma$ the undetermined position-dependent correction $D_\sigma^{0,0,0,1}$ is a constant and we solve the periodicity relations, total energy and total momentum conditions to fix the four momenta and $D_\sigma^{0,0,0,1}$. We include an additional relation between the wave functions,
\begin{align}
    \frac{\psi_{11}(1,0,0,0)}{\psi_{11}(0,0,0,1)}=\kappa^{-1} \; ,
\end{align}
which is coming from the observation \eqref{eq:expl5}. This way, we obtain the solution of the momentum variables which matches the brute force diagonalization by $7$ digits. Additionally, this solution requires the $D_\sigma^{0,0,0,1}$ to be $0.0852016\, -0.103424 i$ which effectively sets the wave function $\psi_{11}(0,0,0,1)$ to be a CBA-like solution shifted by a constant scattering coefficient. The results for two distinct values of $\kappa$ and both twisted and untwisted states of $L=5$, normalized such that $\psi_{11}(0,0,0,1)=1$, are given in Table \ref{tab:L5table}. The solution of the momentum variables for $E_4=0$ and $\kappa=0.7$ is given as
\begin{align}
    \{p_1=0.327198\, -2.38999 i\;,\;\;p_2=-3.05876+0.639858 i\;,\;\;p_3=1.90333\, +1.128 i\} \; ,
\end{align}
while for $E_4=2\kappa+2\kappa^{-1}$ is given as,
\begin{align} \{p_1=-1.36434 \;i\;,p_2=3.14159\, -0.176988 i,p_3=1.35999 i\} \; ,
\end{align}
in this case, we obtain the relative normalization factor from $T\psi_{22}(2,3,4,5)$, while the rest of the wave functions can be calculated by using $\mathbb Z_2$ conjugation.
\begin{table}
    \centering
    \begin{tabular}{c|c||c|c|c|c|c|c}
        $E_4$&$\kappa$ &$\psi_{11}(0,0,0,1)$&$\psi_{11}(1,0,0,0)$&$\psi_{11}(0,1,0,0)$&$\psi_{11}(0,0,1,0)$&$T\psi_{22}(0,0,0,1)$& \\ \hline $0$&$0.7$&$1.$&$1.42857$&$1.$&$1.42857$&$1.$& \\ \hline 
         $0$&$0.4$&1.&2.5&1.&2.5&1.&\\ \hline
         $2\kappa+2\kappa^{-1}$&$0.7$&$1.$&$-0.7$&$1.$&$-0.7$&$1.$&\\ \hline
         $2\kappa+2\kappa^{-1}$&$0.4$&1.&-0.4&1.&-0.4&1.&
    \end{tabular}
    \caption{Numerical Solution for the periodicity relations between the position states.}
    \label{tab:L5table}
\end{table}
The exact result coming from the brute force diagonalization for $E_4=0$ untwisted state is 
\begin{align}
\ket{\Psi(p_1,p_2,p_3,p_4)}_{L=5,+}&=\textbf{(}\phi(1)\textbf{)}_{11}+\kappa^{-1}\textbf{(}\phi(2)\textbf{)}_{11}+\textbf{(}\phi(3)\textbf{)}_{11}+\kappa^{-1}\textbf{(}\phi(4)\textbf{)}_{11}+\textbf{(}\phi(5)\textbf{)}_{11}\nonumber\\
    &+\kappa^{-1}\textbf{(}\phi(1)\textbf{)}_{22}+\textbf{(}\phi(2)\textbf{)}_{22}+\kappa^{-1}\textbf{(}\phi(3)\textbf{)}_{22}+\textbf{(}\phi(4)\textbf{)}_{22}+\kappa^{-1}\textbf{(}\phi(5)\textbf{)}_{22} \; ,
\end{align}
and for $\kappa=0.7$, $\kappa^{-1}\approx1.42857$ which is the result we obtain from the numerical solution. Similarly, the eigenstate for the eigenvalue $E_4=2\kappa+2\kappa^{-1}$ is the twisted state of $L=5$ and it is given as
\begin{align}
\ket{\Psi(p_1,p_2,p_3,p_4)}_{L=5,-}&=\textbf{(}\phi(1)\textbf{)}_{11}-\kappa\textbf{(}\phi(2)\textbf{)}_{11}+\textbf{(}\phi(3)\textbf{)}_{11}-\kappa\textbf{(}\phi(4)\textbf{)}_{11}+\textbf{(}\phi(5)\textbf{)}_{11}\nonumber\\
    &-\kappa\textbf{(}\phi(1)\textbf{)}_{22}+\textbf{(}\phi(2)\textbf{)}_{22}-\kappa\textbf{(}\phi(3)\textbf{)}_{22}+\textbf{(}\phi(4)\textbf{)}_{22}-\kappa\textbf{(}\phi(5)\textbf{)}_{22} \; .
\end{align}
These two states are the only states that correspond to single-trace operators for $L=5$, and we obtain them using the long-range Bethe ansatz and this computation can be performed for any $0<\kappa<1$. However, solving the periodic boundary conditions within this framework is methodologically demanding and not readily tractable. A key source of this complexity is the necessity of working with momentum variables instead of rapidities, which leads to highly intricate wave function expressions. Nevertheless, this solution shows that the long-range Bethe ansatz is compatible with the periodic boundary conditions and it gives the spectrum of the Hamiltonian. We perform the same method for $L=6$ to gain more insight into the problem.
\subsubsection*{$L=6$}
We compute the eigenvectors of the $L=6$ closed spin chains by applying the method we outlined for $L=5$. The momentum states for the $L=6$ spin chains take the form as expected,
\begin{align}
    \ket{\Psi(p_1,p_2,p_3,p_4)}_{L=6,\pm}=\textbf{(}\Psi(p_1,p_2,p_3,p_4)\textbf{)}_{11} + T_\pm(p_1,p_2,p_3,p_4,\kappa)\textbf{(}\Psi(p_1,p_2,p_3,p_4)\textbf{)}_{22}
\end{align}
such that there are fifteen distinct position states for each color index, denoted as
\begin{align}
    \textbf{(}\Psi(p_1,p_2,p_3,p_4)\textbf{)}_{11}=\sum_{n,m,r=0}^2\psi_{11}(\Vec{p},n,m,r,2-n-m-r)\textbf{(}\phi(i)\phi(j)\textbf{)}_{11}\;.
\end{align}
The eigenvalue problem brings variations of four magnon interaction equations \eqref{fourint},
\begin{align}
    \left(\mathcal{H}-E_4\right)\ket{\Psi(p_1,p_2,p_3,p_4)}_{L=6}=\sum_{i=1}^{15} \alpha_{ij} \textbf{(}\phi(i)\phi(j)\textbf{)}_{11}+\sum_{i=1}^{15} \tilde\alpha_{ij} \textbf{(}\phi(i)\phi(j)\textbf{)}_{22}\;.\label{eq:eigL6}
\end{align}
As in the $L=5$ case, we use the short-hand notation, $\textbf{(}\phi(i)\phi(j)\textbf{)}_{11}$ to refer to the length six position state which is characterized by the $\phi_i$ state in the $i$th and $j$th position. Depending on the configuration of the $\phi$ fields, the eigenvalue equations are a variation of the four-magnon interaction (above symbolically we denoted all with $\alpha_{ij}$), the three-magnon interaction or the two two-magnon interaction equations. We refer to the states that have different types of fields at the beginning and the end of the spin chain state as \emph{boundary eigenvalue equations}, and these are 
\begin{align}
    \{\alpha_{12},\alpha_{56},\beta_{13},\beta_{46},\beta_{15},\beta_{26},\gamma_{14},\gamma_{36}\} \;,
\end{align}
while the \emph{bulk eigenvalue equations} are listed as
\begin{align}
    \{\alpha_{16},\alpha_{23},\alpha_{34},\alpha_{45},\beta_{24},\beta_{35},\gamma_{25}\} \;.
\end{align}
In addition to these equations, we also use the eigenvalue equations coming from the open infinite states which we transform under the map \eqref{eq:opentoclose} which are listed as
\begin{align}
    \{\alpha^c,\beta^c_l(1),\beta^c_r(1),\gamma^c_{rl}(1)\} \; .
\end{align}
First, we observe that two of the distinct eigenvalue equations match,
\begin{align}
    \alpha_{16}=\alpha^c \; ,
\end{align}
which supports the expectation that, as we increase the length of the periodic chain, the closed finite and open infinite problems look more and more alike, and the special solution given in Section \ref{sec:special} becomes more relevant. Then, we combine the bulk eigenvalue equations with the open infinite eigenvalue equations and bring all of them to a form similar to \eqref{alpha2last}, which means that the scattering coefficients for each permutation are accompanied by a factor of $\kappa-\kappa^{-1}$ and $\omega(p_{\sigma(2)},p_{\sigma(3)})$ as defined in \eqref{eq:omega}. To achieve this form we used the 2L periodicity relations \eqref{eq:per4ma2} when it is necessary. This way, we obtain as many equations as the number of distinct position-dependent corrections,
\begin{align}
    \{\alpha_{16}=\alpha^c,\alpha_{23},\alpha_{34},\alpha_{45},\beta_{24},\beta_{35},\gamma_{25},\beta^c_l(1),\beta^c_r(1),\gamma^c_{rl}(1)\} \; .
\end{align}
First, to check that this set of equations is indeed compatible with the brute-force diagonalization, we solve the linear set of equations generated by these equations by taking the wave functions unknowns. Then, for $E_4=0$, normalised such that $\psi_{11}(0,0,0,2)=1$, we obtain,
\begin{multline}
    \{\psi_{11}(0,0,1,1),\psi_{11}(0,0,2,0),\psi_{11}(0,1,0,1),\psi_{11}(0,1,1,0),\psi_{11}(0,2,0,0),\psi_{11}(1,0,0,1),\\\psi_{11}(1,0,1,0),\psi_{11}(1,1,0,0),\psi_{11}(2,0,0,0),
\psi_{11}(0,0,0,2),\psi_{11}(0,0,1,1),\psi_{11}(0,1,0,1),\\\psi_{11}(1,0,0,1)\}=\left\{\frac{1}{\kappa },\frac{1}{\kappa ^2},1,\frac{1}{\kappa },1,\frac{1}{\kappa },\frac{1}{\kappa ^2},\frac{1}{\kappa },\frac{1}{\kappa ^2},1,\frac{1}{\kappa },1,\frac{1}{\kappa }\right\}\;.\label{eq:l6things}
\end{multline}
This perfectly agrees with the brute-force diagonalization. We solve for the distinct position-dependent corrections as a function of momenta and $\kappa$.\footnote{In order to reduce the complexity of the analytic part of the computation, we transform the position-dependent corrections to scaled position-dependent corrections as given in \eqref{scaledd} and also impose the partial factorization \eqref{partialfac} from the beginning. Note that due to the extra factors of $\omega(p_{\sigma(2)},p_{\sigma(3)})$ we deduce that the partial factorization is manifest in the solution even if we do not impose it by hand. However, imposing from the beginning of the analytical computation simplifies the expressions drastically and both the scaling of the position-dependent corrections and the partial factorization condition prove to be computationally useful for closed finite spin chains.} Therefore, we obtain the analytic description of the position-dependent corrections and they exhibit the same properties as the special function and the $L=5$ solution. Finally, we impose the 2L periodicity conditions,
\begin{align}
    &\psi_{11}(0,0,0,2)=\psi_{11}(0,2,0,0)\;,\quad\quad \psi_{11}(0,0,2,0)=\psi_{11}(2,0,0,0)\;,\\
    &\psi_{11}(1,1,0,0)=\psi_{11}(0,0,1,1)\;,\quad\quad \psi_{11}(0,1,1,0)=\psi_{11}(1,0,0,1)\;.
\end{align}
Together with the $E_4=0$ condition and the vanishing total momentum condition, we obtain the solution 
\begin{align}
    \{p_1=-0.0218877 i,p_2=0.501353 i\;,p_3=-0.501353 i\;,p_4=0.0218877i\;\}
\end{align}
which correctly matches the brute force diagonalization with a relative error of $0.05$. Same can be done for the twisted state with $E_4=2\kappa+\frac{2}{\kappa}$. We think it is possible to improve the numerics further by following the same technique as $L=5$ and choosing to prioritize what we learn from \eqref{eq:l6things}. However, we postpone this to future work and hope to gain a deeper understanding about the parameter space before and we plan to attempt to solve this problem analytically. For now, we take this as a proof of concept for long-range Bethe ansatz and move on to discover the connection between the solutions of various numbers of excitations.

\section{Conclusion}
\label{sec:conclusion}
This paper continues our investigation \cite{Bozkurt:2024tpz} into a long-range generalization of the CBA for spin chain models originating from marginally deformed orbifolds of $\mathcal{N}=4$ SYM. This way we incorporate the non-local effects that are encoded in the Hamiltonian \eqref{Hamiltonian} into position-dependent corrections. Building on the analysis of the $XZ$ sector, we extend our results for the three-magnon sector to the case of four magnons. This approach provides a family of solutions to the four-magnon eigenvalue equation~\eqref{eq:eig4} with additive energy, $E_4=\sum_i E(p_i)$.

In Section~\ref{sec:special}, we begin to fix the coefficients that remained undetermined after solving the eigenvalue equations. The family of solutions is reduced by imposing additional constraints, guided by the symmetries of the problem and following an approach analogous to the three-magnon case. This procedure leads to a special solution characterized by factorized scattering coefficients $A_\sigma$, given in equation \eqref{specialscat}, and position-dependent corrections $D^{n,m,r}_\sigma$, all proportional to $(\kappa - \kappa^{-1}) A_\sigma$ for each permutation $\sigma \in \mathcal{S}_4$. The presence of this factor for all $n, m, r \in \mathbb{Z}{\geq 0}$ ensures that at the orbifold point ($\kappa = 1$) all these position-dependent corrections to the ordinary CBA vanish. The remaining undetermined $D^{n,m,r}_\sigma$'s, after applying these constraints, can be systematically counted and organized, as summarized in Table~\ref{tab:count}.

In Section~\ref{sec:smearing}, we introduced a physically motivated limiting procedure that relates eigenstates with different magnon numbers. By averaging over all possible positions of one magnon in an $(M+1)$-magnon eigenstate and setting its momentum to zero, we showed that the resulting state reduces to the corresponding $M$-magnon eigenstate. This procedure, which can be interpreted as smearing a magnon, arises naturally within the standard CBA framework (Subsection~\ref{sec:CBAsmearing}) and provides a systematic method for connecting spin chain solutions across successive magnon sectors.

Taken together, these results reveal a coherent underlying structure in the spectrum of eigenstates, with different magnon sectors linked through a well-defined limiting process. This work lays the groundwork for further exploration of long-range CBA and the structure of their eigenstates beyond the sectors considered here.

Notably, for the four-magnon states, the asymptotic relations derived in Section~\ref{sec:smearing},
\begin{align}
\tilde{G}^{(4)}_\sigma(x, y, z) &\sim \frac{1}{1 - e^{i p_{\sigma(4)}} z} \tilde{G}^{(3)}_\sigma(x, y) \;, \quad\quad\quad z \to 1 \; ,\\
\tilde{G}^{(4)}_\sigma(x, y, z) &\sim \frac{1}{1 - e^{-i p_{\sigma(1)}} x} \tilde{G}^{(3)}_\sigma(y, z) \;, \quad\quad\;\; x \to 1 \; ,
\end{align}
as well as
\begin{align}
\tilde{G}^{(4)}_\sigma(x, y, z) \sim \mbox{finite} \;, \quad\quad\quad y \to 1 \; ,
\end{align}
are essential for reducing the four-magnon special solution constructed in Section~\ref{sec:special} to the corresponding three-magnon case. These relations provide additional constraints that enable the complete determination of the four-magnon solution. The resulting position-dependent corrections are encoded in the generating function~\eqref{finalform} and can be expressed explicitly in terms of the known three-magnon solutions.
An important technical observation in this process was the matching of initial conditions for the four-magnon wavefunction---obtained by solving the minimal separation equations~\eqref{eq:first}---with those of the three-magnon case.
This allowed us to construct a unified generating function that incorporates the three-magnon corrections and systematically organizes the four-magnon corrections across all permutation sectors, as presented in~\eqref{eq:finalf}.

In the final expression~\eqref{finalform}, we observe that the coefficients accompanying the plane wave contributions in the wave function admit a unified description in terms of permutations of momenta:
\begin{align}
\sum_{\sigma \in \mathcal{S}_4} \left(A_\sigma + D^{n,m,r}_\sigma\right) e^{i \vec{p}_\sigma \cdot \vec{l}} \xrightarrow[\text{special solution}]{} \sum_{\sigma \in \mathcal{S}_4} A_\sigma \left(1 + (\kappa - \kappa^{-1}) f(\vec{p}_\sigma; n, m, r, \kappa)\right) e^{i \vec{p}_\sigma \cdot \vec{l}} \;.
\end{align}
Since the scattering coefficients factorize and are completely separated from the position-dependent corrections, we interpret this as evidence for a novel basis of functions---replacing the plane waves of the standard CBA---that is suitable for describing the eigenvectors of this model, 
\begin{align}
 e^{i \vec{p}_\sigma \cdot \vec{l}} \longrightarrow \left(1 + (\kappa - \kappa^{-1}) f(\vec{p}_\sigma; n, m, r, \kappa)\right) e^{i \vec{p}_\sigma \cdot \vec{l}} \;.
\end{align}

Although the full mathematical significance of obtaining a common function $f(\vec{p}_\sigma; n, m, r, \kappa)$ for the position-dependent corrections at each $(n, m, r)$ remains to be understood, the resulting structure has a striking implication: the eigenstate wave function can be expressed entirely as a sum over permutation indices, with position-dependent coefficients accompanying the plane wave contributions in~\eqref{eq:integrable-wave}.

Remarkably, although the two-magnon scattering coefficients defined in~\eqref{eq:defscat} and~\eqref{eq:defscat2} do not satisfy the Yang--Baxter equation in the conventional sense, we nonetheless find that a consistent and permutation-symmetric construction is possible. To demonstrate this, we applied the Yang operator formalism to the special solution of the four-magnon problem and uncovered a tower of modified Yang--Baxter equations---one for each $(n,m,r)$---mirroring the structure previously observed in the three-magnon case.

Finally, in Section~\ref{sec:periodic}, we examined the compatibility of the long-range Bethe ansatz with periodic boundary conditions. We began by solving the two-magnon Bethe equations and then extended the analysis to the four-magnon case. Already in the two-magnon sector, a novel feature emerges: by introducing a relative normalization factor between $\mathbb{Z}_2$-conjugate states, we are able to recover all eigenvectors—both for untwisted and twisted boundary conditions—via a generalized version of the Bethe equations. In contrast, the four-magnon problem does not admit a solution within the standard CBA framework, making it infeasible to directly apply Bethe equations to determine the quantized momenta. To address this challenge, we adopted an alternative strategy: we formulated and solved the periodicity relations at the level of wavefunctions for closed spin chains of finite length ($L = 4, 5, 6$). This construction offers a concrete and practical method for incorporating periodic boundary conditions within the long-range Bethe ansatz framework.

Nevertheless, our current approach is not yet fully scalable to arbitrary chain lengths $L$, due to the absence of a systematic method for efficiently implementing periodic boundary conditions within this approach. We believe that a key obstacle lies in the necessity of working with momentum variables, rather than rapidities. Identifying an appropriate rapidity parametrization may allow the position-dependent corrections to be reinterpreted as a coordinate transformation. This, in turn, could lead to a derivation of the correct form of the long-range Bethe equations under periodic boundary conditions.

\bigskip

We conclude by outlining several promising directions for future research:
\begin{itemize}
 \item
  As emphasized throughout this work, it would be highly desirable to identify the correct rapidity description of the model by comparing the $\phi$-vacuum states studied here to the $Q$-vacuum states of~\cite{Pomoni:2021pbj}. We expect that expressing the wave function in terms of rapidities—rather than momenta—will simplify the position-dependent corrections and allow a more efficient treatment of periodic boundary conditions.

\item Given our expectation that this spin chain model is elliptic based on a groupoid symmetry, the rapidity variables should likely involve elliptic functions. We intend to explore this further by building on the elliptic parametrization discussed in~\cite{Pomoni:2021pbj}, potentially establishing a connection to an underlying Felder-like $R$-matrix similar to~\cite{Felder:2020tct,Ren:2023mtn} but for Cyclic SOS  (CSOS) models instead of RSOS.

    \item A suitable rapidity parametrization may also clarify the correct basis transformation from the plane wave basis discussed in Section~\ref{sec:yang}, and help refine the structure of the modified Yang–Baxter tower into a single, unified YBE-type equation.

\item In parallel, we aim to extend the present results by using the four-magnon solution—expressed in terms of the three-magnon solution—to construct explicit $M$-magnon eigenstates via the generating function. We believe that the insights from Section~\ref{sec:smearing} will make this construction feasible~\cite{future}.

\item Another important direction is the search for higher conserved charges, following the approach of~\cite{Bernard:1993va,DeLeeuw:2019gxe,deLeeuw:2020ahe,deLeeuw:2020xrw}. We are already working on generalizing the boost formalism to Felder-like $R$-matrices, with the goal of obtaining the higher charges of the model \cite{future2}. The restricted Hilbert space, arising from the groupoid structure~\cite{Corcoran:2024ofo}, may simplify this task, particularly if one focuses on permutationally symmetric states.

\item It would also be valuable to carry out a detailed numerical analysis of the spectrum of this model, similar to the studies in~\cite{McLoughlin:2022jyt} for spin chains in $\mathcal{N}=4$ SYM and its $\mathcal{N}=1$ Leigh–Strassler deformations. In particular, counting spectral degeneracies—as done for small lengths in~\cite{Liendo:2011wc}—could shed light on the integrable versus chaotic behavior of the system.

\item An exciting recent development has been the extension of the hexagon formalism \cite{Basso:2015zoa} to orbifold theories~\cite{Ferrando:2025qkr,lePlat:2025eod}. A promising future direction to generalize the the hexagon approach beyond the orbifold point, and to extend the analysis of~\cite{lePlat:2025eod} to compute three-point functions of operators with four magnon excitations, using the results obtained in this paper.

  \item Finally, in light of recent developments on the gauge theory side, we are motivated to revisit the string theory dual. The recent supergravity construction of~\cite{Skrzypek:2023fkr}, which resolves the $AdS_5 \times S^5/\mathbb{Z}_2$ orbifold singularity, provides a timely opportunity to study the corresponding two-dimensional worldsheet QFT and explore the string dual of the marginally deformed theory.

\end{itemize}

\section*{Acknowledgements}
 We are grateful to Gleb  Arutyunov, Costas Zoubos, Didina Serban, Shota Komatsu and Konstantin Zarembo for very useful discussions throughout the course of this work.
 The authors have benefited from the German Research Foundation DFG under Germany’s Excellence Strategy – EXC 2121 Quantum Universe – 390833306 as well as the CRC 1624 Higher Structures, Moduli Spaces and Integrability.
  EP is supported by ERC-2021-CoG - BrokenSymmetries 101044226. DB and EP would like to express their gratitude to the organizers of the Program ``Black hole physics from strongly coupled thermal dynamics'' and to the Simons Center for Geometry and Physics (SCGP) for their hospitality and support in the final stage of this project.
\appendix
\section{$3$-Magnon Solution}
\label{sec:three}
\subsection{$S_\kappa$ Solution for $Q_{12}Q_{21}Q_{12}$}
\label{sec:q12}
In this section we review the three-magnon special solution given in \cite{Bozkurt:2024tpz} with scattering coefficients factorized with respect to $S_\kappa$ for the state which has $Q_{12}Q_{21}Q_{12}$. The long-range Bethe ansatz for that configuration is given by
\begin{align}
    \ket{\Psi(p_1,p_2,p_3)}_{12}&=\sum_{l_1<l_2<l_3}\Psi(p_1,p_2,p_3;l_1,l_2,l_3)\ket{Q_{12}(l_1),Q_{21}(l_2),Q_{12}(l_3)} \;, \\
    \Psi(p_1,p_2,p_3;l_1,l_2,l_3)&=\sum_{\sigma\in\mathcal{S}_3
    }\left(A_\sigma+D_\sigma^{l_2-l_1-1,l_3-l_2-1}\right)e^{i\Vec{p}_\sigma^{\,3}\cdot\Vec{l}^{\,3}}\\
    &=e^{il_1\sum_i^3p_i}\sum_{\sigma\in\mathcal{S}_3}\left(A_\sigma+D_\sigma^{n,m}\right)e^{i(n+1) (p_{\sigma(2)}+p_{\sigma(3)})+i(m+1)p_{\sigma(3)}} \;.
\end{align}
Similarly to \eqref{scaledd}, the scaling of the three-magnon three-magnon position-dependent corrections is defined as
\begin{align}
       \tilde D_{\sigma}^{n,m}=e^{-inp_{\sigma(1)}+imp_{\sigma(3)}}D_{\sigma}^{n,m}\;.\label{eq:scaledDs}
   \end{align}
In this case the three-magnon generating function, defined as
\begin{align}
    \tilde{G}_\sigma^{(3)}(x,y)=\sum_{n,m=0}^{\infty} \tilde D_\sigma^{n,m}x^ny^m \; ,
\end{align}
admits the following form when we impose the non-interacting eigenvalue equations,
   \begin{align}
       \tilde{G}^{(3)}_\sigma(x,y)=\frac{F^{(3)}_\sigma(x,y)}{Q_3(x,y)}\; ,\label{genfunc3}
   \end{align}
   where the denominator is given by the polynomial
   \begin{align}
       Q_3(x,y)=\EEt x y-\invPPt x^2-\PPt y^2-x-y-x^2 y- y^2 x\;,\label{q3}
   \end{align}
   and the numerator depends on four single-variable functions, which are partially fixed by the remaining eigenvalue equations,\footnote{Notice that there is a non-trivial relation between the notation of \cite{Bozkurt:2024tpz} and this one, such that $g_{1,\sigma}(x)=\tilde{G}_\sigma(x,0)$ and $g_{3,\sigma}(x)=\tilde{G}_\sigma(0,y)$ however
\begin{align}
    g_{2,\sigma}(x)+\tilde D_\sigma^{0,1}=\partial_y \tilde{G}_\sigma(x,0)\;,\quad\quad g_{4,\sigma}(y)+\tilde D_\sigma^{1,0}=\partial_x \tilde{G}_\sigma(0,y)
\end{align}}
\begin{align}
   F^{(3)}_\sigma(x,y)=\left[Q_3(x,y)+\PPt y^2\left(1 +\invPPt x \right)\right]\tilde G^{(3)}_{\sigma}(x,0)+\left[Q_3(x,y)+\invPPt x^2\left( 1 +\PPt y\right) \right]\tilde G^{(3)}_{\sigma}(0,y) \nonumber \\
       -xy\left(1+\invPPt x\right)\partial_y\tilde G^{(3)}_{\sigma}(x,0)-xy\left(1+\PPt y\right)\partial_x\tilde G^{(3)}_{\sigma}(0,y)+xy\left(\tilde D^{0,1}_{\sigma}+ \tilde D^{1,0}_{\sigma}\right)\label{eq:generating-non-int3}\; .
   \end{align}
   To ensure analyticity, we make use of the observation that the denominator of the generating function factorizes as
   \begin{align}
        Q(x,y)=(x-\rho^{(3)}_+(y,p))(x-\rho^{(3)}_-(y,p))\;,
    \end{align}
   such that
  \begin{align}
       \rho^{(3)}_\pm(y,p)=\frac{-\PPt\left( y^2-\EEt  y+1\right)\pm\sqrt{e^{2i\mathcal{P}_3}\left(y^2-\EEt  y+1\right)^2-4\PPt y (1+\PPt y)^2}}{2 (1+\PPt y)}\;.\label{eq:elpara}
   \end{align}
   With respect to the curve defined above, demanding analyticity of $\tilde{G}^{(3)}_\sigma$ at $x=y=0$ is equivalent to demanding
   \begin{align}
       F^{(3)}_\sigma(\rho_+(y),y)=0\label{analythree}\;,
   \end{align}
   as only the branch $\rho_+$ contains the point $x=y=0$. This condition can be imposed before or after solving the interaction equations. Moreover, the two-magnon interaction equations imply the following relations between the single-variable functions appearing in $F^{(3)}$. For the left two-magnon interaction, we have
   \begin{align}
       &\left(e^{ip_i}\tilde G^{(3)}_{jik}(0,y)+e^{ip_j}\tilde G^{(3)}_{ijk}(0,y)\right)\left(2\kappa - \EEt+ \frac{1}{y} +y\right)\nonumber\\&\quad\quad\quad\quad+\left(e^{ip_i}\partial_x\tilde G^{(3)}_{jik}(0,y)+e^{ip_j}\partial_x\tilde G^{(3)}_{ijk}(0,y)\right) \left(1+\PPt y\right)=C_{12}(y)\label{eq:two-body-rec12}\;,
   \end{align}
   such that
   \begin{align}
       &C_{12}(y)=\left(e^{ip_1}\tilde D_{213}^{0,1}+e^{ip_2}\tilde D_{123}^{0,1}\right)+\left(e^{ip_1}\tilde D_{213}^{1,0}+e^{ip_2}\tilde D_{123}^{1,0}\right)\nonumber\\&\quad\quad\quad\quad\quad\quad\quad+ \frac{y}{1-e^{ip_3}y} \left[\left(e^{ip_1}\tilde D_{213}^{0,2}+e^{ip_2}\tilde D_{123}^{0,2}\right)\right.+\left(e^{ip_1}\tilde D_{213}^{1,1}+e^{ip_2}\tilde D_{123}^{1,1}\right)\nonumber\\&\quad\quad\quad\quad\quad\quad\quad\left.+\PPt\left(e^{ip_1}\tilde D_{213}^{1,0}+e^{ip_2}\tilde D_{123}^{1,0}\right)-(\EEt - 2\kappa)\left(e^{ip_1}\tilde D_{213}^{0,1}+e^{ip_2}\tilde D_{123}^{0,1}\right) \right]\label{eq:const12}\;.
   \end{align}
   And for the right two-magnon interaction equations we obtain,
   \begin{align}
       &\left(e^{ip_j}\tilde G^{(3)}_{ikj}(x,0)+e^{ip_k}\tilde G^{(3)}_{ijk}(x,0)\right)\left(2\kappa^{-1} - \EEt+ \frac{1}{x} +x\right)\nonumber\\&\quad\quad\quad\quad+\left(e^{ip_j}\partial_y\tilde G^{(3)}_{ikj}(x,0)+e^{ip_k}\partial_y\tilde G^{(3)}_{ijk}(x,0)\right)\left(1+\invPPt x\right)=\tilde C_{12}(x)\label{eq:two-body-rec21}\;,
   \end{align}
   such that
   \begin{align}
       &\tilde C_{12}(x)=\left(e^{ip_1}\tilde D_{321}^{1,0}+e^{ip_2}\tilde D_{312}^{1,0}\right)+\left(e^{ip_1}\tilde D_{321}^{0,1}+e^{ip_2}\tilde D_{312}^{0,1}\right)\nonumber\\&\quad\quad\quad\quad\quad\quad\quad+ \frac{x}{1-e^{-ip_3}x} \Big[\left(e^{ip_1}\tilde D_{321}^{2,0}+e^{ip_2}\tilde D_{312}^{2,0}\right)- (\EEt - 2\kappa^{-1})\left(e^{ip_1}\tilde D_{321}^{1,0}+e^{ip_2}\tilde D_{312}^{1,0}\right)\nonumber\\&\quad\quad\quad\quad\quad\quad\quad+\left(e^{ip_1}\tilde D_{321}^{1,1}+e^{ip_2}\tilde D_{312}^{1,1}\right)+\invPPt\left(e^{ip_1}\tilde D_{321}^{0,1}+e^{ip_2}\tilde D_{312}^{0,1}\right) \Big]\;.\label{eq:const21}
   \end{align}
   We can obtain the rest of the two-magnon interaction equations by permuting the indices of the generating functions and the momentum variables.

   The process of constructing the three-magnon special solution are very similar to the one described in section \ref{sec:special}, so we only provide here the final results. We start by imposing the factorization of scattering coefficients,
   \begin{align}
    &\frac{A_{213}}{A_{123}}=S_{\kappa}(p_1,p_2)\;,\quad  \frac{A_{312}}{A_{132}}=S_{\kappa}(p_1,p_3)\;,\quad\frac{A_{321}}{A_{231}}=S_{\kappa}(p_2,p_3)\;, \label{eq:skappafac} \\ &\frac{A_{132}}{A_{123}}=S_{\kappa}(p_2,p_3)\;,\quad
    \frac{A_{231}}{A_{213}}=S_{\kappa}(p_1,p_3)\;,\quad 
    \frac{A_{321}}{A_{312}}=S_{\kappa}(p_1,p_2)\;,\label{eq:skappafac-corr}
\end{align}
together with the partial factorization
\begin{align}
    \tilde D_{jik}^{n,m}= S_{\kappa}(p_i,p_j)  \tilde{D}_{ijk}^{n,m} \;,
\end{align}
which translates to the following expression at the level of the generating function
\begin{align}
    G_{jik}(x,y)=S_\kappa(p_i,p_j)G_{ijk}(x,y)\label{parfac3}\;.
\end{align}
Then, we choose the following initial conditions for the position-dependent corrections of minimally separated magnons
\begin{align}
    \tilde D_\sigma^{0,1}=\tilde D_\sigma^{1,0}&=\left(\kappa-\frac{1}{\kappa}\right)A_\sigma\;,
\end{align}
which solve the three-magnon interaction equation. The two-magnon interaction equation with minimum separation can be solved if we choose
\begin{gather}
       \tilde D_\sigma^{1,1}=0\;,\quad\quad \tilde D_\sigma^{0,2}=\left(\kappa-\kappa^{-1}\right)\left(\EEt-\PPt-2\kappa\right)A_\sigma \;,\\
       \tilde D_\sigma^{2,0}=-\left(\kappa-\kappa^{-1}\right)\kappa^{-1} f_{20}\left(p_{\sigma(3)}\right)A_\sigma\;,
   \end{gather}
   such that  \begin{align}
       f_{20}(p_3)=\left[e^{-i p_3} \left(\EEt+\PPt +\invPPt -2 \kappa \right) -\kappa \left(3\EEt-2 e^{ip_3} +\invPPt  -4 \kappa -2\kappa^{-1} \right)\right]\;.
   \end{align}
Solving the two-magnon interaction equations gives us
\begin{align}
       &\tilde G^{(3)}_{\sigma}(0,y)\left(2\kappa - \EEt+ \frac{1}{y} +y\right)=-\partial_x\tilde G^{(3)}_{\sigma}(0,y)\left(1+\PPt y\right)+\tilde D_{\sigma}^{0,1}+\tilde D^{1,0}_{\sigma}\;,\label{sol3y}
\end{align}
and
\begin{align}
       \tilde G^{(3)}_{\sigma}(x,0)\left(2\kappa^{-1} - \EEt+ \frac{1}{x} +x\right)=-\partial_y\tilde G^{(3)}_{\sigma}(x,0)\left(1+\invPPt x\right)+c_{\sigma}(x)\label{sol3x}\;,
   \end{align}
where $c_{\sigma}(x)$ is given by
\begin{multline}
    c_{123}(x)=\frac{(\kappa -\kappa^{-1}) x}{ \left(x-e^{i p_1}\right) \left(x-e^{i p_2}\right) \left(x-e^{i p_3}\right)\left(\kappa -2 \kappa  x \EEt+x (\kappa  x+2)\right) }\Bigg(\frac{e^{i (p_1+p_2)}-2 \kappa  x+1}{ e^{i p_3} (\kappa -\kappa^{-1})}\\
    +\PPt  \left[x +x \left( e^{i p_1}-2 \kappa \right) \left( e^{i p_2}-2 \kappa \right)-2 \kappa\right]+
    x \left(e^{i p_1}+e^{i p_2} -2 \kappa \right) \left(e^{i (p_1+p_2)}-2 \kappa  x+1\right)\Bigg)\label{cijkx}\;.
\end{multline}
Once we implement all the minimum separation equations and the solution of interacting equations given above to the generating function given in \eqref{genfunc3} and impose analyticity as stated in \eqref{analythree} we obtain the final form of three-magnon solution. Note that in this solution $\partial_y G_\sigma(x,0)$ is not fixed which we will use in Section \ref{sec:specialspecial}.

By using the final form of this solution we can define the function,
\begin{align}
    f^{(3)}(p_1,p_2,p_3;n,m,\kappa)=-\oint_\Sigma \frac{dx dy}{4\pi^2}\frac{G_{123}(x,y)}{x^{n+1}y^{m+1}}\;,\label{knownfunction}
\end{align}
which gives us a common description of all three-magnon position-dependent corrections,
\begin{align}
\label{eqn:SpecialDs}
    D_{123}^{n,m}&=f(p_1,p_2,p_3;n,m;\kappa)\;,\\
    D_{132}^{n,m}&=S_\kappa(p_2,p_3)f(p_1,p_3,p_2;n,m;\kappa)\;,\\
    D_{213}^{n,m}&=S_\kappa(p_1,p_2)f(p_2,p_1,p_3;n,m;\kappa)\;,\\
    D_{231}^{n,m}&=S_\kappa(p_1,p_3)S_\kappa(p_1,p_2)f(p_2,p_3,p_1;n,m;\kappa)\;,\\
    D_{312}^{n,m}&=S_\kappa(p_1,p_3)S(p_2,p_3)f(p_3,p_1,p_2;n,m;\kappa)\;,\\
    D_{321}^{n,m}&=S_{\kappa}(p_2,p_3)S_\kappa(p_1,p_3)S_\kappa(p_1,p_2)f(p_3,p_2,p_1;n,m;\kappa)\;.\label{eqn:SpecialDs1}
\end{align}
Finally, for this solution the Yang operators take the following form,
\begin{align}
   Y_j^{n,m}(p_1,p_2,p_3)= \begin{pmatrix}
        S_\kappa(p_1,p_2)\frac{1+f(p_2,p_1,p_3;n,m;\kappa)}{1+f(p_1,p_2,p_3;n,m;\kappa)} & 0 \\ 0 & S_\kappa(p_2,p_1)\frac{1+f(p_3,p_1,p_2;m,n;\kappa)}{1+f(p_3,p_2,p_1;m,n;\kappa)}\end{pmatrix} \; ,\label{eq:yangskappa}
\end{align}
and for all $n,m$ they satisfy the modified tower of YBEs
\begin{multline}
    Y_j^{n,m}(p_2,p_3,p_1)Y_{j+1}^{n,m}(p_1,p_3,p_2)Y_j^{n,m}(p_1,p_2,p_3)\\
    =Y_{j+1}^{n,m}(p_1,p_2,p_3)Y_j^{n,m}(p_1,p_3,p_2)Y_{j+1}^{n,m}(p_2,p_3,p_1)\label{eq:towerofYBE}\;.
\end{multline}
\subsection{$S_\kappa$ Solution for $Q_{21}Q_{12}Q_{21}$}
\label{sec:q21}
For the ordering $Q_{21}Q_{12}Q_{21}$, the long-range Bethe ansatz is given by
\begin{align}
    \ket{\Psi(p_1,p_2,p_3)}_{21}&=\sum_{l_1<l_2<l_3}\Psi(p_1,p_2,p_3;l_1,l_2,l_3)\ket{Q_{21}(l_1),Q_{12}(l_2),Q_{21}(l_3)} \;, \\
    \Psi(p_1,p_2,p_3;l_1,l_2,l_3)&=e^{il_1\sum_i^3p_i}\sum_{\sigma\in\mathcal{S}_3}\left(A_\sigma+D_\sigma^{n,m}\right)e^{i(n+1) (p_{\sigma(2)}+p_{\sigma(3)})+i(m+1)p_{\sigma(3)}} \;.
\end{align}
Then, to distinguish from the previous case we denote the generating function as $H_\sigma(x,y)$ which has the same description for the non-interacting equations with the previous case \eqref{eq:generating-non-int3} and we impose the factorization of scattering coefficients, as it is given in \eqref{eq:skappafac} and \eqref{eq:skappafac-corr}. Instead of the partial factorization property of the previous case \eqref{parfac3}, we impose
\begin{align}
    H_{ikj}(x,y)=S_\kappa(p_j,p_k)H_{ijk}(x,y)\;.
\end{align}
Then we obtain the following relations coming from the two-magnon interaction equations,
\begin{align}
       &\tilde H^{(3)}_{\sigma}(0,y)\left(2\kappa^{-1} - \EEt+ \frac{1}{y} +y\right)=-\partial_x\tilde H^{(3)}_{\sigma}(0,y)\left(1+\PPt y\right)+c_\sigma(y)\;,\label{eq:confunc211}
\end{align}
and
\begin{align}
       \tilde H^{(3)}_{\sigma}(x,0)\left(2\kappa - \EEt+ \frac{1}{x} +x\right)=-\partial_y\tilde H^{(3)}_{\sigma}(x,0)\left(1+\invPPt x\right)+\tilde D_{\sigma}^{0,1}+\tilde D^{1,0}_{\sigma}\label{eq:confunc212}
   \end{align}
   where the constant function is given in \eqref{cijkx}.

\section{Additional Term in the Four-Magnon Generating Function} \label{sec:cijkl}

The function, $C_{ijkl}(x,y,z)$ given in \eqref{finalform} is a combination of the four-magnon single-variable generating functions fixed by the three-magnon problem such as \eqref{eq:solz2} and \eqref{eq:solx2} and the analyticity conditions that we need to impose to finalize the form of the four-magnon equation. Crucially, the final form of $C_{ijkl}(x,y,z)$ is also given in terms of the single-variable generating functions of three-magnon problem but, unlike the rest of the generating function $\tilde G_{ijkl}^{(4)}(x,y,z)$ given in \eqref{finalform}, the factors inside $C_{ijkl}(x,y,z)$ are evaluated on curves such as $y_+(z)$ in \eqref{curve1} and curves coming from the analyticity condition of non-interacting equations like $z_+(y,z)$ in \eqref{curvez}. Therefore, $C_{ijkl}(x,y,z)$ is required to fulfill the technical requirements of the generating function method for position-dependent corrections.

Here we present the form of the $C_{ijkl}(x,y,z)$ function before imposing the analyticity conditions \eqref{eq:first-analy} and \eqref{eq:second-analy} for the non-interacting equations, however we have already imposed the analyticity conditions for the two-magnon interaction equations to fix $\tilde G_{ijkl}(0,0,z)$ and $\tilde G_{ijkl}(x,0,0)$ as given in \eqref{eq:solz2} and \eqref{eq:solx2}. After giving this form we will step by step show how to impose the final set of analyticity conditions explained in Section \ref{sec:analy},
\begin{align}
    C_{ijkl}&(x,y,z)=-\frac{y z (1+\PP y-2 \kappa  x)\tilde f_3(y,z)}{y z (-\EE+2 \kappa +z)+y^2+y+z^2}\nonumber\\&-\frac{x y (1+y-2 \kappa  z)\tilde g_3(x,y)}{(\PP y+1) \left(\PP y+x^2\right)-\PP x y (\EE-2 \kappa )}\nonumber\\
    +\Bigg(&\frac{z^2 (1-2 \kappa  x) (2 y-\kappa  z) \left(x \left(2 y (\kappa  z (-\EE+\kappa +z)+\kappa +z)+\kappa  y^2+\kappa  z^2\right)+\kappa  x^2 y z+\kappa  y z\right)}{\kappa  (\kappa  (x z (-\EE+x+z)+x+z)+2 x z) \left(y z (-\EE+2 \kappa +z)+y^2+y+z^2\right)}\nonumber\\
    &+\frac{\PP y^2z^2\left(2y\kappa^{-1}-z\right)}{y z (-\EE+2 \kappa +z)+y^2+y+z^2}+\frac{\invPP x^2 z^2\left(2x\kappa-1\right)}{x z (-\EE+x+z)+x+z+2 x z\kappa^{-1}}\Bigg)\tilde G_{ijkl}(0,0,z)\nonumber\\
    +\Bigg(&\frac{\left(x^2-2 \PP x y\kappa^{-1}\right) \left(\EE x y z-\invPP x^2 z-\PP y^2 z-x^2 y z-x y^2-x y-y z\right)}{(\PP y+1) \left(\PP y+x^2\right)-\PP x y (\EE-2 \kappa )}\nonumber\\&+\frac{(x-2 \kappa  x z) \left(\EE x y z-\invPP x^2 z-x^2yz-x y z^2-x y-x z^2-y z\right)}{x z (-\EE+x+z)+2 x z\kappa^{-1}+x+z}\nonumber\\
    &+x\Big[y+ \invPP x z-2 y z(\kappa +\kappa^{-1} ) \Big]\Bigg)\tilde G_{ijkl}(x,0,0)\label{cijkl}\;.
\end{align}
Finally, we ensure analyticity of the entire four-magnon generating function by ruling out the over-counting of eigenvalue equations. Practically, we impose equations \eqref{eq:first-analy} and \eqref{eq:second-analy} to fix the only leftover freedom, given in terms of $\tilde f_3(y,z)$ and $\tilde g_3(x,y)$. The resulting solution satisfies all eigenvalue equations and asymptotic behavior required by the smearing limit. 

To impose the analyticity conditions, it is convenient to separate the contributions proportional to $\tilde{f}_3$ and $\tilde{g}_3$ and define,
\begin{align}
    \tilde C_{ijkl}(x,y,z)=C_{ijkl}(x,y,z)&+\frac{y z (1+\PP y-2 \kappa  x)\tilde f_3(y,z)}{y z (-\EE+2 \kappa +z)+y^2+y+z^2}\nonumber\\&\;+\frac{x y (1+y-2 \kappa  z)\tilde g_3(x,y)}{(\PP y+1) \left(\PP y+x^2\right)-\PP x y (\EE-2 \kappa )}\;.
\end{align}
Final step is to impose analyticity conditions given in \eqref{eq:first-analy} and \eqref{eq:second-analy}. First we use \eqref{eq:first-analy} condition to solve for the expression $\tilde g_3(x,y)$, keeping $\tilde f_3(y,z)$ undetermined, and \eqref{eq:second-analy} condition to solve for $\tilde f_3(x,y)$, keeping $\tilde g_3(y,z)$ undetermined. With this, we can solve for $\tilde f_3(x,y)$ order by order by demanding the compatibility of these two equations.

Recall that in \eqref{finalform} we have the following expression once we cancel the $Q_4(x,y,z)$ from the denominator,
\begin{align}
    \tilde F_{ijkl}^{(4)}(x,y,z)=&\frac{r_1\,\tilde G_{ijk}^{(3)}(x,0)+r_2\,\tilde G_{ijk}^{(3)}(0,y)+r_3\,\partial_y\tilde G_{ijk}^{(3)}(x,0)+r_4\,\partial_x\tilde G_{ijk}^{(3)}(0,y)+x y z\, c_{ijk}(x)}{(1-e^{ip_{l}}z)}\nonumber\\
    &+\frac{r_5\,\tilde H_{jkl}^{(3)}(y,0)+r_6\,\tilde H_{jkl}^{(3)}(0,z)+r_7\,\partial_z\tilde H_{jkl}^{(3)}(y,0)+r_8\,\partial_y\tilde H_{jkl}^{(3)}(0,z)+x y z\, c_{jkl}(z)}{(1-e^{-ip_{i}}x)}\nonumber\\
    &+\frac{2x y z A_{ijkl} \left(\kappa-\kappa^{-1}\right) \left(1- e^{i p_l}z-e^{-i p_i}x\right)}{(1-e^{ip_{l}}z)(1-e^{-ip_{i}}x)}-\frac{y z (1+\PP y-2 \kappa  x)\tilde f_3(y,z)}{y z (-\EE+2 \kappa +z)+y^2+y+z^2}\nonumber\\&-\frac{x y (1+y-2 \kappa  z)\tilde g_3(x,y)}{(\PP y+1) \left(\PP y+x^2\right)-\PP x y (\EE-2 \kappa )}+\tilde C_{ijkl}(x,y,z)\;.
\end{align}
such that the first analyticity condition \eqref{eq:first-analy}, given as
\begin{align}
    \tilde F_{ijkl}^{(4)}(x,y,z_+(y,z))=0\;,
\end{align}
implies that
\begin{align}
    \tilde g_3(x,y)=\frac{(\PP y+1) \left(\PP y+x^2\right)-\PP x y (\EE-2 \kappa )}{xy(1+y-2 \kappa  z_+(y,z))}\Bigg[\frac{r_1\,\tilde G_{ijk}^{(3)}(x,0)+\cdots+r_4\,\partial_x\tilde G_{ijk}^{(3)}(0,y)}{(1-e^{ip_{l}}z)}\nonumber\\+\cdots+\tilde C_{ijkl}(x,y,z)-\frac{y z (1+\PP y-2 \kappa  x)\tilde f_3(y,z)}{y z (-\EE+2 \kappa +z)+y^2+y+z^2}\Bigg]_{z\to z_+(x,y)} \;. \label{solg3}
\end{align}
Then, we can use this result and impose the second analyticity condition \eqref{eq:second-analy}, given as,
\begin{align}
    \tilde F_{ijkl}^{(4)}(x_+(y,z),y,z)=0 \;,
\end{align}
which implies
\begin{align}
    \tilde f_3(y,z)=\frac{y z (-\EE+2 \kappa +z)+y^2+y+z^2}{y z (1+\PP y-2 \kappa   x_+(y,z))}\Bigg[\frac{r_1\,\tilde G_{ijk}^{(3)}(x,0)+\cdots+r_4\,\partial_x\tilde G_{ijk}^{(3)}(0,y)}{(1-e^{ip_{l}}z)}\nonumber\\+\cdots+\tilde C_{ijkl}(x,y,z)\frac{x y (1+y-2 \kappa  z)\tilde g_3(x,y)}{(\PP y+1) \left(\PP y+x^2\right)-\PP x y (\EE-2 \kappa )}\Bigg]_{x\to x_+(y,z)} \;,\label{solf3}
\end{align}
such that we input the solution of $\tilde g_3$ as given in \eqref{solg3}. Inserting the solution of $\tilde g_3$ into the above expression results in having $\tilde f_3$ in both sides of this equation. However, the $\tilde f_3$ function on the right-hand-side will be evaluated on the curve $z_+(x,y)$. We solve this equation order by order in $y$ and $z$. After imposing the analyticity condition we conclude that the solution of the four-magnon problem is given in terms of the three-magnon solution pieces however imposing analyticity adds extra factors to the generating function which only depends on the three-magnon pieces but evaluated on the curves which are defined by the vanishing of the denominator \eqref{eq:def-q4}.

In this process, indeed all position-dependent corrections are fixed. Remarkably, all polynomial expressions that constitute $r_i$s have only total energy, $\EE$ and total momentum $\PP$ as their coefficients apart from factors $\left(1-e^{-ip_{i}}x\right)$ and $\left(1-e^{ip_{4}}z\right)$ and final form of $C_{ijkl}$, additionally only includes scattering coefficients $A_{ijkl}$.
\newpage
	 \bibliographystyle{JHEP}
	 \bibliography{longchains}

\end{document}